\documentclass[12pt]{article}

  \usepackage{graphicx}
  \usepackage{epsfig}
  \usepackage{amssymb}
  \usepackage{subfigure}
  \usepackage{axodraw}
\evensidemargin - 1.truecm
\begin{document}

\newcommand{\be}{\begin{equation}}
\newcommand{\ee}{\end{equation}}
\newcommand{\nn}{\nonumber}
\newcommand{\bea}{\begin{eqnarray}}
\newcommand{\eea}{\end{eqnarray}}
\newcommand{\bfig}{\begin{figure}}
\newcommand{\efig}{\end{figure}}
\newcommand{\bc}{\begin{center}}
\newcommand{\ec}{\end{center}}
\newcommand{\bd}{\begin{displaymath}}
\newcommand{\ed}{\end{displaymath}}

\begin{titlepage}
\nopagebreak
{\flushright{
        \begin{minipage}{5cm}
        Freiburg-THEP 04/07\\
 UCLA/04/TEP/18\\
 CERN-PH-TH/2004-087\\
        {\tt hep-ph/0405275}\\
        \end{minipage}        }

}
\vspace*{-1.5cm}                        
\vskip 3.5cm
\begin{center}
\boldmath
{\Large \bf Two-Loop  $N_{F}=1$  
 QED Bhabha Scattering \\ [2mm] Differential Cross Section}\unboldmath
\vskip 1.2cm
{\large  R.~Bonciani$\rm \, ^{a, \,}$\footnote{Email: 
{\tt Roberto.Bonciani@physik.uni-freiburg.de}}}
{\large A.~Ferroglia$\rm \, ^{a, \,}$\footnote{Email: 
{\tt Andrea.Ferroglia@physik.uni-freiburg.de}}},
{\large P.~Mastrolia$\rm \, ^{b, \,}$\footnote{Email: 
{\tt mastrolia@physics.ucla.edu}}}, \\[2mm] 
{\large E.~Remiddi$\rm \, ^{c, \, d, \,}$\footnote{Email: 
{\tt Ettore.Remiddi@bo.infn.it}}},
and  {\large  J. J. van der Bij$\rm \, ^{a, \,}$\footnote{Email: 
{\tt jochum@physik.uni-freiburg.de}}}
\vskip .7cm
{\it $\rm ^a$ Fakult\"at f\"ur Mathematik und Physik, 
Albert-Ludwigs-Universit\"at
Freiburg, \\ D-79104 Freiburg, Germany} 
\vskip .3cm
{\it $\rm ^b$ Department of Physics and Astronomy, UCLA,
Los Angeles, CA 90095-1547} 
\vskip .3cm
{\it $\rm ^c$ Physics Department, Theory Division, CERN, CH-1211 Geneva 23, 
Switzerland} 
\vskip .3cm
{\it $\rm ^d$ Dipartimento di Fisica dell'Universit\`a di Bologna, and
INFN, Sezione di Bologna, I-40126 Bologna, Italy} 
\end{center}
\vskip 1.5cm


\begin{abstract} 
We calculate the two-loop virtual, UV renormalized corrections at
order $\alpha^4 (N_F=1)$ in QED to the Bhabha scattering differential 
cross section, for arbitrary values of the squared c.m. energy $s$ 
and momentum transfer $t$, and on-shell electrons and positrons of 
finite mass $m$. The calculation is carried out within the dimensional
regularization scheme; the remaining IR divergences appear as polar 
singularities in $(D-4)$. The result is presented in terms of 1- and  
2-dimensional harmonic polylogarithms, of maximum weight 3.

\vskip .7cm 
\flushright{
        \begin{minipage}{12.3cm}
{\it Key words}: Feynman
diagrams, Multi-loop calculations,  Box diagrams, \hspace*{2cm} Bhabha 
scattering 
        \end{minipage}        }
\end{abstract}
\vfill
\end{titlepage}

\section{Introduction \label{Intro}}

The Bhabha scattering,  $e^+ e^- \rightarrow e^+ e^-$,   plays a special role
in the phenomenology of particle physics, since it is employed to determine the
luminosity of $e^+ e^-$ colliders operating  at both high ($\sim 100 \,
\mbox{GeV}$) and intermediate  ($\sim 1-10 \, \mbox{GeV}$) center of mass
energies. The luminosity of such facilities is essentially determined by
calculating the (inverse) ratio of the theoretical Bhabha
scattering cross section to the number of Bhabha scattering events observed.
The Bhabha scattering is a  process particularly suited for luminosity
monitoring purposes, since its cross section is large and QED dominated (this is
true  in specific kinematic regions: the small scattering angle region for what
concerns  high energy colliders, the large scattering angle configuration for
machines operating in the  $\sim 1-10 \, \mbox{GeV}$ center of mass energy
range). As a consequence, this process can be calculated in perturbation  theory
with a high degree  of accuracy. Since the theoretical error on the Bhabha
scattering cross section  affects directly the precision of the
luminosity of an $e^+ e^-$  collider, a remarkable amount of
work has been devoted to  the study of the radiative corrections to this
process, both virtual and real,  in the last three decades. For an extensive
list of references  on the subject  we refer the interested reader to the
review in~\cite{mont}.

In this paper, we focus our attention on the technical problem of the
diagrammatic calculation of the \emph{virtual} corrections to the   Bhabha
scattering cross section. The one-loop radiative corrections to  this process,
together with the corresponding real corrections  involving the emission of a
soft photon, calculated within the framework of the full electroweak  Standard
Model, are well known   \cite{Bhabha1loop}. At the two-loop level, the
calculation of the virtual corrections becomes much more involved. Even in pure
QED, the complete (non approximated) two-loop differential cross section, i.~e.
the two-loop amplitude interfered with the tree level matrix element and
summed  over all spins, has not yet been calculated.  
A large amount of work was devoted to the study of the contributions 
enhanced by factors of $\ln(s/m_e^2)$ \cite{TUTTI}. In recent years, the
complete two-loop QED virtual cross section has been obtained in the limit 
of zero electron mass  \cite{Bhabha2loop}. The IR divergent structure of 
this result has been studied in detail in \cite{Bas}.

The main technical challenge encountered in calculating  the complete,
non approximated, set of two-loop virtual correction is represented by the
evaluation of the box-diagrams. Recently, we proved that the calculation of 
the two-loop box diagrams involving a closed fermion loop can be reduced to 
the calculation of a set of 14 scalar Master Integrals (MIs) \cite{us}. 
This result has been obtained by means of the (by now standard) Laporta 
algorithm \cite{laporta}, for the systematic exploitation of the 
Integration-By-Parts (IBPs) \cite{Tkac}, Lorentz invariance (LIs) \cite{Rem3}, 
and general symmetry \cite{RoPieRem1} identities. The MIs involved in the 
calculation have  been evaluated in \cite{us, RoPieRem1}, within the context 
of $D$-dimensional regularization scheme \cite{DimReg}, and by employing 
the method of the differential equations in the external kinematical variables
\cite{Kot, Rem1, Rem2}. The result was conveniently expressed in terms of 
1- and 2-dimensional harmonic polylogarithms (HPLs, 2dHPLs) 
\cite{Polylog,Polylog3,Polylog2,Polylog4,Polylog5}, of maximal weight 3.

The results in \cite{us, RoPieRem1} completed the calculation of all the MIs
that are necessary in order to evaluate the two-loop Bhabha scattering
differential cross section in pure QED, if we limit our analysis to diagrams
involving a closed fermion loop (conventionally referred to as the  $\alpha^4
(N_F = 1)$ differential cross section). In fact, the relevant  vacuum
polarization diagrams have been calculated in \cite{paolo}.  The two-loop
vertex  corrections involving a closed electron loop have been obtained in
\cite{BMR,RoPieRem2}, where the authors present a complete calculation 
of the vertex form factors in QED, without neglecting the electron mass. 

The purpose of the present paper is to provide the  expression  of the
contribution of the virtual corrections to the Bhabha scattering  differential
cross section at order $\alpha^4 (N_F =1)$ in QED.  The UV renormalization of
all  the two-loop diagrams is carried out within the $D$-continuous 
dimensional regularization in the on-shell renormalization 
scheme. The residual IR divergence in the final result appears as a pole in
$(D-4)$, to be canceled, as usual, by the corresponding divergence due to the
real emission of soft photons in the complete expression of the 
physical cross section. 

The paper is organized as follows. In Section~\ref{Stl} we describe the various
steps of the calculation,  introducing our notations and conventions.  In
Section~\ref{Ren} we recall the main points of the  UV on-shell
renormalization scheme. In Section~\ref{S1l} we re-derive, for completeness, 
the one-loop virtual corrections and the Bhabha scattering differential cross
section at order $\alpha^3$. The main result of this paper, the virtual
differential cross section at order $\alpha^4 (N_F =1)$, is presented in
Section~\ref{S2l}.  In Appendix \ref{AF} we collect the explicit expressions of
the auxiliary functions introduced in the paper in terms of HPLs, and we
briefly review the analytical continuation of the HPLs from the Euclidean 
to the physical kinematic region. In Appendix \ref{EXPANSIONS} we give the
expansions of the auxiliary functions in different kinematical regimes.

\section{The Bhabha scattering \label{Stl}} 

In this Section we summarize the strategy followed in carrying out the
calculation. 
In order to fix the notation and conventions employed throughout the paper,
we start by discussing the tree-level Bhabha scattering
differential cross section.

We consider the following scattering process mediated by photon exchange:
\be
e^{-}(p_1) + e^{+}(p_2) \rightarrow e^{-}(p_3) + e^{+}(p_4) \, ,
\ee

The incoming electron, incoming positron, outgoing electron and outgoing 
positron have momenta $p_1,p_2,p_3$ and $p_4$, respectively, and finite 
mass $m$. The process is well described in terms of the Mandelstam 
kinematic invariants $s$, $t$ and $u$, defined by the relations:
\bea
s &=& - P^2 \equiv - (p_1 + p_2)^2  = 4 E^2 \, , \\
t &=& - Q^2 \equiv - (p_1 - p_3)^2  = 
- 4 (E^2 - m^2) \sin^2{\frac{\theta}{2}} \, , \\
u &=& - V^2 \equiv - (p_1 - p_4)^2  = 
- 4 (E^2 - m^2) \cos^2{\frac{\theta}{2}}  \, ,
\eea
with
\be
s + t + u = 4 \, m^2 \, ,
\ee
and where $E$ is the particle energy in the center of mass frame of 
reference and $\theta$ is the scattering angle in the same frame. 

The differential cross section for the Bhabha scattering process can be
written, in perturbation theory, as an expansion in the fine-structure 
constant $\alpha$:
\be
\frac{d \sigma(s,t,m^2)}{d \Omega} =
                  \frac{d \sigma_{0}(s,t,m^2)}{d \Omega} 
                + \left( \frac{\alpha}{\pi} \right) 
           \frac{d \sigma_{1}(s,t,m^2)}{d \Omega}
                + \left( \frac{\alpha}{\pi} \right)^2 
    \frac{d \sigma_{2}(s,t,m^2)}{d \Omega}
 + {\mathcal O} 
        \left( \Biggl( \frac{\alpha}{\pi} \Biggr)^3 \right) ,
\ee
where $\sigma_{0}(s,t,m^2)$ is the tree-level (Born) cross section and 
$\sigma_{i}(s,t,m^2)$ ($i=1,2$) are the higher order contributions.
\begin{figure}[t]
\vspace*{.6cm}
\[\vcenter{
\hbox{
  \begin{picture}(0,0)(0,0)
\SetScale{1.5}
  \SetWidth{.3}
\ArrowLine(-45,20)(-25,20)
\ArrowLine(-25,20)(-5,20)
\ArrowLine(-25,-20)(-45,-20)
\ArrowLine(-5,-20)(-25,-20)
\Photon(-25,20)(-25,-20){2}{8}
\LongArrow(-60,16)(-45,16)
\LongArrow(-60,-15)(-45,-15)
\LongArrow(-5,16)(10,16)
\LongArrow(-5,-15)(10,-15)
\Text(-28,12)[cb]{{\footnotesize $e^-$}}
\Text(-28,-5)[cb]{{\footnotesize $e^+$}}
\Text(3,12)[cb]{{\footnotesize $e^-$}}
\Text(3,-5)[cb]{{\footnotesize $e^+$}}
\Text(-13,-18)[cb]{{\footnotesize $t$-channel }}
\Text(-35,-9)[cb]{{\footnotesize $p_2$}}
\Text(-35,7)[cb]{{\footnotesize $p_1$}}
\Text(10,-9)[cb]{{\footnotesize $p_4$}}
\Text(10,7)[cb]{{\footnotesize $p_3$}}
\end{picture}}  
}
\hspace{3.6cm}
  \vcenter{
\hbox{
  \begin{picture}(0,0)(0,0)
\SetScale{1.5}
  \SetWidth{.3}
\ArrowLine(20,0)(10,-20)
\ArrowLine(10,20)(20,0)
\Photon(20,0)(40,0){2}{6}
\ArrowLine(40,0)(50,20)
\ArrowLine(50,-20)(40,0)
\LongArrow(5,15)(10,5)
\LongArrow(5,-15)(10,-5)
\LongArrow(50,5)(55,15)
\LongArrow(50,-5)(55,-15)
\Text(0,12)[cb]{{\footnotesize $e^-$}}
\Text(0,-12)[cb]{{\footnotesize $e^+$}}
\Text(33,12)[cb]{{\footnotesize $e^-$}}
\Text(33,-12)[cb]{{\footnotesize $e^+$}}
\Text(17,-18)[cb]{{\footnotesize $s$-channel }}
\Text(0,-7)[cb]{{\footnotesize $p_2$}}
\Text(0,5)[cb]{{\footnotesize $p_1$}}
\Text(33,-7)[cb]{{\footnotesize $p_4$}}
\Text(33,5)[cb]{{\footnotesize $p_3$}}
\end{picture}}
}\]
\vspace*{.8cm}
\caption[]{\it Tree-level diagrams.}
\label{figT}
\end{figure}
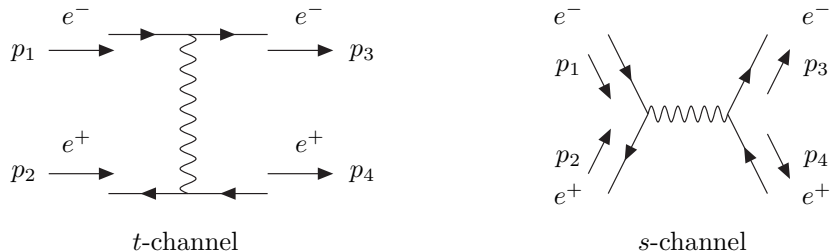

The $s$- and $t$-channel diagrams contributing to the scattering 
process at the tree level are shown in Fig.~\ref{figT}, where, 
as usual, the wavy lines represent photons and the arrows on the 
fermionic (solid) lines follow the flow of the negative charge. 

The tree-level amplitude, ${\mathcal M}_0$, is given by
\be
{\mathcal M}_0 = {\mathcal A}_t - {\mathcal A}_s \, , 
\label{MET}
\ee
where the matrix element of the $t$-channel ($s$-channel) diagram in 
Fig.~\ref{figT} is indicated by ${\mathcal A}_t$ (${\mathcal A}_s$).  In
Eq.~(\ref{MET}) we indicated explicitly the reciprocal negative  sign between
the $s$- and $t$- channel diagrams; in the following, the expressions of the
contribution of the $s$-channel virtual diagrams to the differential cross
section  are given including this extra minus sign.

After averaging the squared amplitude at order $\alpha^2$ over the spins of 
the initial states and summing over the spins of the final states, the 
differential Born cross section reads: 
\bea 
\frac{d \sigma_{0}(s,t,m^2)}{d \Omega} & = & 
                    \frac{\alpha^2}{s} \Bigg\{ \frac{1}{s^2} 
      \left[s t + \frac{s^2}{2} + (t - 2 m^2)^2 \right] 
  + \frac{1}{t^2} \left[ s t + \frac{t^2}{2} 
  + (s - 2 m^2)^2 \right]   \nn\\  
& & + \frac{1}{s t}\left[(s + t)^2 - 4 m^4\right] \Bigg\} \, .
\label{TL} 
\eea  
The first, second, and third term between curly brackets represent the 
contribution of the squared $s$-channel diagram, the squared $t$-channel 
diagram, and the $s$-$t$-diagram interference, respectively. 
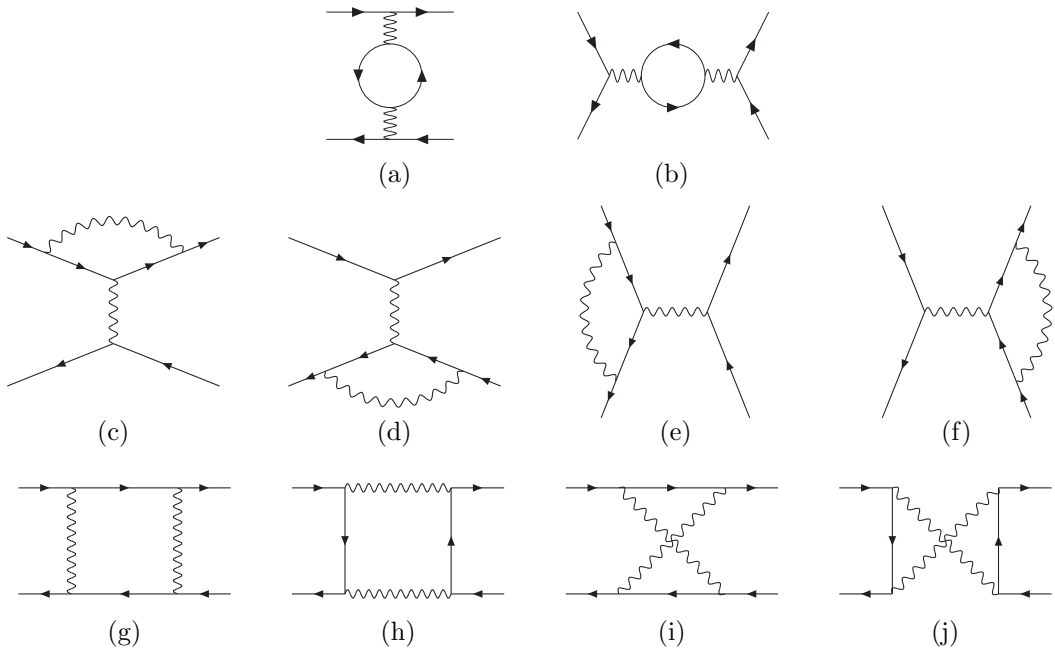
\begin{figure}[ht]
\vspace*{.5cm}
\[ \hspace*{-3mm}
\vcenter{
\hbox{
  \begin{picture}(0,0)(0,0)
\SetScale{1.2}
  \SetWidth{.3}
\ArrowLine(-45,20)(-25,20)
\ArrowLine(-25,20)(-5,20)
\ArrowLine(-25,-20)(-45,-20)
\ArrowLine(-5,-20)(-25,-20)
\Photon(-25,20)(-25,10){2}{4}
\Photon(-25,-10)(-25,-20){2}{4}
\ArrowArc(-25,0)(10,90,270)
\ArrowArc(-25,0)(10,270,90)
\Text(-10,-15)[cb]{{\footnotesize (a)}}
\end{picture}}  
}
\hspace{1.3cm}
  \vcenter{
\hbox{
  \begin{picture}(0,0)(0,0)
\SetScale{1.2}
  \SetWidth{.3}
\ArrowLine(10,0)(0,-20)
\ArrowLine(0,20)(10,0)
\Photon(10,0)(20,0){2}{3}
\Photon(40,0)(50,0){2}{3}
\ArrowLine(50,0)(60,20)
\ArrowLine(60,-20)(50,0)
\ArrowArc(30,0)(10,0,180)
\ArrowArc(30,0)(10,180,0)
\Text(12.5,-15)[cb]{{\footnotesize (b)}}
\end{picture}}
}
\]
\vspace*{1.7cm}
\[
  \vcenter{
\hbox{
  \begin{picture}(0,0)(0,0)
  \SetScale{0.8}
  \SetWidth{.5}
\ArrowLine(-50,35)(-30,27) 
\ArrowLine(-30,27)(0,15) 
\ArrowLine(0,15)(35,29) 
\ArrowLine(35,29)(50,35) 
\Photon(0,15)(0,-15){2}{5}
\PhotonArc(0,0)(43,41,139){2}{10}
\ArrowLine(0,-15)(-50,-35) 
\ArrowLine(50,-35)(0,-15) 
\Text(0,-18)[cb]{{\footnotesize (c)}}
\end{picture}}
}
\hspace{3.6cm}
  \vcenter{
\hbox{
\begin{picture}(0,0)(0,0)
  \SetScale{.8}
  \SetWidth{.5}
\ArrowLine(-50,35)(0,15) 
\ArrowLine(0,15)(50,35) 
\Photon(0,15)(0,-15){2}{5}
\PhotonArc(0,0)(43,-139,-41){2}{10}
\ArrowLine(-30,-27)(-50,-35) 
\ArrowLine(0,-15)(-30,-27) 
\ArrowLine(35,-29)(0,-15) 
\ArrowLine(50,-35)(35,-29) 
\Text(0,-18)[cb]{{\footnotesize (d)}}
\end{picture}}  
}
\hspace{3.6cm}
  \vcenter{\hbox{
\begin{picture}(0,0)(0,0)
  \SetScale{.8}
  \SetWidth{.5}
\ArrowLine(35,-50)(15,0) 
\ArrowLine(15,0)(35,50) 
\PhotonArc(0,0)(43,131,229){2}{10}
\Photon(15,0)(-15,0){2}{5}
\ArrowLine(-15,0)(-27,-30) 
\ArrowLine(-27,-30)(-35,-50) 
\ArrowLine(-35,50)(-27,30) 
\ArrowLine(-27,30)(-15,0) 
\Text(0,-18)[cb]{{\footnotesize (e)}}
\end{picture}}}
\hspace{3.6cm}
  \vcenter{\hbox{
\begin{picture}(0,0)(0,0)
  \SetScale{0.8}
  \SetWidth{.5}
\ArrowLine(27,-30)(15,0) 
\ArrowLine(35,-50)(27,-30) 
\ArrowLine(27,30)(35,50) 
\ArrowLine(15,0)(27,30) 
\PhotonArc(0,0)(43,-49,49){2}{10}

\Photon(15,0)(-15,0){2}{5}
\ArrowLine(-15,0)(-35,-50) 
\ArrowLine(-35,50)(-15,0) 
\Text(0,-18)[cb]{{\footnotesize (f)}}
\end{picture}}}
\]
\vspace*{1.8cm}
\[
\vcenter{\hbox{
  \begin{picture}(0,0)(0,0)
\SetScale{.8}
  \SetWidth{.5}
\ArrowLine(-50,25)(-25,25)
\ArrowLine(-25,25)(25,25)
\ArrowLine(25,25)(50,25)
\Photon(-25,25)(-25,-25){2}{10}
\Photon(25,25)(25,-25){2}{10}
\ArrowLine(-25,-25)(-50,-25)
\ArrowLine(25,-25)(-25,-25)
\ArrowLine(50,-25)(25,-25)
\Text(0,-14)[cb]{{\footnotesize (g)}}
\end{picture}}}
\hspace{3.5cm}
\vcenter{\hbox{
  \begin{picture}(0,0)(0,0)
\SetScale{.8}
  \SetWidth{.5}
\ArrowLine(-50,25)(-25,25)
\ArrowLine(25,25)(50,25)
\Photon(-25,-25)(25,-25){2}{10}
\Photon(-25,25)(25,25){2}{10}
\ArrowLine(-25,-25)(-50,-25)
\ArrowLine(50,-25)(25,-25)
\ArrowLine(-25,25)(-25,-25)
\ArrowLine(25,-25)(25,25)
\Text(0,-14)[cb]{{\footnotesize (h)}}
\end{picture}}}
\hspace{3.5cm}
\vcenter{\hbox{
  \begin{picture}(0,0)(0,0)
\SetScale{.8}
  \SetWidth{.5}
\ArrowLine(-50,25)(-25,25)
\ArrowLine(-25,25)(25,25)
\ArrowLine(25,25)(50,25)
\Photon(-25,25)(25,-25){2}{10}
\Photon(25,25)(-25,-25){2}{10}
\ArrowLine(-25,-25)(-50,-25)
\ArrowLine(25,-25)(-25,-25)
\ArrowLine(50,-25)(25,-25)
\Text(0,-14)[cb]{{\footnotesize (i)}}
\end{picture}}}
%
\hspace{3.5cm}
\vcenter{\hbox{
  \begin{picture}(0,0)(0,0)
\SetScale{.8}
  \SetWidth{.5}
\ArrowLine(-50,25)(-25,25)
\ArrowLine(25,25)(50,25)
\Photon(-25,25)(25,-25){2}{10}
\Photon(-25,-25)(25,25){2}{10}
\ArrowLine(-25,-25)(-50,-25)
\ArrowLine(50,-25)(25,-25)
\ArrowLine(-25,25)(-25,-25)
\ArrowLine(25,-25)(25,25)
\Text(0,-14)[cb]{{\footnotesize (j)}}
\end{picture}}}
\]
\vspace*{.6cm}
\caption[]{\it One-loop diagrams.}
\label{fig1ltot}
\end{figure}

If we consider only the virtual contributions at order $\alpha^3$,  
it is necessary to evaluate
the diagrams shown in Fig.~\ref{fig1ltot}. The corresponding
squared matrix element, summed over the spin of the initial particles
and averaged over the spins of the final ones, will be given by the 
interference of each diagram  in Fig.~\ref{fig1ltot} with the tree-level 
diagrams of Fig.~\ref{figT}. The differential cross section assumes the 
following form:
\be
\left(\frac{\alpha}{\pi}\right)\frac{d \sigma_{1}^{V}(s,t,m^2)}{d \Omega} = 
\frac{s}{16}\sum_{\mbox{spin}} \Biggl\{ \Biggl(
\hspace*{1.3cm}
\hbox{
  \begin{picture}(0,0)(0,0)
\SetScale{.8}
  \SetWidth{.5}
\ArrowLine(-45,23)(-25,23)
\ArrowLine(-25,23)(-5,23)
\ArrowLine(-25,-17)(-45,-17)
\ArrowLine(-5,-17)(-25,-17)
\Photon(-25,23)(-25,-17){2}{8}
\end{picture}}  
\hspace*{-.2cm} - \hspace*{-.2cm}
\hbox{
  \begin{picture}(0,0)(0,0)
\SetScale{.8}
  \SetWidth{.5}
\ArrowLine(20,3)(10,-17)
\ArrowLine(10,23)(20,3)
\Photon(20,3)(40,3){2}{6}
\ArrowLine(40,3)(50,23)
\ArrowLine(50,-17)(40,3)
\end{picture}}
\hspace*{1.7cm}
\Biggr)^* \times
\hspace*{0.8cm}\hbox{
  \begin{picture}(0,0)(0,0)
\SetScale{.65}
  \SetWidth{.5}
\ArrowLine(-50,28)(-25,28)
\ArrowLine(-25,28)(25,28)
\ArrowLine(25,28)(50,28)
\Photon(-25,28)(-25,-22){2}{9}
\Photon(25,28)(25,-22){2}{9}
\ArrowLine(-25,-22)(-50,-22)
\ArrowLine(25,-22)(-25,-22)
\ArrowLine(50,-22)(25,-22)
\end{picture}}
\hspace*{0.8cm} +  \mbox{c.c.} + \cdots \Biggr\} \, ,
\label{M2at1l}
\ee
where the superscript $V$ stands for ``virtual'' (and the sum is performed 
on all spins). The contributions of the various terms on the r.h.s. of 
Eq.~(\ref{M2at1l}) to the differential cross section will be discussed 
in Section~\ref{S1l}.

\begin{figure}
\vspace*{.3cm}
\[\vcenter{
\hbox{
  \begin{picture}(0,0)(0,0)
\SetScale{.8}
  \SetWidth{.5}
\ArrowLine(20,-55)(0,-40)
\ArrowLine(0,-40)(-20,-55)
\ArrowLine(0,40)(20,55)
\ArrowLine(-20,55)(0,40)
\Photon(0,-40)(0,-20){2}{4}
\Photon(0,20)(0,40){2}{4}
\ArrowArc(0,0)(20,-90,0)
\ArrowArc(0,0)(20,0,90)
\ArrowArc(0,0)(20,90,180)
\Photon(20,0)(-20,0){2}{8}
\ArrowArc(0,0)(20,180,270)
\Text(0,-20)[cb]{{\footnotesize (a)}}
\end{picture}}  
}
\hspace{4.4cm}
  \vcenter{
\hbox{
  \begin{picture}(0,0)(0,0)
\SetScale{.8}
  \SetWidth{.5}
\ArrowLine(20,-55)(0,-40)
\ArrowLine(0,-40)(-20,-55)
\Photon(0,-40)(0,-20){2}{4}
\Photon(0,20)(0,40){2}{4}
\ArrowLine(0,40)(20,55)
\ArrowLine(-20,55)(0,40)

\ArrowArc(0,0)(20,90,270)
\ArrowArc(0,0)(20,-60,60)
\ArrowArc(0,0)(20,60,90)
\ArrowArc(0,0)(20,-90,-60)
\PhotonArc(20,0)(15,-248,-112){2}{5}
\Text(0,-20)[cb]{{\footnotesize (b)}}
\end{picture}}
}
\hspace{4.4cm}
  \vcenter{
\hbox{
  \begin{picture}(0,0)(0,0)
\SetScale{.8}
  \SetWidth{.5}
\ArrowLine(20,-55)(0,-40)
\ArrowLine(0,-40)(-20,-55)
\Photon(0,-40)(0,-20){2}{4}
\Photon(0,20)(0,40){2}{4}
\ArrowLine(0,40)(20,55)
\ArrowLine(-20,55)(0,40)

\ArrowArc(0,0)(20,-90,90)
\ArrowArc(0,0)(20,90,120)
\ArrowArc(0,0)(20,120,240)
\ArrowArc(0,0)(20,240,270)
\PhotonArc(-20,0)(15,-68,68){2}{5}
\Text(0,-20)[cb]{{\footnotesize (c)}}
\end{picture}}
}\]
\vspace*{1.5cm}
\[\vcenter{
\hbox{
  \begin{picture}(0,0)(0,0)
\SetScale{.8}
  \SetWidth{.5}
\ArrowLine(-55,20)(-40,0)
\ArrowLine(-40,0)(-55,-20)
\ArrowLine(40,0)(55,20)
\ArrowLine(55,-20)(40,0)
\Photon(-40,0)(-20,0){2}{4}
\Photon(20,0)(40,0){2}{4}
\ArrowArc(0,0)(20,0,90)
\ArrowArc(0,0)(20,90,180)
\ArrowArc(0,0)(20,180,270)
\Photon(0,20)(0,-20){2}{8}
\ArrowArc(0,0)(20,270,360)
\Text(0,-12)[cb]{{\footnotesize (d)}}
\end{picture}}  
}
\hspace{4.4cm}
  \vcenter{
\hbox{
  \begin{picture}(0,0)(0,0)
\SetScale{.8}
  \SetWidth{.5}
\ArrowLine(-55,20)(-40,0)
\ArrowLine(-40,0)(-55,-20)
\ArrowLine(40,0)(55,20)
\ArrowLine(55,-20)(40,0)
\Photon(-40,0)(-20,0){2}{4}
\Photon(20,0)(40,0){2}{4}
\ArrowArc(0,0)(20,0,30)
\ArrowArc(0,0)(20,30,150)
\ArrowArc(0,0)(20,150,180)
\PhotonArc(0,20)(15,-158,-22){2}{5}
\ArrowArc(0,0)(20,180,360)
\Text(0,-12)[cb]{{\footnotesize (e)}}
\end{picture}}
}
\hspace{4.4cm}
  \vcenter{
\hbox{
  \begin{picture}(0,0)(0,0)
\SetScale{.8}
  \SetWidth{.5}
\ArrowLine(-55,20)(-40,0)
\ArrowLine(-40,0)(-55,-20)
\Photon(-40,0)(-20,0){2}{4}
\Photon(20,0)(40,0){2}{4}
\ArrowLine(40,0)(55,20)
\ArrowLine(55,-20)(40,0)

\ArrowArc(0,0)(20,0,180)
\ArrowArc(0,0)(20,180,210)
\ArrowArc(0,0)(20,210,330)
\ArrowArc(0,0)(20,330,360)
\PhotonArc(0,-20)(15,22,158){2}{5}
\Text(0,-12)[cb]{{\footnotesize (f)}}
\end{picture}}
}\]
\vspace*{1.3cm}
%

\[
  \vcenter{
\hbox{
  \begin{picture}(0,0)(0,0)
  \SetScale{0.8}
  \SetWidth{.5}
\ArrowLine(-50,35)(-30,27) 
\ArrowLine(-30,27)(0,15) 
\ArrowLine(0,15)(35,29) 
\ArrowLine(35,29)(50,35) 
\Photon(0,15)(0,-15){2}{5}
\PhotonArc(0,0)(43,41,74){2}{4}
\PhotonArc(0,0)(43,106,139){2}{4}
\ArrowLine(0,-15)(-50,-35) 
\ArrowLine(50,-35)(0,-15) 
\ArrowArc(0,43)(12,-10,190)
\ArrowArc(0,43)(12,190,350)
\Text(0,-20)[cb]{{\footnotesize (g)}}
\end{picture}}
}
\hspace{3.4cm}
  \vcenter{
\hbox{
\begin{picture}(0,0)(0,0)
  \SetScale{.8}
  \SetWidth{.5}
\ArrowLine(-50,35)(0,15) 
\ArrowLine(0,15)(50,35) 
\Photon(0,15)(0,-15){2}{5}
\PhotonArc(0,0)(43,-74,-41){2}{4}
\PhotonArc(0,0)(43,-139,-106){2}{4}
\ArrowArc(0,-43)(12,-10,190)
\ArrowArc(0,-43)(12,190,350)
\ArrowLine(-30,-27)(-50,-35) 
\ArrowLine(0,-15)(-30,-27) 
\ArrowLine(35,-29)(0,-15) 
\ArrowLine(50,-35)(35,-29) 
\Text(0,-20)[cb]{{\footnotesize (h)}}
\end{picture}}  
}
\hspace{3.4cm}
  \vcenter{\hbox{
\begin{picture}(0,0)(0,0)
  \SetScale{.8}
  \SetWidth{.5}
\ArrowLine(35,-50)(15,0) 
\ArrowLine(15,0)(35,50) 
\PhotonArc(0,0)(43,131,164){2}{4}
\PhotonArc(0,0)(43,196,229){2}{4}
\Photon(15,0)(-15,0){2}{5}
\ArrowArc(-43,0)(12,70,280)
\ArrowArc(-43,0)(12,280,430)
\ArrowLine(-15,0)(-27,-30) 
\ArrowLine(-27,-30)(-35,-50) 
\ArrowLine(-35,50)(-27,30) 
\ArrowLine(-27,30)(-15,0) 
\Text(0,-20)[cb]{{\footnotesize (i)}}
\end{picture}}}
\hspace{3.4cm}
  \vcenter{\hbox{
\begin{picture}(0,0)(0,0)
  \SetScale{0.8}
  \SetWidth{.5}
\ArrowLine(27,-30)(15,0) 
\ArrowLine(35,-50)(27,-30) 
\ArrowLine(27,30)(35,50) 
\ArrowLine(15,0)(27,30) 
\ArrowArc(43,0)(12,70,280)
\ArrowArc(43,0)(12,280,430)
\PhotonArc(0,0)(43,16,49){2}{4}
\PhotonArc(0,0)(43,-49,-16){2}{4}
\Photon(15,0)(-15,0){2}{5}
\ArrowLine(-15,0)(-35,-50) 
\ArrowLine(-35,50)(-15,0) 
\Text(0,-20)[cb]{{\footnotesize (j)}}
\end{picture}}}
\]
\vspace*{1.8cm}
\[\vcenter{\hbox{
  \begin{picture}(0,0)(0,0)
\SetScale{.8}
  \SetWidth{.5}
\ArrowLine(-50,25)(-25,25)
\ArrowLine(-25,25)(25,25)
\ArrowLine(25,25)(50,25)
\Photon(-25,25)(-25,-25){2}{10}
\Photon(25,25)(25,10){2}{4}
\Photon(25,-25)(25,-10){2}{4}
\ArrowArc(25,0)(10,-90,-270)
\ArrowArc(25,0)(10,-270,-90)
\ArrowLine(-25,-25)(-50,-25)
\ArrowLine(25,-25)(-25,-25)
\ArrowLine(50,-25)(25,-25)
\Text(0,-13)[cb]{{\footnotesize (k)}}
\end{picture}}}
\hspace{3.4cm}
\vcenter{\hbox{
  \begin{picture}(0,0)(0,0)
\SetScale{.8}
  \SetWidth{.5}
\ArrowLine(-50,25)(-25,25)
\ArrowLine(-25,25)(25,25)
\ArrowLine(25,25)(50,25)
\Photon(25,25)(25,-25){2}{10}
\Photon(-25,25)(-25,10){2}{4}
\Photon(-25,-25)(-25,-10){2}{4}
\ArrowArc(-25,0)(10,-90,-270)
\ArrowArc(-25,0)(10,-270,-90)
\ArrowLine(-25,-25)(-50,-25)
\ArrowLine(25,-25)(-25,-25)
\ArrowLine(50,-25)(25,-25)
\Text(0,-13)[cb]{{\footnotesize (l)}}
\end{picture}}}
\hspace{3.4cm}
\vcenter{\hbox{
  \begin{picture}(0,0)(0,0)
\SetScale{.8}
  \SetWidth{.5}
\ArrowLine(-50,25)(-25,25)
\ArrowLine(-25,25)(25,25)
\ArrowLine(25,25)(50,25)
\Photon(-25,25)(25,-25){2}{10}
\Photon(25,25)(7,7){2}{4}
\Photon(-25,-25)(-7,-7){2}{4}
\ArrowArc(0,0)(10,-90,-270)
\ArrowArc(0,0)(10,-270,-90)
\ArrowLine(-25,-25)(-50,-25)
\ArrowLine(25,-25)(-25,-25)
\ArrowLine(50,-25)(25,-25)
\Text(0,-13)[cb]{{\footnotesize (m)}}
\end{picture}}}
\hspace{3.4cm}
\vcenter{\hbox{
  \begin{picture}(0,0)(0,0)
\SetScale{.8}
  \SetWidth{.5}
\ArrowLine(-50,25)(-25,25)
\ArrowLine(-25,25)(25,25)
\ArrowLine(25,25)(50,25)
\Photon(25,25)(-25,-25){2}{10}
\Photon(-25,25)(-7,7){2}{4}
\Photon(25,-25)(7,-7){2}{4}
\ArrowArc(0,0)(10,-90,-270)
\ArrowArc(0,0)(10,-270,-90)
\Text(0,-13)[cb]{{\footnotesize (n)}}
\ArrowLine(-25,-25)(-50,-25)
\ArrowLine(25,-25)(-25,-25)
\ArrowLine(50,-25)(25,-25)
\end{picture}}}
\]
\vspace*{1.5cm}
\[\vcenter{\hbox{
  \begin{picture}(0,0)(0,0)
\SetScale{.8}
  \SetWidth{.5}
\ArrowLine(-50,25)(-25,25)
\ArrowLine(25,25)(50,25)
\Photon(-25,-25)(25,-25){2}{10}
\Photon(-25,25)(-10,25){2}{4}
\Photon(25,25)(10,25){2}{4}
\ArrowArc(0,25)(10,-180,0)
\ArrowArc(0,25)(10,0,180)
\ArrowLine(-25,-25)(-50,-25)
\ArrowLine(50,-25)(25,-25)
\ArrowLine(-25,25)(-25,-25)
\ArrowLine(25,-25)(25,25)
\Text(0,-15)[cb]{{\footnotesize (o)}}
\end{picture}}}
\hspace{3.5cm}
\vcenter{\hbox{
  \begin{picture}(0,0)(0,0)
\SetScale{.8}
  \SetWidth{.5}
\ArrowLine(-50,25)(-25,25)
\ArrowLine(25,25)(50,25)
\Photon(-25,25)(25,25){2}{10}
\Photon(-25,-25)(-10,-25){2}{4}
\Photon(25,-25)(10,-25){2}{4}
\ArrowArc(0,-25)(10,-180,0)
\ArrowArc(0,-25)(10,0,180)
\ArrowLine(-25,-25)(-50,-25)
\ArrowLine(50,-25)(25,-25)
\ArrowLine(-25,25)(-25,-25)
\ArrowLine(25,-25)(25,25)
\Text(0,-15)[cb]{{\footnotesize (p)}}
\end{picture}}}
\hspace{3.5cm}
\vcenter{\hbox{
  \begin{picture}(0,0)(0,0)
\SetScale{.8}
  \SetWidth{.5}
\ArrowLine(-50,25)(-25,25)
\ArrowLine(25,25)(50,25)
\Photon(-25,25)(25,-25){2}{10}
\Photon(-25,-25)(-7,-7){2}{4}
\Photon(25,25)(7,7){2}{4}
\ArrowArc(0,0)(10,-180,0)
\ArrowArc(0,0)(10,0,180)
\ArrowLine(-25,-25)(-50,-25)
\ArrowLine(50,-25)(25,-25)
\ArrowLine(-25,25)(-25,-25)
\ArrowLine(25,-25)(25,25)
\Text(0,-15)[cb]{{\footnotesize (q)}}
\end{picture}}}
\hspace{3.5cm}
\vcenter{\hbox{
  \begin{picture}(0,0)(0,0)
\SetScale{.8}
  \SetWidth{.5}
\ArrowLine(-50,25)(-25,25)
\ArrowLine(25,25)(50,25)
\Photon(25,25)(-25,-25){2}{10}
\Photon(-25,25)(-7,7){2}{4}
\Photon(25,-25)(7,-7){2}{4}
\ArrowArc(0,0)(10,-90,-270)
\ArrowArc(0,0)(10,-270,-90)
\ArrowLine(-25,-25)(-50,-25)
\ArrowLine(50,-25)(25,-25)
\ArrowLine(-25,25)(-25,-25)
\ArrowLine(25,-25)(25,25)
\Text(0,-15)[cb]{{\footnotesize (r)}}
\end{picture}}}
\]
%
%
\vspace*{1.7cm}
\[
  \vcenter{
\hbox{
  \begin{picture}(0,0)(0,0)
  \SetScale{0.8}
  \SetWidth{.5}
\ArrowLine(-50,35)(-30,27) 
\ArrowLine(-30,27)(0,15) 
\ArrowLine(0,15)(35,29) 
\ArrowLine(35,29)(50,35) 
\ArrowArc(0,0)(7,-90,90)
\ArrowArc(0,0)(7,90,270)
\Photon(0,15)(0,7){2}{2}
\Photon(0,-15)(0,-7){2}{2}
\PhotonArc(0,0)(43,41,139){2}{10}
\ArrowLine(0,-15)(-50,-35) 
\ArrowLine(50,-35)(0,-15) 
\Text(0,-20)[cb]{{\footnotesize (s)}}
\end{picture}}
}
\hspace{3.6cm}
  \vcenter{
\hbox{
\begin{picture}(0,0)(0,0)
  \SetScale{.8}
  \SetWidth{.5}
\ArrowLine(-50,35)(0,15) 
\ArrowLine(0,15)(50,35) 
\PhotonArc(0,0)(43,-139,-41){2}{10}
\ArrowArc(0,0)(7,-90,90)
\ArrowArc(0,0)(7,90,270)
\Photon(0,15)(0,7){2}{2}
\Photon(0,-15)(0,-7){2}{2}
\ArrowLine(-30,-27)(-50,-35) 
\ArrowLine(0,-15)(-30,-27) 
\ArrowLine(35,-29)(0,-15) 
\ArrowLine(50,-35)(35,-29) 
\Text(0,-20)[cb]{{\footnotesize (t)}}
\end{picture}}  
}
\hspace{3.6cm}
  \vcenter{\hbox{
\begin{picture}(0,0)(0,0)
  \SetScale{.8}
  \SetWidth{.5}
\ArrowLine(35,-50)(15,0) 
\ArrowLine(15,0)(35,50) 
\PhotonArc(0,0)(43,131,229){2}{10}
\ArrowArc(0,0)(7,0,180)  
\ArrowArc(0,0)(7,180,360)  
\Photon(-15,0)(-7,0){2}{2}
\Photon(15,0)(7,0){2}{2}
\ArrowLine(-15,0)(-27,-30) 
\ArrowLine(-27,-30)(-35,-50) 
\ArrowLine(-35,50)(-27,30) 
\ArrowLine(-27,30)(-15,0) 
\Text(0,-20)[cb]{{\footnotesize (u)}}
\end{picture}}}
\hspace{3.6cm}
  \vcenter{\hbox{
\begin{picture}(0,0)(0,0)
  \SetScale{0.8}
  \SetWidth{.5}
\ArrowLine(27,-30)(15,0) 
\ArrowLine(35,-50)(27,-30) 
\ArrowLine(27,30)(35,50) 
\ArrowLine(15,0)(27,30) 
\PhotonArc(0,0)(43,-49,49){2}{10}
\ArrowArc(0,0)(7,0,180)  
\ArrowArc(0,0)(7,180,360)  
\Photon(-15,0)(-7,0){2}{2}
\Photon(15,0)(7,0){2}{2}
\ArrowLine(-15,0)(-35,-50) 
\ArrowLine(-35,50)(-15,0) 
\Text(0,-20)[cb]{{\footnotesize (v)}}
\end{picture}}}
\]
\vspace*{1.6cm}
\caption[]{\it Two-loop $N_F=1$ diagrams.}
\label{fig2ltot}
\end{figure}
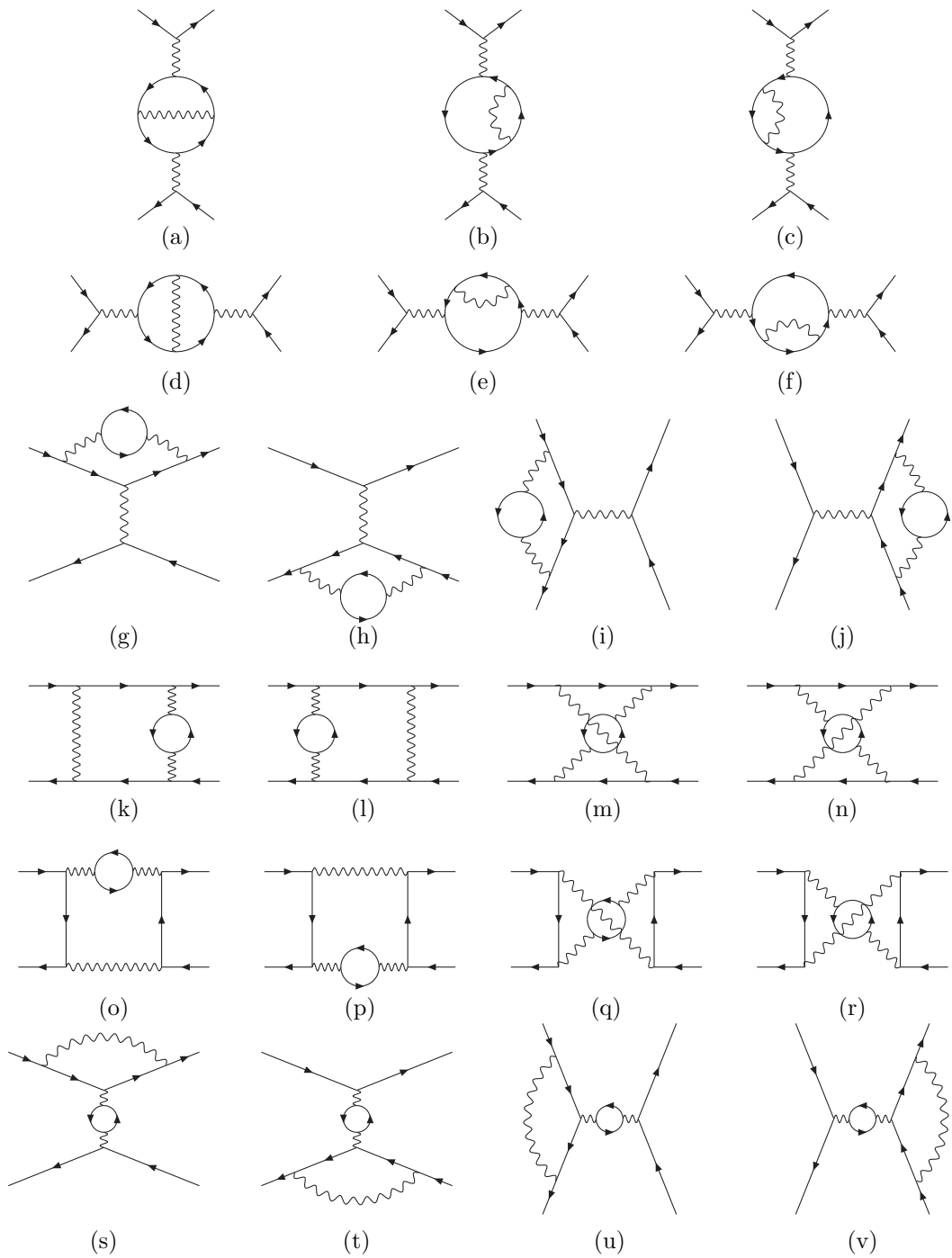

At order $\alpha^4$, if we limit our analysis to the diagrams containing 
a closed electron loop  ($N_F=1$), it is necessary to consider the 
contributions coming from the interference between the diagrams  
shown in Fig.~\ref{fig2ltot} and the tree-level amplitude, as well as the 
interference between the one-loop vacuum polarization diagrams with the 
other one-loop diagrams without a fermionic loop. Diagrammatically, we 
have:
\vspace*{.5cm}
\bea
\left(\frac{\alpha}{\pi}\right)^2  
\frac{d \sigma_{2}^{V}(s,t,m^2)}{d \Omega} \! \! & = & \! \! 
\frac{s}{16}\sum_{\mbox{spin}} \Biggl\{ \Biggl(
\hspace*{1.3cm}
\hbox{
  \begin{picture}(0,0)(0,0)
    \SetScale{.8}
    \SetWidth{.5}
    \ArrowLine(-45,23)(-25,23)
    \ArrowLine(-25,23)(-5,23)
    \ArrowLine(-25,-17)(-45,-17)
    \ArrowLine(-5,-17)(-25,-17)
    \Photon(-25,23)(-25,-17){2}{8}
  \end{picture} 
     }  
\hspace*{-.2cm} - \hspace*{-.2cm}
\hbox{
  \begin{picture}(0,0)(0,0)
\SetScale{.8}
  \SetWidth{.5}
\ArrowLine(20,3)(10,-17)
\ArrowLine(10,23)(20,3)
\Photon(20,3)(40,3){2}{6}
\ArrowLine(40,3)(50,23)
\ArrowLine(50,-17)(40,3)
\end{picture}}
\hspace*{1.7cm}
\Biggr)^* \times
\hspace*{0.8cm}\hbox{
  \begin{picture}(0,0)(0,0)
\SetScale{.65}
  \SetWidth{.5}
\ArrowLine(-50,28)(-25,28)
\ArrowLine(-25,28)(25,28)
\ArrowLine(25,28)(50,28)
\Photon(-25,28)(-25,-22){2}{10}
\Photon(25,28)(25,13){2}{4}
\Photon(25,-22)(25,-7){2}{4}
\ArrowArc(25,3)(10,-90,-270)
\ArrowArc(25,3)(10,-270,-90)
\ArrowLine(-25,-22)(-50,-22)
\ArrowLine(25,-22)(-25,-22)
\ArrowLine(50,-22)(25,-22)
\end{picture}}
\hspace*{1.1cm} + \mbox{c.c.} \nn\\
& &  \nn\\
&  + & 
\! \! \Biggl(
\hspace*{1.3cm}
\hbox{
  \begin{picture}(0,0)(0,0)
\SetScale{.8}
  \SetWidth{.5}
\ArrowLine(-45,23)(-25,23)
\ArrowLine(-25,23)(-5,23)
\ArrowLine(-25,-17)(-45,-17)
\ArrowLine(-5,-17)(-25,-17)
\Photon(-25,23)(-25,13){2}{4}
\Photon(-25,-7)(-25,-17){2}{4}
\ArrowArc(-25,3)(10,90,270)
\ArrowArc(-25,3)(10,270,90)
\end{picture}}  
\hspace*{-.2cm} - \hspace*{-.2cm}
\hbox{ 
  \begin{picture}(0,0)(0,0)
\SetScale{.8}
  \SetWidth{.5}
\ArrowLine(10,3)(0,-17)
\ArrowLine(0,23)(10,3)
\Photon(10,3)(20,3){2}{3}
\Photon(40,3)(50,3){2}{3}
\ArrowLine(50,3)(60,23)
\ArrowLine(60,-17)(50,3)
\ArrowArc(30,3)(10,0,180)
\ArrowArc(30,3)(10,180,0)
\end{picture}}
\hspace*{1.9cm}
\Biggr)^* \times
\hspace*{0.8cm} \hbox{
  \begin{picture}(0,0)(0,0)
\SetScale{.65}
  \SetWidth{.5}
\ArrowLine(-50,28)(-25,28)
\ArrowLine(25,28)(50,28)
\Photon(-25,-22)(25,-22){2}{10}
\Photon(-25,28)(25,28){2}{10}
\ArrowLine(-25,-22)(-50,-22)
\ArrowLine(50,-22)(25,-22)
\ArrowLine(-25,28)(-25,-22)
\ArrowLine(25,-22)(25,28)
\end{picture}}
\hspace*{0.9cm} + \mbox{c.c.} 
+  \cdots \Biggr\} \, .
\label{M2at2l}
\eea

\vspace*{5mm}

The expression of the various contributions to the differential cross
section shown in Eq.~(\ref{M2at2l}) will be given in Section \ref{S2l}.
%
%

The evaluation of the virtual radiative corrections to the Bhabha scattering,
Eqs.~(\ref{M2at1l},\ref{M2at2l}), leads to the calculation of UV- and 
IR-divergent integrals, which need to be regularized.  Both classes
of divergences are regularized within the framework of  Dimensional
Regularization scheme \cite{DimReg}, in which the  regularization parameter is
the dimension of the space-time $D$.  In order to evaluate  the virtual
radiative corrections  we follow a  procedure  consisting of three  main
steps:

\begin{itemize}
\item First, we consider the Dirac structure of the squared matrix 
elements; the sum over the spins generates traces over the Dirac 
indices, that can be easily evaluated in $D$ dimensions.
\item At this stage, the squared amplitude describing the interference 
of a one-loop or two-loop diagram with the tree-level matrix element, 
as well as the interference of a one-loop with another one-loop diagram, 
can be expressed as a sum of a large number of scalar integrals (several 
thousands). According to \cite{us,RoPieRem1},
it is possible to express all these 
integrals as a combination of 14 Master Integrals.
This result was obtained by using the Laporta algorithm~\cite{laporta} 
in order to exploit the information contained in 
various sets of identities (Integration-by-Parts, Lorentz 
invariance and other general symmetry identities) 
relating the scalar integrals among themselves.
Once the reduction to the MIs has been  
carried out, we can expand the squared amplitude, 
which will depend on the dimensional regulator $D$, 
around $D=4$. Using the expressions for the MIs also given in 
\cite{us,RoPieRem1}, we finally obtain an analytic result, expressed 
in terms of 1- and 2-dimensional harmonic polylogarithms, in which both 
UV and IR divergences are regularized by the same parameter $D$ and 
appear as poles in $(D-4)$.
\item The last step concerns the UV renormalization. By subtracting the
appropriate counterterm graphs, the poles in $(D-4)$ corresponding to 
the unphysical UV divergences of the diagrams cancel out. We perform the 
subtraction of the UV  poles in the {\it on-shell}  renormalization 
scheme. The result obtained in this way still contains IR poles in 
$(D-4)$, which are canceled in the physical cross section by the 
contribution of the real soft photon radiation.
\end{itemize}

\section{Renormalization \label{Ren}}

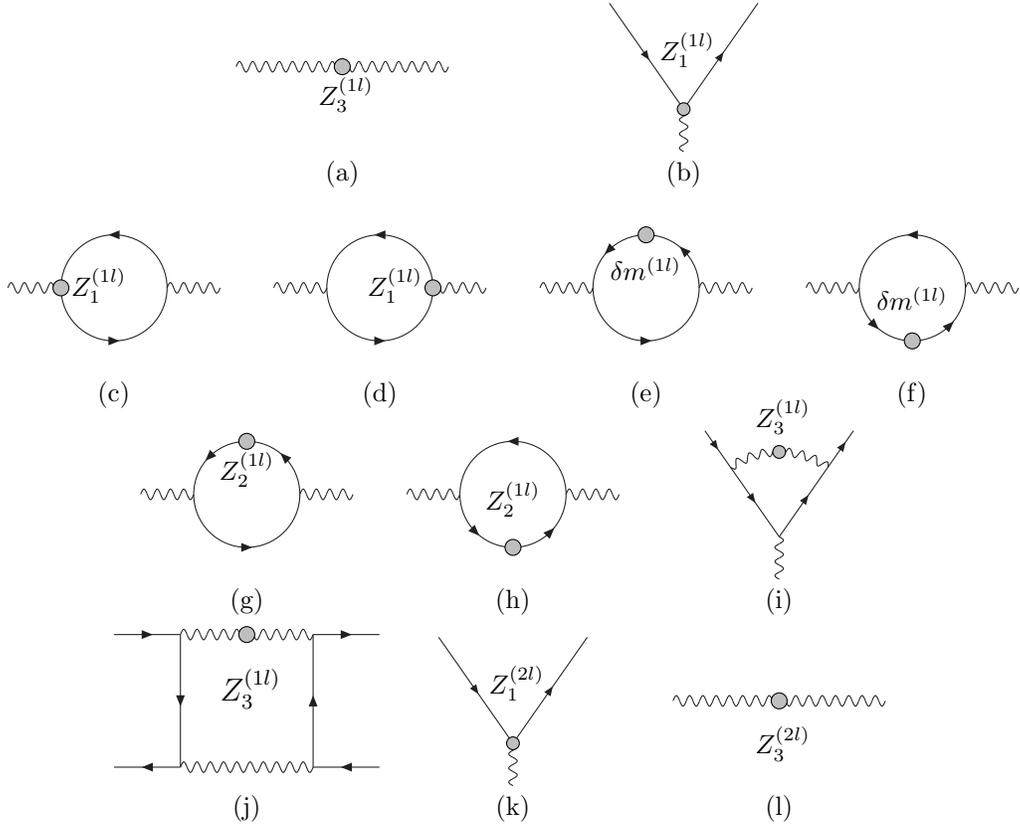
\begin{figure}[ht]
\vspace*{.5cm}
\[\vcenter{
\hbox{
  \begin{picture}(0,0)(0,0)
\SetScale{1}
  \SetWidth{.4}
\Photon(-40,0)(40,0){2}{16}
\GCirc(0,0){3}{0.75}
\Text(0.5,-6)[cb]{{\footnotesize $Z_3^{(1l)}$}}
\Text(0,-16)[cb]{{\footnotesize (a)}}
\end{picture}}  
}
\hspace{4.4cm}
\vcenter{
\hbox{
  \begin{picture}(0,0)(0,0)
\SetScale{.8}
  \SetWidth{.5}
\ArrowLine(-35,30)(0,-20)
\ArrowLine(0,-20)(35,30)
\Photon(0,-20)(0,-40){2}{3}
\GCirc(0,-20){3}{.75}
\Text(.5,0)[cb]{{\footnotesize $Z_1^{(1l)}$}}
\Text(0,-16)[cb]{{\footnotesize (b)}}
\end{picture}}
}
\]
\vspace*{1.5cm}
\[
\vcenter{
\hbox{
  \begin{picture}(0,0)(0,0)
\SetScale{1.}
  \SetWidth{.4}
\Photon(-40,0)(-20,0){2}{4}
\Photon(20,0)(40,0){2}{4}
\ArrowArc(0,0)(20,0,180)
\ArrowArc(0,0)(20,180,360)
\GCirc(-20,0){3}{.75}
\Text(-2,-2)[cb]{{\footnotesize $Z_1^{(1l)}$}}
\Text(0,-16)[cb]{{\footnotesize (c)}}
\end{picture}}  
}
\hspace{3.4cm}
\vcenter{
\hbox{
  \begin{picture}(0,0)(0,0)
\SetScale{1.}
  \SetWidth{.4}
\Photon(-40,0)(-20,0){2}{4}
\Photon(20,0)(40,0){2}{4}
\ArrowArc(0,0)(20,0,180)
\ArrowArc(0,0)(20,180,360)
\GCirc(20,0){3}{.75}
\Text(2,-2)[cb]{{\footnotesize $Z_1^{(1l)}$}}
\Text(0,-16)[cb]{{\footnotesize (d)}}
\end{picture}}  
}
\hspace{3.4cm}
  \vcenter{
\hbox{
  \begin{picture}(0,0)(0,0)
\SetScale{1.}
  \SetWidth{.4}
\Photon(-40,0)(-20,0){2}{4}
\Photon(20,0)(40,0){2}{4}
\ArrowArc(0,0)(20,0,90)
\ArrowArc(0,0)(20,90,180)
\ArrowArc(0,0)(20,180,360)
\GCirc(0,20){3}{.75}
\Text(0,1)[cb]{{\footnotesize $\delta m^{(1l)}$}}
\Text(0,-16)[cb]{{\footnotesize (e)}}
\end{picture}}
}
\hspace{3.4cm}
  \vcenter{
\hbox{
  \begin{picture}(0,0)(0,0)
\SetScale{1.}
  \SetWidth{.4}
\Photon(-40,0)(-20,0){2}{4}
\Photon(20,0)(40,0){2}{4}
\ArrowArc(0,0)(20,0,180)
\ArrowArc(0,0)(20,180,270)
\ArrowArc(0,0)(20,270,360)
\GCirc(0,-20){3}{.75}
\Text(0,-3)[cb]{{\footnotesize $\delta m^{(1l)}$}}
\Text(0,-16)[cb]{{\footnotesize (f)}}
\end{picture}}
}
\]
\vspace*{1.5cm}
\[
  \vcenter{
\hbox{
  \begin{picture}(0,0)(0,0)
\SetScale{1.}
  \SetWidth{.4}
\Photon(-40,0)(-20,0){2}{4}
\Photon(20,0)(40,0){2}{4}
\ArrowArc(0,0)(20,0,90)
\ArrowArc(0,0)(20,90,180)
\ArrowArc(0,0)(20,180,360)
\GCirc(0,20){3}{.75}
\Text(0,1)[cb]{{\footnotesize $Z_2^{(1l)}$}}
\Text(0,-16)[cb]{{\footnotesize (g)}}
\end{picture}}
}
\hspace{3.4cm}
  \vcenter{
\hbox{
  \begin{picture}(0,0)(0,0)
\SetScale{1.}
  \SetWidth{.4}
\Photon(-40,0)(-20,0){2}{4}
\Photon(20,0)(40,0){2}{4}
\ArrowArc(0,0)(20,0,180)
\ArrowArc(0,0)(20,180,270)
\ArrowArc(0,0)(20,270,360)
\GCirc(0,-20){3}{.75}
\Text(0,-3)[cb]{{\footnotesize $Z_2^{(1l)}$}}
\Text(0,-16)[cb]{{\footnotesize (h)}}
\end{picture}}
}
\hspace{3.4cm}
\vcenter{
\hbox{
  \begin{picture}(0,0)(0,0)
\SetScale{.8}
  \SetWidth{.5}
\ArrowLine(-35,30)(-25,15.7143)
\ArrowLine(-25,15.7143)(0,-20)
\ArrowLine(0,-20)(25,15.7143)
\ArrowLine(25,15.7143)(35,30)
\PhotonArc(0,-20)(40,55,125){2}{8}
\Photon(0,-20)(0,-40){2}{3}
\GCirc(0,20){3}{.75}
\Text(.5,8)[cb]{{\footnotesize $Z_3^{(1l)}$}}
\Text(0,-16)[cb]{{\footnotesize (i)}}
\end{picture}}  
}
\]
\vspace*{15mm}
\[
  \vcenter{
\hbox{
  \begin{picture}(0,0)(0,0)
\SetScale{1.}
  \SetWidth{.4}
\ArrowLine(-50,25)(-25,25)
\ArrowLine(25,25)(50,25)
\Photon(-25,25)(25,25){2}{10}
\Photon(-25,-25)(25,-25){2}{10}
\GCirc(0,25){3}{.75}
\ArrowLine(-25,-25)(-50,-25)
\ArrowLine(50,-25)(25,-25)
\ArrowLine(-25,25)(-25,-25)
\ArrowLine(25,-25)(25,25)
\Text(0.5,-1)[cb]{$Z_3^{(1l)}$}
\Text(0,-16)[cb]{{\footnotesize (j)}}
\end{picture}}
}
\hspace{3.4cm}
\vcenter{
\hbox{
  \begin{picture}(0,0)(0,0)
\SetScale{.8}
  \SetWidth{.5}
\ArrowLine(-35,30)(0,-20)
\ArrowLine(0,-20)(35,30)
\Photon(0,-20)(0,-40){2}{3}
\GCirc(0,-20){3}{.75}
\Text(.5,0)[cb]{{\footnotesize $Z_1^{(2l)}$}}
\Text(0,-16)[cb]{{\footnotesize (k)}}
\end{picture}}
}
\hspace{3.4cm}
\vcenter{
\hbox{
  \begin{picture}(0,0)(0,0)
\SetScale{1}
  \SetWidth{.4}
\Photon(-40,0)(40,0){2}{16}
\GCirc(0,0){3}{0.75}
\Text(0.5,-8)[cb]{{\footnotesize $Z_3^{(2l)}$}}
\Text(0,-16)[cb]{{\footnotesize (l)}}
\end{picture}}  
}
\]
\vspace*{.8cm}
\caption[]{\it Counterterm diagrams at one and two loops.}
\label{figCTer}
\end{figure}

We perform the UV renormalization in the {\it on-shell} scheme. 
The counterterm diagrams for the renormalization of the one-loop 
and two-loop cross section are shown in Fig.~\ref{figCTer}. 
%

The counterterms give rise to the following Feynman rules: 
\bea
\hbox{
  \begin{picture}(0,0)(0,0)
\SetScale{1}
  \SetWidth{.4}
\Photon(-40,3)(40,3){2}{16}
\GCirc(0,3){3}{0.75}
\Text(0.5,-6)[cb]{{\footnotesize $Z_3^{(1l)}$}}
\end{picture}} \hspace*{16mm} & = & - \, \left( \frac{\alpha}{\pi} \right) 
\, Z_3^{(1l)} \,  ( p^2 \delta_{\mu \nu} - p_{\mu} p_{\nu} )
\, , \\
& & \nn\\
\hbox{
  \begin{picture}(0,0)(0,0)
\SetScale{1}
  \SetWidth{.4}
\Photon(-40,3)(40,3){2}{16}
\GCirc(0,3){3}{0.75}
\Text(0.5,-6)[cb]{{\footnotesize $Z_3^{(2l)}$}}
\end{picture}} \hspace*{16mm} & = & - \, \left( \frac{\alpha}{\pi} \right)^2 
\, Z_3^{(2l)} \, ( p^2 \delta_{\mu \nu} - p_{\mu} p_{\nu} )
\, , \\
& & \nn\\
\hbox{
  \begin{picture}(0,0)(0,0)
\SetScale{1}
  \SetWidth{.4}
\ArrowLine(-40,3)(0,3)
\ArrowLine(0,3)(40,3)
\GCirc(0,3){3}{0.75}
\Text(0.5,-6)[cb]{{\footnotesize $Z_2^{(1l)}$}}
\end{picture}} \hspace*{16mm} & = & - \, \left( \frac{\alpha}{\pi} \right) 
\, Z_2^{(1l)} \, ( i \not{\! p} + m )
\, , \\
& & \nn\\
\hbox{
  \begin{picture}(0,0)(0,0)
\SetScale{1}
  \SetWidth{.4}
\ArrowLine(-40,3)(0,3)
\ArrowLine(0,3)(40,3)
\GCirc(0,3){3}{0.75}
\Text(0.5,-6)[cb]{{\footnotesize $\delta m^{(1l)}$}}
\end{picture}} \hspace*{16mm} & = & - \, \left( \frac{\alpha}{\pi} \right) 
\, \delta m^{(1l)} \, , \\
& & \nn\\
& & \nn\\
\hbox{
  \begin{picture}(0,0)(0,0)
\SetScale{.8}
  \SetWidth{.5}
\ArrowLine(-35,30)(0,-20)
\ArrowLine(0,-20)(35,30)
\Photon(0,-20)(0,-40){2}{3}
\GCirc(0,-20){3}{.75}
\Text(.5,0)[cb]{{\footnotesize $Z_1^{(1l)}$}}
\end{picture}} \hspace*{16mm} & = & \left( \frac{\alpha}{\pi} \right) 
\, Z_1^{(1l)} \, , \\
& & \nn\\
& & \nn\\
\hbox{
  \begin{picture}(0,0)(0,0)
\SetScale{.8}
  \SetWidth{.5}
\ArrowLine(-35,30)(0,-20)
\ArrowLine(0,-20)(35,30)
\Photon(0,-20)(0,-40){2}{3}
\GCirc(0,-20){3}{.75}
\Text(.5,0)[cb]{{\footnotesize $Z_1^{(2l)}$}}
\end{picture}} \hspace*{16mm} & = & \left( \frac{\alpha}{\pi} \right)^2 
\, Z_1^{(2l)} \, .
\eea

\vspace*{10mm}

The one- and two-loop renormalization constants are:
\bea
Z_{1}^{(1l)}(D) & = &  \frac{3}{2(D-4)} - 1 + (D-4) 
+ {\mathcal O} \left( (D-4)^2 \right)
\label{Z11l} 
 \ , \\
 & & \nonumber \\
Z_{2}^{(1l)}(D)  &=& Z_{1}^{(1l)}(D) \qquad (\mbox{\rm Ward identity}) 
\ , 
\label{Z21l} \\ 
Z_{3}^{(1l)}(D) & = & 
\frac{2}{3(D-4)}  
\label{Z31l} 
 \ , \\
\frac{\delta m^{(1l)}(D,m)}{m} & = & Z_{1}^{(1l)}(D) \ ,  
\label{Dem} 
\eea

\bea
Z_{1}^{(2l,N_F=1)}(D) & = & - \frac{1}{8(D-4)}  + \frac{947}{288}
- 2 \zeta(2) + {\mathcal O} (D-4) \, . 
\label{Z12l} 
\eea

\bea
Z_{3}^{(2l)}(D) & = & \frac{1}{4(D-4)}  -\frac{15}{16}
       + {\mathcal O} (D-4) \, .
\label{Z32l} 
\eea

The renormalization constants are defined in such a way that the 
UV-renormalized quantity is obtained by adding  to the UV-divergent 
diagrams the corresponding counterterm graphs.

At the one-loop level, it is necessary  to renormalize the photon vacuum
polarization graphs, shown in Fig.~\ref{fig1ltot} (a) and (b), and the 
$\gamma f \bar{f}$-vertex graphs, shown in Fig.~\ref{fig1ltot} (c)--(f). 
In the first case, we
add the UV-divergent diagram of Fig.~\ref{fig1ltot} (a) (or (b)) and the
corresponding counterterm of  Fig.~\ref{figCTer} (a), ending up with the
renormalized expression for the photon self-energy given in the 
next Section. For the one-loop vertex  it is necessary to add the UV-divergent diagram
of Fig.~\ref{fig1ltot} (c) (or (d)--(f)) with the corresponding counterterm of 
Fig.~\ref{figCTer} (b). The UV-renormalized form factors found in this way are
given in Eqs.~(\ref{F1ren},\ref{F2ren}).

In the two-loop case, we have to renormalize the UV divergences occurring in
the diagrams of Fig.~\ref{fig2ltot}.

The three two-loop vacuum polarization diagrams have to be added with the
corresponding seven counterterm diagrams shown in Fig.~\ref{figCTer} (c)--(h)
and (l). Note that, because of the QED Ward identity, $Z_1(D)=Z_2(D)$ and the 
contributions of diagrams (c), (d), (g) and (h) in Fig.~\ref{figCTer} cancel 
among themselves. The renormalization of  the two-loop vacuum polarization 
consists, therefore, only in the one-loop electron mass renormalization  and  
in the subsequent cancellation of genuine two-loop UV divergence  against  
the counterterm diagram (l). For the two-loop vacuum polarization vertices, 
Figs.~\ref{fig2ltot} (g)--(j), we have to add the counterterms (i) and (k) of 
Fig.~\ref{figCTer}: the first one cancels the subdivergences due to the 
one-loop vacuum polarization insertion, while the second cancels the overall 
two-loop divergence. The box graphs, shown in Figs.~\ref{fig2ltot}
(k)--(r), do not present two-loop UV divergences and their subgraph divergence
is  renormalized  by adding the corresponding counterterm diagrams 
of the kind shown in Fig.~\ref{figCTer} (j). Finally, the diagrams in 
Fig.~\ref{fig2ltot} (s)--(v) are renormalized by adding to them the 
corresponding counterterm graphs subtracting the one-loop divergences
of the vacuum polarization insertion and of the  vertex correction.

Note that, even after the UV renormalization procedure,  the Bhabha scattering
differential cross section still  shows poles in $(D-4)$.  This is due to
the fact that IR divergences are still present and  they are regularized 
by the same parameter $D$ that regularizes the UV ones. These 
divergences are not physical and will be removed  adding
to  the process also the corresponding soft real emission.

The contributions to the differential cross section appearing in the 
remainder of this work are all UV-renormalized.
In the following, whenever discussing the contribution of a UV divergent
one- or two-loop diagram, we assume as understood that the 
appropriate counterterm graphs have  already been added 
and therefore the quantities under study are UV finite.

\section{One-Loop Differential Cross Section \label{S1l}}

For completeness, we report in this Section the contribution to the 
cross section given by the interference between  the one-loop virtual 
graphs and the tree-level amplitude. The one-loop diagrams 
contributing to the Bhabha scattering process are of three kinds: 
vacuum polarization diagrams, vertex correction diagrams, and box 
diagrams. Their contribution are discussed separately in the following 
subsections.

\subsection{One-Loop Vacuum Polarization Corrections}

The two one-loop diagrams containing a vacuum polarization subgraph can 
be obtained by inserting a photon self energy correction in the photon 
propagator, as shown in Fig.~\ref{fig1ltot} (a) and (b). The photon 
self energy can be written as
\be
\Pi^{(1l)}_{\mu \nu}(-p^2) = \left( \frac{\alpha}{\pi} \right) 
(\delta_{\mu \nu} p^2 - p^{\mu} p^{\nu}) 
\Pi^{(1l)}_0 (-p^2) \, ,
\ee
where $p$ indicates the generic 4-momentum entering the  
self-energy graph. The UV renormalized $\Pi^{(1l)}_0 (p^2)$ has the 
following Laurent expansion: 
\be 
 \Pi^{(1l)}_0 (-p^2) = \Pi^{(1l,0)}_0 (-p^2) 
                     + (D-4)\ \Pi^{(1l,1)}_0 (-p^2) 
                     + {\mathcal{O}} \Big((D-4)^2\Big) \ ; 
\label{1lsLaur} 
\ee 
the explicit expression of the $\Pi^{(1l,i)}_0 (-p^2), i=0,1$ is 
given in Eqs.~(\ref{A1lS},\ref{A1ls1}). 

The self-energy diagrams in Fig.~\ref{fig1ltot} (a) and (b)
factorize in the product of the tree-level amplitude times the 
function $\Pi^{(1l)}_0(-p^2)$, so that one finds:
\be
{\mathcal A}^{(1l,S)}_s = {\mathcal A}^{(0)}_s \, 
\left( \frac{\alpha}{\pi} \right) \, \Pi^{(1l)}_0(s) \, , \quad
{\mathcal A}^{(1l,S)}_t = {\mathcal A}^{(0)}_t \, 
\left( \frac{\alpha}{\pi} \right) \, \Pi^{(1l)}_0(t) \, . 
\ee
The superscript ``$S$'' in the equation above stands for 
``self-energy''.

This set of virtual QED corrections is IR finite; 
therefore, the contribution of these diagrams to the differential
cross section at order $\alpha^3$ is:
\bea
\frac{d \sigma_{1}^{V}(s,t,m^2)}{d \Omega} \Big|_{(1l,S)} &=&  
\frac{\alpha^2}{s} \Bigg\{
\frac{1}{s^2}\left[s t + \frac{s^2}{2} + (t - 2 m^2)^2\right] 
2 \mbox{Re}\Pi^{(1l,0)}_0(s) \nn\\  
& & + \frac{1}{t^2}\left[s t + \frac{t^2}{2} + (s - 2 m^2)^2\right]  
2 \Pi^{(1l,0)}_0(t) \nn\\ 
& & + \frac{1}{s t}\left[(s + t)^2 - 4 m^4\right]
 \left(\mbox{Re}\Pi^{(1l,0)}_0(s) +  \Pi^{(1l,0)}_0(t)\right)
\Bigg\} \, , 
\label{amp1lS} 
\eea

\subsection{One-Loop Vertex Corrections}

At the  one-loop level, there are four vertex correction diagrams 
(see Fig.~\ref{fig1ltot}~(c), (d), (e) and (f)). At first, we consider 
the diagram in Fig~\ref{fig1ltot}~(c). The vertex correction to the
electron current can be written as
\be
\Gamma^{\mu}(p_1,p_3) = \left(\frac{\alpha}{\pi}\right) \left[F_1^{(1l)}(t)\gamma^{\mu} 
+ \frac{1}{2 m} F_2^{(1l)}(t)
\sigma^{\mu \nu} (p_1 - p_3)_{\nu}\right] \, ,  
\label{1lV}
\ee  
where $\sigma^{\mu \nu} = -i/2 \left[\gamma^\mu, \gamma^\nu\right]$ 
and $t=-(p_1-p_3)^2$ (space-like) ; 
the electron and positron spinors have been omitted from 
Eq.~(\ref{1lV}). The functions $F_1^{(1l)}(t)$ and $F_2^{(1l)}(t)$ are the UV
renormalized form factors; their Laurent expansion reads 
\bea 
 F_1^{(1l)}(t) &=& \frac{1}{(D-4)}F_1^{(1l,-1)}(t) + F_1^{(1l,0)}(t) 
                + {\mathcal{O}} \Big((D-4)\Big) \ , \label{1lFFLaurU} \\ 
 F_2^{(1l)}(t) &=& F_2^{(1l,0)}(t) 
                + {\mathcal{O}} \Big((D-4)\Big) \ . \label{1lFFLaurD} 
\eea 
The renormalized form factor $F_1^{(1l)}(t)$ still shows an IR polar 
divergence in $(D-4)$; the explicit expressions in terms 
of harmonic polylogarithms are given in 
Eqs.~(\ref{F1renp},\ref{F1ren},\ref{F2ren}) of Appendix~\ref{AF}.

The contribution  of the diagram in Fig.~\ref{fig1ltot}~(c) to the 
Bhabha differential cross section at order $\alpha^3$ can be written as 
\be
\frac{d \sigma_{1}^{V}(s,t)}{d \Omega} \Big|_{(V,(c))} = 
     \frac{\alpha^2}{s} \left[\frac{1}{s t} V_1^{(1l)}(s,t) 
                            + \frac{1}{t^2}V_2^{(1l)}(s,t)\right] \, , 
\label{1laV}
\ee
where 
\bea 
V_1^{(1l)}(s,t) &=&  \Biggl[ 
2 \left( s t  + \frac{1}{2} s^2  + \frac{1}{2} t^2  
- 2  m^4 \right)  + \frac{1}{2} (D-4) ( st  - 2sm^2 + s^2 \nn\\
& & - \, 2 t m^2 +  t^2 ) \Biggr]  \mbox{Re} F_1^{(1l)} (t) 
+ 2 \Biggl( s t - \frac{4}{3} t m^2 + \frac{3}{4} t^2 \Biggr) \mbox{Re}
  F_2^{(1l)}(t) \, , \label{VsU} \\
V_2^{(1l)}(s,t) & = & \left[2 \left(s t  
- 4 s m^2 + s^2   + \frac{1}{2} t^2  + 4 m^4 \right) 
+ \frac{1}{2} (D - 4) t^2 \right] \mbox{Re} F_1^{(1l)} (t)\nn\\ 
& & + \, 2 \left(t m^2 + \frac{1}{2} t^2\right) 
 \mbox{Re} F_2^{(1l)}(t)  \, . \label{VsD}
\eea
Due to the IR pole present in $F_1^{(1l)} (t)$, the two functions 
$V_i^{(1l)}(s,t), i=1,2$ have the Laurent expansions 
\be 
 V_i^{(1l)}(s,t) = \frac{1}{(D-4)}V_i^{(1l,-1)}(s,t) 
                 + V_i^{(1l,0)}(s,t) 
                 + {\mathcal{O}} \Big((D-4)\Big) \ . 
\label{ViLaur} 
\ee 
The explicit values of the Laurent coefficients can then be expressed in 
terms of the Laurent coefficients of the form factors $F_i^{(1l)} (t)$ 
in Eqs.~(\ref{1lFFLaurU},\ref{1lFFLaurD}) : 
\bea 
 V_i^{(1l,-1)}(s,t) &=& c_{i,1}(s,t) \mbox{Re}F_1^{(1l,-1)}(t) 
\label{ViLaur1U} \\ 
 V_i^{(1l,0)}(s,t) &=& c_{i,1}(s,t) \mbox{Re}F_1^{(1l,0)}(t) 
                     + c_{i,2}(s,t) \mbox{Re}F_1^{(1l,-1)}(t)  \nn\\ 
                    && + c_{i,3}(s,t) \mbox{Re}F_2^{(1l,0)}(t) \ , 
\label{ViLaur1D} 
\eea 
with 
\bea 
c_{1,1}(s,t) &=& (s+t)^2 - 4m^4 \ ,
\label{ViLaur2U} \\ 
c_{1,2}(s,t) &=& \frac{1}{2} ( st  - 2sm^2 + s^2 - 2 t m^2 +  t^2 ) \ ,
\label{ViLaurD} \\ 
c_{1,3}(s,t) &=& 2 \Biggl( s t - \frac{4}{3} t m^2 
                 + \frac{3}{4} t^2 \Biggr) \ ,
\label{ViLaur2T} \\ 
c_{2,1}(s,t) &=& 2\left( s t - 4 s m^2 + s^2 
                 + \frac{1}{2} t^2  + 4 m^4 \right) \ ,
\label{ViLaur2Q} \\ 
c_{2,2}(s,t) &=& \frac{1}{2}\ t^2 \ ,
\label{ViLaur2C} \\ 
c_{2,3}(s,t) &=& t(t+2m^2) \ . 
\label{ViLaur2} 
\eea 

For convenience of later use in the above equations we write
$\mbox{Re} F_i^{(1l)} (t)$ ($i=1,2$) even if the $F_i^{(1l)} (t)$ are 
real functions for space-like $t$. 
The first term between square brackets in Eq.~(\ref{1laV}) comes from the 
interference between the tree level $s$-channel diagram and the diagram 
in Fig.~\ref{fig1ltot}~(c); the second term originates in the interference 
between the same vertex correction and the tree level $t$-channel. 
The terms proportional to $(D-4)$ in $V_1(s,t)$ and $V_2(s,t)$ give rise, 
when multiplied by the IR pole in $F_1^{(1l)}(t)$, to the finite 
contributions proportional to $c_{i,2}(s,t)$ in Eq.~(\ref{ViLaur1D}). 
We observe that in $V_1(s,t)$, the term $c_{1,1}(s,t)$ of 
Eqs.~(\ref{ViLaur1U},\ref{ViLaur1D}) 
is identical to the coefficient of the $s$-$t$-interference term in the 
tree-level, Eq.~(\ref{TL}), while the term $c_{2,1}(s,t)$ is twice 
the coefficient of the $1/t^2$ term in Eq.~(\ref{TL}). 
Those relations are expected since $F_1^{(1l)}(t)$ appears as a
multiplicative factor of the tree-level QED vertex in the amplitude 
to be squared. 

The diagram in Fig.~\ref{fig1ltot} (d) can be obtained from the diagram in
Fig.~\ref{fig1ltot} (c) with the transformations  $p_2 \leftrightarrow -p_3$
and $p_4 \leftrightarrow  - p_1$. Since  these  transformations
leave the  Mandelstam invariants $s$ and $t$ unchanged, the contribution of
the diagram (d) to the cross section is equal to the 
contribution of diagram (c).

The contributions of diagrams (e) and (f) in Fig.~\ref{fig1ltot} are also 
identical, since the latter can be obtained from the former by exchanging 
$p_1 \leftrightarrow - p_4$ and $p_2 \leftrightarrow -p_3$, which leave 
$s$ and $t$ unchanged. Moreover, diagram (e) 
can be obtained from diagram (c) by the replacement 
$p_2 \leftrightarrow - p_3$ which exchanges $s\leftrightarrow t$, so that 
the contribution of diagram (e) to the differential cross 
section at order $\alpha^3$ will be
\be
\frac{d \sigma_{1}^{V}(s,t)}{d \Omega} \Big|_{(V,(e))} = 
\frac{\alpha^2}{s} \left[\frac{1}{s^2} V_2^{(1l)}(t,s) + 
\frac{1}{s t}V_1^{(1l)}(t,s)\right] 
\, .  \label{1laVc}
\ee
According to the definitions in Eqs.~(\ref{VsU},\ref{VsD}), 
the expression above involves the functions $F_i^{(1l)}(s)$ ($i=1,2$); 
above threshold ($s>4m^2$) they develop an imaginary part, which obviously 
cancels out in the cross section at this order. In view of that, 
in writing Eqs.~(\ref{VsU},\ref{VsD}) we explicitly indicated that only the real part
of the functions $F_i^{(1l)}$ contributes to the cross section. 

The total contribution of the four  vertex correction diagrams of 
Fig.~\ref{fig1ltot}~(c)--(f) to the differential cross section is
\bea
\frac{d \sigma_{1}^{V}(s,t)}{d \Omega} \Big|_{(1l,V)}
& = & 2 \frac{\alpha^2}{s} \Biggl[ \frac{1}{s^2} V_2^{(1l)}(t,s) + 
\frac{1}{t^2} V_2^{(1l)}(s,t) \nn\\
& & \hspace{7mm} + \frac{1}{s t} \left( V_1^{(1l)}(s,t) 
+ V_1^{(1l)}(t,s) \right) \Biggr]  \, .
\label{1lVtot}
\eea

\subsection{One-Loop Box Corrections \label{sec1lbox}}

The one-loop virtual corrections to Bhabha scattering in QED include four 
box diagrams: the direct box diagrams in the $t$- and $s$-channel 
(shown in Fig.~\ref{fig1ltot}~(g) and (h), respectively), and the  
corresponding crossed graphs (shown in  Fig.~\ref{fig1ltot}~(i) and (j)). 
The interference of the diagram in Fig.~\ref{fig1ltot}~(g) with the 
tree-level matrix element is given by
\be
\frac{d \sigma_{1}^{V}(s,t)}{d \Omega} \Big|_{(B,(g))} = 
\frac{\alpha^2}{4s} \left[ \frac{m^2}{s} \mbox{Re}B_1^{(1l)}(s,t)  + 
\frac{m^2}{t} \mbox{Re}B_2^{(1l)}(s,t) \right] \, ,
\label{1lBg} 
\ee
where the quantities $B_1^{(1l)}(s,t)$ and $B_2^{(1l)}(s,t)$ are functions of
the $s$ and $t$ Mandelstam invariants (and of the electron mass); 
they are UV finite, but develop an IR pole in $(D-4)$ (to be canceled out 
in the physical cross section once the real soft radiation is accounted 
for) so that their Laurent expansion reads 
\be 
  B_i^{(1l)}(s,t) = \frac{1}{(D-4)} B_i^{(1l,-1)}(s,t) 
                  + B_i^{(1l,0)}(s,t) 
                  + {\mathcal{O}} \Big((D-4)\Big) \ . 
\label{BiLaur} 
\ee 
The explicit expression of the various functions is given by 
Eqs.~(\ref{1lB1p},\ref{1lB1},\ref{1lB2p},\ref{1lB2}) 
in Appendix~\ref{AF}. 

The diagram (h) in Fig.~\ref{fig1ltot} is the same as diagram 
(g), in the same figure, once  we
exchange $p_2 \leftrightarrow -p_3$, which  is equivalent to the exchange 
$s \leftrightarrow t$. As a consequence, the contribution of
diagram (h) to the  differential cross section at order $\alpha^3$ is: 
\be
\frac{d \sigma_{1}^{V}(s,t)}{d \Omega} \Big|_{(B,(h))} = 
\frac{\alpha^2}{4s} \left[ \frac{m^2}{s} \mbox{Re}B_2^{(1l)}(t,s)  + 
\frac{m^2}{t} \mbox{Re}B_1^{(1l)}(t,s) \right] \, .
\label{1lBh} 
\ee

Interfering the diagram (i) in Fig.~\ref{fig1ltot} with the tree-level 
graphs one finds:
\be
\frac{d \sigma_{1}^{V}(s,t)}{d \Omega} \Big|_{(B,(i))} = 
\frac{\alpha^2}{4s} \left[ \frac{m^2}{s} B_3^{(1l)}(u,t)  - 
\frac{m^2}{t} B_2^{(1l)}(u,t) \right] \, , 
\label{boxb} 
\ee
where $B_3^{(1l)}(u,t)$ is another auxiliary function, with the same 
Laurent expansion as Eq.~(\ref{BiLaur}), whose explicit expression is 
given in Eqs.~(\ref{1lB3p},\ref{1lB3}) of Appendix \ref{AF}. 
In Eq.~(\ref{boxb}),
the functions $B_3^{(1l)}(u,t)$ and $B_2^{(1l)}(u,t)$ do not develop an
imaginary part in the physical region, so that 
$\mbox{Re}B_3^{(1l)}(u,t) = B_3^{(1l)}(u,t)$, 
and $\mbox{Re}B_2^{(1l)}(u,t) = B_2^{(1l)}(u,t)$. 

The diagram (j) of Fig.~\ref{fig1ltot} can be obtained from diagram 
(i) of the same figure, by exchanging  $p_2 \leftrightarrow -p_3$, 
so that $s \leftrightarrow t$; 
the contribution of this diagram to the cross section can then be 
obtained from Eq.~(\ref{boxb}) by replacing $t$ with $s$ in the 
arguments of the functions $B_2^{(1l)}$ and $B_3^{(1l)}$: 
\be
\frac{d \sigma_{1}^{V}(s,t)}{d \Omega} \Big|_{(B,(j))} = 
-\frac{\alpha^2}{4s} \left[ \frac{m^2}{s} \mbox{Re}B_2^{(1l)}(u,s)  - 
\frac{m^2}{t} \mbox{Re}B_3^{(1l)}(u,s) \right] \, .
\ee

Summarizing,  the contribution of the interference of the 4 
diagrams in Fig.~\ref{fig1ltot}~(g)--(j) with the tree-level matrix
element to the Bhabha scattering differential cross section is given by: 
\bea
 \frac{d \sigma_{1}^{V}(s,t)}{d \Omega} \Big|_{(1l,B)} & = & 
\frac{\alpha^2}{4s} \Biggl[ \frac{m^2}{s} \Bigl(\mbox{Re}B_1^{(1l)}(s,t)  +
\mbox{Re}B_2^{(1l)}(t,s) + B_3^{(1l)}(u,t)\nn\\ 
 &  & - \mbox{Re}B_2^{(1l)}(u,s)  
\Bigr)  + \frac{m^2}{t} \Biggl(\mbox{Re}B_2^{(1l)}(s,t) +
\mbox{Re}B_1^{(1l)}(t,s)\nn\\ 
 &  &  - B_2^{(1l)}(u,t) + \mbox{Re}B_3^{(1l)}(u,s)  
\Biggr)
\Biggr] . 
\label{1lBamp}
\eea

We checked the expression of the IR pole for the direct $t$-channel diagram
of Fig.~\ref{fig1ltot}~(g) against the corresponding result given in 
\cite{VN}, where the author calculates the one-loop virtual corrections to 
electron-muon scattering, without neglecting the two fermion masses 
and using a small photon mass $\lambda$ for regularizing the IR divergences. 
Setting the mass of the muon equal to the mass of the electron in \cite{VN}, 
we find full agreement with our result provided that the dimensional IR 
pole is replaced by 
\be
\frac{1}{(D-4)} \rightarrow \log{\left( \frac{m}{\lambda} \right)} \, .
\label{corr}
\ee

We compared also the contribution of the $s$-channel diagrams 
to the differential cross section (pole and finite part), in the 
limit $m \rightarrow 0$, with the results in \cite{BOX1l}, where 
the $\lambda$-regularization is used and the Dirac algebra is in 4 
dimensions. If we calculate the contribution of the diagrams in 
Fig.~\ref{fig1ltot}~(h) and (j) to the cross section using 
$D$-continuous dimensions for the MIs but 4 dimensions for the Dirac 
algebra, in the limit $m \rightarrow 0$ we find again agreement 
with \cite{BOX1l}, provided the correspondence of Eq.~(\ref{corr}) 
is used. When calculating all the Dirac traces also in $D$ dimensions, 
we obtain, compared to the results in \cite{BOX1l}, an extra finite 
term proportional to the coefficient of the IR pole in $(D-4)$.

Let us recall here that the expression of the one-loop scalar box integral 
in the $D$-regularization, given in \cite{us} and used in this paper, 
is the same as the result of Appendix E~(b) of \cite{Scalar} (where a 
small photon mass is still used), once the replacement (\ref{corr}) is 
carried out.

\boldmath
\section{Two-Loop $N_F=1$ Differential Cross Section \label{S2l}}
\unboldmath

In this Section we discuss the  virtual diagrams contributing to the 
Bhabha scattering differential cross section at order $\alpha^4$ 
(summed over the spin of the final states and averaged over the spin of 
the initial ones). We limit our analysis to the diagrams that include a 
closed electron loop. Beside the genuine two-loop diagrams, we have to 
consider the interference of the one-loop vacuum polarization graphs with 
the one-loop vertex and box corrections. 

\subsection{Two-Loop Vacuum Polarization Corrections}

The two-loop photon self-energy can be written in a way completely 
analogous to the one-loop case:
\be
\Pi^{(2l)}_{\mu \nu}(-p^2) = \left( \frac{\alpha}{\pi} \right)^2 
\left(p^\mu p^\nu - 
\delta_{\mu \nu} p^2\right) \Pi^{(2l)}_{0}(-p^2) \, .
\ee

The explicit expression of the UV-renormalized function $\Pi^{(2l)}_0(-p^2)$ 
in terms of HPLs is given in Eq.~(\ref{2lPi0}) of Appendix \ref{AF}.
 
The contribution of the two-loop vacuum polarization diagrams 
in Fig.~\ref{fig2ltot} (a)--(f) can be obtained by replacing 
$\Pi^{(1l)}_0$ by $\Pi^{(2l)}_0$  in Eq.~(\ref{amp1lS}):
\bea
\frac{d \sigma_{2}^{V}(s,t,m^2)}{d \Omega} \Big|_{ (2l,S)} &=& 
\frac{\alpha^2}{s} \Bigg\{
\frac{1}{s^2} \left[ s t + \frac{s^2}{2} + (t - 2 m^2)^2 \right] 
2 \mbox{Re}\Pi^{(2l)}_0(s) 
\nn\\  
& & + \frac{1}{t^2} \left[ s t + \frac{t^2}{2} + (s - 2 m^2)^2 \right]  
2 \Pi^{(2l)}_0(t) \nn\\ 
& & + \frac{1}{s t}\left[(s + t)^2 - 4 m^4\right]
 \left(\mbox{Re}\Pi^{(2l)}_0(s) +  \Pi^{(2l)}_0(t)\right)
\Bigg\} \, . 
\label{amp2lS}
\eea

\subsection{Two-Loop Vertex Corrections}

The two-loop $N_F =1$ virtual corrections to Bhabha scattering include the
four diagrams shown in Fig.~\ref{fig2ltot}~(g)--(j) (and the corresponding 
counter terms), obtained by including a vacuum polarization correction in the 
photon propagator of the one-loop vertex diagram. The Lorentz structure of 
these corrections is identical to the one-loop case, Eq.~(\ref{1lV}), 
with the form factors $F^{(1l)}_i(-p^2)$ ($i=1,2; -p^2 = s,t$) 
replaced by $F^{(2l)}_i(-p^2)$ ($i=1,2; -p^2 = s,t$). 
The explicit expressions of the UV renormalized form factors 
$F^{(2l)}_i(-p^2)$ are given in 
Eqs.~(\ref{2lF1},\ref{2lF2}) of Appendix~\ref{AF}; at variance with 
the one loop case, the two-loop contributions of the graphs 
in Fig.~\ref{fig2ltot}~(g)--(j) are infrared finite, so that no 
expansion in $(D-4)$ is needed. 

The full set of two-loop $N_F = 1$ vertex corrections to the squared amplitude 
can be written, similarly to Eq.~(\ref{1lVtot}), as 
\bea
\frac{d \sigma_{2}^{V}(s,t,m^2)}{d \Omega} \Big|_{(2l,V)}
& = & 2 \frac{\alpha^2}{s} \Biggl[ \frac{1}{s^2} V^{(2l)}_1(t,s) + 
\frac{1}{t^2} V^{(2l)}_2(s,t) \nn\\
& & \hspace{7mm}
+ \frac{1}{s t} \left( V_1^{(2l)}(s,t) \! + \! V_2^{(2l)} (t,s)\right)
\Biggr]  ,
\label{tot2lvertex}
\eea
where the functions $V_1^{(2l)}$ and $V_2^{(2l)}$ are given by 
\bea 
V_1^{(2l)}(s,t) & = & c_{1,1}(s,t) \mbox{Re}F_1^{(2l)}(t)
                  + c_{1,3}(s,t) \mbox{Re}F_2^{(2l)}(t) \, , 
\label{V1V22lU}\\ 
V_2^{(2l)}(s,t) & = & c_{2,1}(s,t) \mbox{Re}F_1^{(2l)}(t)
                  + c_{2,3}(s,t) \mbox{Re}F_2^{(2l)}(t) \, , 
\label{V1V22l}
\eea
and the coefficients $c_{1,1}(s,t)$ etc. are the same as in 
Eqs.~(\ref{ViLaur2U}--\ref{ViLaur2}).

\subsection{Two-Loop Box Corrections}

There are 8 two-loop box diagrams including a closed fermion loop, 
shown in Fig.~(\ref{fig2ltot})~(k)--(r).

Let us consider the diagram~(l) in Fig.~\ref{fig2ltot}; once we transform 
the external momenta according to $p_4 \leftrightarrow -p_1$ and $p_2 
\leftrightarrow - p_3$, this diagram becomes identical to diagram~(k). 
Since this set of transformations leaves unchanged the Mandelstam variables 
$s$ and $t$, the contributions of diagrams (k) and~(l) to the differential 
cross section at order $\alpha^4 (N_F =1)$ is identical. Similar 
considerations hold for the pairs of diagrams~(m)-(n), (o)-(p), and (q)-(r).  
The diagrams (k), (m), (o), and~(q) in Fig.~\ref{fig2ltot} are linked by the
same symmetry relations that connect the four one-loop box diagrams
in Fig.~\ref{fig1ltot}~(g), (i), (h), and~(j), respectively (see the
discussion in Section~\ref{sec1lbox}). 
Consequently, in strict analogy to the one-loop case, the contributions 
of the two-loop box diagrams in  Fig.~(\ref{fig2ltot}) to the Bhabha 
scattering differential cross section can be written, similarly to 
Eq.~(\ref{1lBamp}), as 
\bea
\frac{d \sigma_{2}^{V}(s,t,m^2)}{d \Omega} \Big|_{(2l,B)}  & = & 
 - 2 \frac{\alpha^2}{4s}
\Bigl[ \frac{m^2}{s} \Bigl(\mbox{Re}B_1^{(2l)}(s,t) +
\mbox{Re}B_2^{(2l)}(t,s)+ B_3^{(2l)}(u,t) \nn\\ 
 & &  - \mbox{Re}B_2^{(2l)}(u,s)  
\Bigr)  + \frac{m^2}{t} \Bigl(
\mbox{Re}B_2^{(2l)}(s,t) +
\mbox{Re}B_1^{(2l)}(t,s) \nn\\ 
 & &- B_2^{(2l)}(u,t) + \mbox{Re}B_3^{(2l)}(u,s)  
\Bigr) 
\Bigr] \, .
\label{2lBamp}
\eea
We choose to factor out in Eq.~(\ref{2lBamp}) an overall 
multiplicative factor $-2$ accounting for the minus sign coming from the 
closed fermion loops and the identity of the contributions from the 
pairs of diagrams (k)-(l), (m)-(n), (o)-(p), and (q)-(r) of 
Fig.~\ref{fig2ltot}.

The Laurent expansion of the UV-renormalized functions $B_{i}^{(2l)}$ reads:
\be
B_i^{(2l)}(s,t) = \frac{1}{(D-4)} B_i^{(2l,-1)}(s,t) 
                  + B_i^{(2l,0)}(s,t) 
                  + {\mathcal{O}} \Big( (D-4) \Big) \ . 
\label{BiLaur2l} 
\ee

The expressions of the coefficients of the Laurent expansion, $B_i^{(2l,j)}$
($j=-1,0$), are given in Eqs.~(\ref{B12lM1}--\ref{2lB3}) in Appendix \ref{AF}.

\subsection{Two-Loop Reducible Corrections}

The complete set of two-loop diagrams to the Bhabha scattering process
containing a closed electron loop includes the four 1-loop reducible 
diagrams, shown in Fig.~\ref{fig2ltot}~(s)--(v). They can be obtained from 
the one-loop  vertex diagrams by inserting a one-loop photon self-energy 
correction in  the photon
propagator connecting the two fermionic currents. Since the  vacuum
polarization corrections amount to a multiplicative factor 
$\Pi_0^{(1l)}(-p^2)$, the contribution of such diagrams to the the 
differential
cross section at order $\alpha^4$ is easily obtained by appending the 
appropriate $\Pi_0^{(1l)}(-p^2)$ ($p^2 = -s, -t$) to the various terms in
Eq.~(\ref{1lVtot}) and properly accounting for the $(D-4)$ expansion. 
In close analogy to Eq.~(\ref{1lVtot}) we write therefore 
\bea
\frac{d \sigma_{2}^{V}(s,t,m^2)}{d \Omega} \Big|_{(2l,R)} 
&=& 2 \frac{\alpha^2}{s} \Biggl[\frac{1}{s^2} V^R_2(t,s)  + 
\frac{1}{t^2} V^R_2(s,t)  \nn \\
& & + \frac{1}{s t}\left(V^R_1(s,t) + V^R_1 (t,s)\right)
\Biggr] \ , \label{2lRtot} 
\eea
where the $V_i^R$ are to be Laurent expanded as in Eq.~(\ref{ViLaur}). 
The coefficients of the Laurent expansions are 
\bea 
 V_i^{(R,-1)}(s,t) & = &  c_{i,1}(s,t) \ 
          \mbox{Re}\Big(\ F_1^{(1l,-1)}(t)\ \Pi^{(1l,0)}_0(t) \ \Big) 
                                                            \ , \\ 
 V_i^{(R,0)}(s,t) & = &  c_{i,1}(s,t) \ 
          \mbox{Re}\Big(\  F_1^{(1l,0)}(t)\ \Pi^{(1l,0)}_0(t) 
                        + F_1^{(1l,-1)}(t)\ \Pi^{(1l,1)}_0(t) \ \Big) \\ 
                    & & + c_{i,2}(s,t) \ 
          \mbox{Re}\Big(\ F_1^{(1l,-1)}(t)\ \Pi^{(1l,0)}_0(t) \ \Big) \\ 
                    & & + c_{i,3}(s,t) \ 
          \mbox{Re}\Big(\ F_2^{(1l,0)}(t)\ \Pi^{(1l,0)}_0(t) \ \Big) \ , 
\label{2lR} 
\eea 
where the coefficients $c_{i,j}(s,t)$ are given by 
Eqs.~(\ref{ViLaur2U}--\ref{ViLaur2}), 
the Laurent coefficients of $\Pi^{(1l)}_0(t)$ and of the $F_i^{(1l)}(t)$ 
by the Eqs.~(\ref{A1lS}-\ref{F2ren}). 

\subsection{Products of Two One-Loop Corrections}

Finally, to conclude our catalog of contributions to the Bhabha 
scattering cross section at order $\alpha^4 (N_F =1)$, we need to discuss 
the interference between the one-loop vacuum polarization diagrams 
of Fig.~\ref{fig1ltot}~(a), (b) and the one-loop vertex and box graphs 
of Fig.~\ref{fig1ltot}~(c)--(f) and (g)--(j). Since the one-loop vacuum 
polarization insertion factorizes in the tree-level diagram and the 
multiplicative factor $\Pi^{(1l)}_0(-p^2)$ ($-p^2 = s,t$), the contribution 
of these interferences among one-loop diagrams is easily obtained by replacing 
the tree-level photon propagators $1/s$ and $1/t$ in Eqs.~(\ref{1lVtot}) 
and~(\ref{1lBamp}) by $\mbox{Re} \Pi^{(1l)}_0(s)/s$ and 
$\Pi^{(1l)}_0(t)/t$, respectively. 

In this way one obtains for the interference of the self mass 
graphs in Fig.~(\ref{fig1ltot})~(a), (b) and the vertex graphs 
in Fig.~(\ref{fig1ltot})~(c)--(f) 
\bea
\frac{d \sigma_{2}^{V}(s,t,m^2)}{d \Omega} \Big|_{(S,V)} &= & 
   2 \frac{\alpha^2}{s} \mbox{Re} \Bigg[ 
      \frac{1}{s^2} V_2^{(1l)}(t,s)\bigg( \Pi^{(1l)}_0(s) \bigg)^* \nn\\ 
  && \hspace{13mm} 
    + \frac{1}{t^2} V_2^{(1l)}(s,t)\bigg( \Pi^{(1l)}_0(t) \bigg)^* \nn\\ 
  && \hspace{13mm} 
    + \frac{1}{st} V_1^{(1l)}(s,t)\bigg( \Pi^{(1l)}_0(s) \bigg)^* \nn\\ 
  && \hspace{13mm} 
    + \frac{1}{st} V_1^{(1l)}(t,s)\bigg( \Pi^{(1l)}_0(t) \bigg)^* \Bigg] 
                                \ , \label{1lS1lV} 
\eea 
where the functions $V_i^{(1l)}(s,t)$ are the same as in 
Eqs.~(\ref{1laV},\ref{VsU},\ref{VsD}). As they have the Laurent expansion of 
Eq.~(\ref{ViLaur}), the above cross section can be Laurent expanded in the 
same way. One has for instance, in the notation of Eqs.~(\ref{ViLaur}) and 
(\ref{1lsLaur}):
\bea 
  V_2^{(1l)}(t,s)\bigg( \Pi^{(1l)}_0(s) \bigg)^* &=& \frac{1}{(D-4)} 
  V_2^{(1l,-1)}(t,s)\bigg( \Pi^{(1l,0)}_0(s) \bigg)^* \nn\\ 
 & & + V_2^{(1l,0)}(t,s)\bigg( \Pi^{(1l,0)}_0(s) \bigg)^* 
  + V_2^{(1l,-1)}(t,s)\bigg( \Pi^{(1l,1)}_0(s) \bigg)^* \nn\\ 
  & & + {\mathcal{O}} \Big((D-4)\Big) \ , 
\label{1lS1lVLaur} 
\eea 
and similar expressions for all the other terms of Eq.~(\ref{1lS1lV}). 

For the interference of the self mass graphs in Fig.~\ref{fig1ltot}~(a), (b) 
and the box graphs in Fig.~\ref{fig1ltot}~(g)--(j) one finds, in analogy with 
Eq.~(\ref{1lBamp}), 
\bea
\frac{d \sigma_{2}^{V}(s,t,m^2)}{d \Omega} \Big|_{(S,B)} &= & 
    \frac{\alpha^2}{4s} \mbox{Re} \Bigg[ 
    \frac{1}{s} \bigg( B_1^{(1l)}(s,t) + B_2^{(1l)}(t,s) \nn\\  
& & \hspace{10mm}
                + B_3^{(1l)}(u,t) - B_2^{(1l)}(u,s) \bigg) 
           \bigg( \Pi^{(1l)}_0(s) \bigg)^*           \nn\\ 
& &  \hspace{10mm} 
    + \frac{1}{t} \bigg( B_2^{(1l)}(s,t) + B_1^{(1l)}(t,s) \nn\\
& &  \hspace{10mm} 
                - B_2^{(1l)}(u,t) + B_3^{(1l)}(u,s)  \bigg) 
           \bigg( \Pi^{(1l)}_0(t) \bigg)^* 
    \Biggr]                                 \ , \label{1lS1lB} 
\eea 
where the functions $B_i^{(1l)}(s,t)$ are the same as in 
Eqs.~(\ref{1lBg},\ref{boxb}), with the Laurent expansion of 
Eq.~(\ref{BiLaur}). A similar expansion holds for Eq.~(\ref{1lS1lB}); 
in analogy with Eq.~(\ref{1lS1lVLaur}), one has for instance:
\bea 
  B_1^{(1l)}(s,t)\bigg( \Pi^{(1l)}_0(s) \bigg)^* &=& \frac{1}{(D-4)} 
  B_1^{(1l,-1)}(s,t)\bigg( \Pi^{(1l,0)}_0(s) \bigg)^* \nn\\ 
 & & + B_1^{(1l,0)}(s,t)\bigg( \Pi^{(1l,0)}_0(s) \bigg)^* 
  + B_1^{(1l,-1)}(s,t)\bigg( \Pi^{(1l,1)}_0(s) \bigg)^* \nn\\ 
  & & + {\mathcal{O}} \Big((D-4)\Big) \ , 
\label{1lS1lBLaur} 
\eea 
and similar expressions for all the other terms of Eq.~(\ref{1lS1lB}).

\section{Conclusions}

In this paper, we considered the virtual radiative  corrections to the Bhabha
scattering differential cross section  (summed over the spins of the final 
states and averaged over the spins of the initial ones) in pure QED; we 
calculated  the contribution of all the one-loop diagrams (order $\alpha^3$)
and of the subset of two-loop diagrams 
involving an electron loop (order $\alpha^4 (N_F=1)$). 

The provided result does not rely on any kind of approximation; 
the full dependence of the cross section on the electron mass $m$ and on
the Mandelstam invariants $s$ and $t$ was retained.

Both UV and IR divergences were regularized within the framework of
dimensional regularization. The UV renormalization was performed
in the {\it on-shell} renormalization scheme.

The reduction of the integrals to the Master Integrals was carried out
employing the Laporta algorithm. 
The MIs involved in the calculation of the cross 
section were evaluated in two previous works \cite{us,RoPieRem1}, by means of
the differential equation method.  

All the results are written in terms of 1- and 2-dimensional HPLs of 
maximum weight 3.

The differential cross sections at the one- and two-loop level can be 
obtained adding up the various contributions discussed in the paper.

In particular,  the one-loop virtual cross section is the sum of 
the contributions
given in Eqs.~(\ref{amp1lS},\ref{1lVtot},\ref{1lBamp}) respectively:
\be
\frac{d \sigma_{1}^{V}(s,t,m^2)}{d \Omega} = 
\frac{d \sigma_{1}^{V}(s,t,m^2)}{d \Omega} \Big|_{(1l,S)} +
\frac{d \sigma_{1}^{V}(s,t,m^2)}{d \Omega} \Big|_{(1l,V)} +
\frac{d \sigma_{1}^{V}(s,t,m^2)}{d \Omega} \Big|_{(1l,B)} 
\, ;
\ee
 the two-loop differential cross section is the sum of the expressions given
in Eqs.~(\ref{amp2lS},\ref{tot2lvertex}, \ref{2lBamp},\ref{2lRtot},
\ref{1lS1lV},\ref{1lS1lB} ):
\bea
\! \! \frac{d \sigma_{2}^{V}(s,t,m^2)}{d \Omega} \! \! & = & \, 
\frac{d \sigma_{2}^{V}(s,t,m^2)}{d \Omega} \Big|_{(2l,S)} \! \! + \! 
\frac{d \sigma_{2}^{V}(s,t,m^2)}{d \Omega} \Big|_{(2l,V)} \! \! + \! 
\frac{d \sigma_{2}^{V}(s,t,m^2)}{d \Omega} \Big|_{(2l,B)} \nn\\
\! \! \! & & \! \! \! + \, \frac{d \sigma_{2}^{V}(s,t,m^2)}{d \Omega} \Big|_{(2l,R)} 
\! \! \! + \! 
\frac{d \sigma_{2}^{V}(s,t,m^2)}{d \Omega} \Big|_{(S,V)} \! \! \! + \! 
\frac{d \sigma_{2}^{V}(s,t,m^2)}{d \Omega} \Big|_{(S,B)} \! \! .
\eea

The differential cross section still includes single poles in $(D-4)$, 
of IR origin, to be canceled by the inclusion of the contribution of the 
diagrams involving the emission of a real soft photon, not discussed 
in this paper.

\section{Acknowledgments}

We are grateful to J.~Vermaseren for his kind assistance in the use
of the algebra manipulating program {\tt FORM}~\cite{FORM}, by which
all our calculations were carried out. R.~B. and A.~F. wish to thank 
B.~Tausk for very useful discussions. We kindly acknowledge discussions 
with M. Argeri for the vacuum polarization in terms of harmonic 
polylogarithms \cite{Mario}. The work of R.~B. was supported by the 
European Union under contract HPRN-CT-2000-00149. 
The work of A. F. was partially supported by the DFG-Forschergruppe 
``\emph{Quantenfeldtheorie, Computeralgebra und
 Monte-Carlo-Simulation}''.
The work of P.~M. was supported by the USA DoE under the grant 
DE-FG03-91ER40662, Task J. 

We would like to thank C.M.~Carloni~Calame for pointing out a sign mistake
in the one-loop vacuum polarization generating some sign mistakes in the 
formulas in appendices of the {\tt hep-ph/0405275v1} version of the present 
paper; these mistakes have been corrected in {\tt hep-ph/0405275v2}. Useful 
discussions with F.~Piccinini are also gratefully acknowledged \cite{CP}.


\appendix

\section{Auxiliary Functions \label{AF}}

In this Appendix, we give the explicit expressions of the auxiliary 
functions used throughout the paper. They involve 1- and 2-dimensional 
Harmonic PolyLogarithms (HPLs), introduced and discussed 
in~\cite{Polylog,Polylog3,Polylog2,Polylog4,Polylog5,us}, of 
arguments $x,y$ related to the kinematical Mandelstam variables 
$s,t$ by the relations 
\bea
& & 
 -s = (p_1+p_2)^2= P^2 = m^2 \frac{(1-x)^2}{x} \, , \qquad
x = \frac{\sqrt{P^2+4m^2}- \sqrt{P^2}}{\sqrt{P^2+4m^2}+\sqrt{P^2}} \, , 
                                                      \label{defx} \\ 
& & 
-t  = (p_1-p_3)^2= Q^2  = m^2 \frac{(1-y)^2}{y} \, , \qquad
y = \frac{\sqrt{Q^2+4m^2}- \sqrt{Q^2}}{\sqrt{Q^2+4m^2}+\sqrt{Q^2}} \, . 
                                                      \label{defy} 
\eea 
For completeness, we introduce also 
\be 
 -u = (p_1-p_4)^2= V^2 = m^2 \frac{(1-z)^2}{z} \, , \qquad
z  = \frac{\sqrt{V^2+4m^2}- \sqrt{V^2}}{\sqrt{V^2+4m^2}+\sqrt{V^2}} \, ; 
\label{defz} 
\ee 
the variable $z$ is indeed related to $x$ and $y$ through the relation
\be
- u = s+ t -4 m^2 = m^2 \frac{1}{x}(x+y) \left( x+ \frac{1}{y} \right) = 
m^2 \frac{(1-z)^2}{z} 
\, .
\ee
If $s=-P^2$ is space-like ($s$ negative, $P^2$ positive), $x$ as given by 
Eq.~(\ref{defx}) is real and positive, varying from $x=1$ at $s=0$ to $x=0$ 
at $s=-\infty$, and similarly for the variable $y$. 
\par 
In the explicit formulas which will follow, the one-dimensional HPLs are 
written, according to~\cite{Polylog}, as $H(0,1;x), H(-1,0;y)$ etc.; 
examples of two-dimensional HPLs are $G(-1/y;x), G(-y,0,0;x)$ etc., 
for their definition see~\cite{us}. 
\par 
The arguments of 
the auxiliary functions will be taken to be space-like, ($-P^2$ in the 
case of a single argument, $(-P^2,-Q^2)$ if the functions depend on 
two variables) so that the corresponding 
$x$ and $y$ vary both in the $(0,1)$ range and all HPLs are real. 
The time-like regions can be recovered by analytic continuation. 

Let us consider, for definiteness, the variable $s=-P^2$. The continuation 
to the time-like region is performed, according to the usual 
$i \epsilon$-prescription, by setting 
\be                          
P^2 = - s - i \epsilon \, ,             
\ee
when $s>0$. For $0<s<4m^2$, $\sqrt{-s-i\epsilon}=-i \sqrt{s}$ 
and Eq.~(\ref{defx}) gives $x=r$, where $r$ is the 
phase factor 
\be
r = \frac{\sqrt{4 m^2-s} + i\sqrt{s}}{\sqrt{4 m^2-s} 
- i\sqrt{s}} = e^{i 2 \phi} \, ,
\ee
with 
\be
\phi = \arctan{\sqrt{\frac{s}{4m^2-s}}} \, .
\ee
Above threshold, $s>4m^2$, we define 
\be
  x' = \frac{\sqrt{s}-\sqrt{s-4m^2}}{\sqrt{s}+\sqrt{s-4m^2}} \, ,
\ee 
with $x'=1$ at $s=4m^2$ and $x'=0$ at $s=\infty$ and the continuation 
in $x$ is performed by the replacement 
\be 
   x = - x' + i \epsilon \ . 
\ee 
Remembering that
\[  H(0;x) = \ln{x} \ , \] 
the continuation from space-like $s$ to $s$ above threshold is then 
performed by 
\be
   H(0;x) = H(0;-x'+i\epsilon) = H(0;x')+i \pi \, .
\ee
Another less trivial example is 
\bea
H(-1,0;x) &=& H(-1;x) H(0;x) - H(0,-1;x) \nn \\
          &=&  H(-1;-x' + i \epsilon) H(0;-x' + i \epsilon) 
                           - H(0,-1; -x'+ i \epsilon )    \nn \\
   &=&  - H(1;x') \bigg( H(0;x') + i \pi \bigg) + H(0,1;x') \, , 
\eea 
where an algebraic property of the 1-dimensional HPLs (shuffle algebra) 
has been used in the first line and the relation reversing the sign 
of the argument in the last line. 
For more details on the analytic continuation of the HPLs we refer to 
the already quoted papers. 


The contribution of the two-loop box diagrams to the differential cross
section includes a subset of the 2-dimensional HPLs introduced in \cite{us};
this subset can be expressed in terms of Nielsen
polylogarithms (whose argument depends however on both $x$ and $y$,
i.e. on both $s$ and $t$), explicitly given in Eqs.~(C.21,C.22,C.27-C.30)
of \cite{us}. As an example of analytic continuation which applies to
those cases, consider for instance $\mbox{Li}_2(-x/y)$; if $x,y$ correspond
to $s,t$ when they are both spacelike, the continuation to physical
(timelike) $s$
and spacelike $t$ implies $x \to -x' +i\epsilon,\ $ with $x'<y$, so that
one has simply $ \mbox{Li}_2(-x/y) \to \mbox{Li}_2(x'/y) $.
The case of timelike $t$, spacelike $s$, on the other hand, implies
$y\to -y' +i\epsilon,\ $ with $y'<x$, so that the continuation is given by
\be
  \mbox{Li}_2\left(-\frac{x}{y}\right) \to
  \mbox{Li}_2\left(\frac{x}{y'}+i\epsilon\right) =
  - \mbox{Li}_2\left(\frac{y'}{x}\right)
  - \frac{1}{2}\ln^2\left(\frac{y'}{x}\right) + 2 \zeta(2)
  - i\pi \ln\left(\frac{y'}{x}\right) \, .
\ee

Similarly for $\mbox{Li}_3$ we have:
\bea
  \mbox{Li}_3\left(-\frac{x}{y}\right) \to
\mbox{Li}_3 \left( \frac{x}{y'} + i \epsilon \right) & = &
\mbox{Li}_3 \left( \frac{y'}{x} \right) + \frac{1}{6}\ln^3{\left( \frac{y'}{x} \right)} 
- 2 \zeta(2) \ln{\left( \frac{y'}{x} \right)} \nn\\
& & + i \pi \frac{1}{2} \ln^2{\left( \frac{y'}{x} \right)} \, .
\eea

In the following subsections we provide the explicit expressions of the 
auxiliary functions.

\subsection{Auxiliary Functions for the One-Loop Cross Section}


\bea 
\Pi^{(1l,0)}_0 (-P^2) & = & -\Bigg\{
            \frac{5}{9}
          - \frac{4}{3 (1-x)^2}
          + \frac{4}{3 (1-x)}
          - \Biggl[ \frac{1}{3} 
          + \frac{4}{3 (1-x)^3}  \nn\\
\! & & 
          - \frac{2}{(1-x)^2} \Biggr] H(0;x)\Bigg\} \, , 
\label{A1lS} \\ 
\Pi^{(1l,1)}_0 (-P^2) & = & -\Bigg\{
             \frac{14}{27} \! 
          -  \! \frac{16}{9 (1-x)^2} \! 
          +  \! \frac{16}{9 (1-x)} \! 
          +  \! \Biggl[ \frac{2}{3 (1-x)^3} \! 
          -  \! \frac{1}{(1-x)^2}
          \! + \! \frac{1}{6} \Biggr] \zeta(2) \nn\\
\! & & 
          - \Biggl[ \frac{5}{18} \! 
          + \! \frac{16}{9 (1-x)^3}\! 
          - \! \frac{8}{3 (1-x)^2}\! 
          + \! \frac{1}{3 (1-x)} \Biggr] H(0;x) \nn\\
\! & & 
         \!  - \Biggl[ \frac{1}{6}
          + \frac{2}{3 (1-x)^3}
          - \frac{1}{(1-x)^2} \Biggr] H(0,0;x) \nn\\ 
\! & & 
          + \Biggl[ \frac{1}{3}
          + \frac{4}{3 (1-x)^3} 
         \!  - \frac{2}{ (1-x)^2} \Biggr] H(-1,0;x)\Bigg\} \ , 
\label{A1ls1} 
\eea

\bea
F^{(1l,-1)}_{1}(-P^2) & = &   
1 - \biggl[ 1 - \frac{1}{(1-x)} - \frac{1}{(1+x)} \biggr] H(0;x) \ , 
\label{F1renp} \\
F^{(1l,0)}_{1}(-P^2) & = & - 1 
          - \frac{1}{2} \biggl[ 
     \frac{1}{2} \! 
   - \! \frac{1}{(1\! +\! x)}
          \biggr] H(0;x) \nn\\ 
& & - \frac{1}{2} \biggl[ 1 \! - \! 
   \frac{1}{(1\! -\! x)} \! - \! \frac{1}{(1\! +\! x)} \biggr] \bigl[ 
   \zeta(2)  - 2 H(0;x) - H(0,0;x)  \nn\\
& &   + 2 H(-1,0;x) \bigr] 
\label{F1ren}
\, , \\ 
F^{(1l,0)}_{2}(-P^2) & = &  - \frac{1}{2} \left[ \frac{1}{(1-x)} -
\frac{1}{(1+x)} \right] H(0;x) \label{F2ren} \, .
\eea

\bea 
B_1^{(1l,-1)}(-P^2,-Q^2) &=&  \Biggl(
          - 48
          - \frac{8}{ x^2 (1-y)^2}
          \! + \! \frac{ 8}{ x^2 (1-y)}
          \! + \! \frac{ 32}{ x (1-y)^2}
          - \frac{ 32}{ x (1-y)} \nn\\
& & 
          - \frac{ 16}{ x}
          - \frac{ 32 x}{ (1-y)^2}
          \! + \! \frac{ 32 x}{ (1-y)}
          + 16 x
          \! + \! \frac{ 8 x^2}{ (1-y)^2}
          - \frac{ 8 x^2}{ (1-y)} \nn\\
& & 
          - \frac{ 8}{ y (1+x)}
          - \frac{ 8}{ y (1-x)}
          \! + \! \frac{ 8}{ y}
          - \frac{ 8 y}{ (1+x)}
          - \frac{ 8 y}{ (1-x)}
          + 8 y \nn\\
& & 
          - \frac{ 96}{ (1+x) (1-y)^2}
          \! + \! \frac{ 96}{ (1+x) (1-y)}
          \! + \! \frac{ 80}{ (1+x)} \nn\\
& & 
          \! + \! \frac{ 32}{ (1-x) (1-y)^2}
          - \frac{ 32}{ (1-x) (1-y)}
          \! + \! \frac{ 16}{ (1-x)}\nn\\
& & 
          \! + \! \frac{ 32}{ (1-y)^2} 
          - \frac{ 32}{ (1-y)}
          \Biggr) H(0;x) \, ,
\label{1lB1p} 
\eea

\bea
B_1^{(1l,0)}(-P^2,-Q^2) &=&  \zeta(2) \Biggl( 
            \frac{ 8}{ x^2 (1+y)}
          - \frac{ 8}{ x^2 (1-y)}
          - \frac{ 32}{ x (1+y)^3}
           +  \frac{ 48}{ x (1+y)^2} \nn\\
& &
          - \frac{ 88}{ x (1+y)}
           +  \frac{ 40}{ x (1-y)}
           +  \frac{ 12}{ x}
          - \frac{ 32 x}{ (1+y)^3}
           +  \frac{ 48 x}{ (1+y)^2} \nn\\
& &
          - \frac{ 88 x}{ (1+y)}
           +  \frac{ 40 x}{ (1-y)}
          + 20 x
           +  \frac{ 8 x^2}{ (1+y)}
          - \frac{ 8 x^2}{ (1-y)} \nn\\
& &
          - \frac{ 4}{ y (1-x)}
          - \frac{ 6}{ y}
          - \frac{ 4 y}{ (1-x)}
          + 10 y
           +  \frac{ 32}{ (1+x)} \nn\\
& &
           +  \frac{ 64}{ (1+y)^3}
          - \frac{ 96}{ (1+y)^2}
           +  \frac{ 208}{ (1+y)}
          - \frac{ 48}{ (1-y)}
          -  80 \Biggr) \nn\\
& &
       +   \Biggl(
          - \frac{ 4}{ x^2 (1-y)^2}
           +  \frac{ 4}{ x^2 (1-y)}
           +  \frac{ 8}{ x (1-y)^2}
          - \frac{ 8}{ x (1-y)}
          - \frac{ 2}{ x} \nn\\
& &
          - \frac{ 8 x}{ (1-y)^2}
           +  \frac{ 8 x}{ (1-y)}  
          +  2 x
           +  \frac{ 4 x^2}{ (1-y)^2}
          - \frac{ 4 x^2}{ (1-y)} \nn\\
& &
           +  \frac{ 4}{ y (1+x)}
          - \frac{ 8 }{y (1-x)}
           +  \frac{ 2 }{y}
           +  \frac{ 4 y}{ (1+x)}  
          - \frac{ 8 y }{(1-x)} \nn\\
& &
          +  2 y 
          - \frac{ 32}{ (1+x) (1-y)^2}
          + \frac{ 32}{ (1+x) (1-y)}
          + \frac{ 16}{ (1-y)^2}   \nn\\
& &
          - \frac{ 16}{ (1-y)}
          \Biggr) H(0;x)
       +   \Biggl(
            4
          - \frac{ 8}{ x (1+y)^2}
           +  \frac{ 8}{ x (1+y)} \nn\\
& &
          - \frac{ 2}{ x}  
          - \frac{ 8 x}{ (1+y)^2}
           +  \frac{ 8 x}{ (1+y)}
          -  2 x
          - \frac{ 2}{ y}
          -  2 y
          + \frac{ 16}{ (1+y)^2} \nn\\
& &
          - \frac{ 16}{ (1+y)}
          \Biggr) H(0;y)  
       +   \Biggl(
            8
          - \frac{ 16}{ x (1+y)^2}
          + \frac{ 16}{ x (1+y)} \nn\\
& &
          - \frac{ 4}{ x}
          - \frac{ 16 x }{(1+y)^2}
           +  \frac{ 16 x}{ (1+y)}
          -  4 x
          - \frac{ 4}{ y}
          -  4 y
           +  \frac{ 32}{ (1+y)^2} \nn\\
& &
          - \frac{ 32}{ (1+y)}
          \Biggr) H(1;y)
       +   \Biggl(
          -  32
          - \frac{ 8}{ x}  
          +  8 x 
          - \frac{ 8}{ y (1-x)}
           +  \frac{ 4}{ y} \nn\\
& &
          - \frac{ 8 y}{ (1-x)}
          +  4 y
           +  \frac{ 64}{ (1+x)}
          \Biggr) H(-1,0;x)  
       +   \Biggl(
            8
           +  \frac{ 4}{ x^2 (1-y)^2} \nn\\
& &
          - \frac{ 4}{ x^2 (1-y)}
          - \frac{ 16}{ x (1-y)^2}
           +  \frac{ 16}{ x (1-y)}
           +  \frac{ 4}{ x}  
           +  \frac{ 16 x }{(1-y)^2} \nn\\
& &
          - \frac{ 16 x}{ (1-y)}
          -  4 x
          - \frac{ 4 x^2}{ (1-y)^2}
           +  \frac{ 4 x^2}{ (1-y)}
           +  \frac{ 4}{ y (1+x)}
          - \frac{ 2}{ y}   \nn\\
& &
           +  \frac{ 4 y}{ (1+x)}
          -  2 y
           +  \frac{ 48}{ (1+x) (1-y)^2}
          - \frac{ 48}{ (1+x) (1-y)} \nn\\
& &
          - \frac{ 8 }{(1+x)}  
          - \frac{ 16}{ (1-x) (1-y)^2}
           +  \frac{ 16}{ (1-x) (1-y)}
          - \frac{ 8}{ (1-x)} \nn\\
& &
          - \frac{ 16 }{(1-y)^2}
           +  \frac{ 16 }{(1-y)}
          \Biggr) H(0;x) H(0;y)
       +   \Biggl(
            16
           +  \frac{ 8}{ x^2 (1-y)^2} \nn\\
& &
          - \frac{ 8}{ x^2 (1-y)}
          - \frac{ 32}{ x (1-y)^2}
           +  \frac{ 32}{ x (1-y)}
           +  \frac{ 8}{ x}
           +  \frac{ 32 x}{ (1-y)^2} \nn\\
& &
          - \frac{ 32 x}{ (1-y)}
          -  8 x
          - \frac{ 8 x^2}{ (1-y)^2}
           +  \frac{ 8 x^2}{ (1-y)}
           +  \frac{ 8}{ y (1+x)}
          - \frac{ 4}{ y} \nn\\
& &
           +  \frac{ 8 y}{ (1+x)}
          - 4 y
           +  \frac{ 96}{ (1+x) (1-y)^2}
          - \frac{ 96}{ (1+x) (1-y)} \nn\\
& &
          - \frac{ 16}{ (1+x)}
          - \frac{ 32}{ (1-x) (1-y)^2}
           +  \frac{ 32}{ (1-x) (1-y)}
          - \frac{ 16 }{(1-x)} \nn\\
& &
          - \frac{ 32}{ (1-y)^2}
           +  \frac{ 32}{ (1-y)}
          \Biggr) H(0;x) H(1;y)
       +   \Biggl(
            16 
          +  \frac{ 4}{ x} 
          -  4 x \nn\\
& &
          +  \frac{ 4}{ y (1-x)} 
          - \frac{ 2}{ y}
          +  \frac{ 4 y}{ (1-x)}
          - 2 y
          - \frac{ 32}{ (1+x)}
          \Biggr) H(0,0;x)  \nn\\
& & 
       +   \Biggl(
          -  16
           +  \frac{ 2}{ x^2 (1+y)}
          - \frac{ 2}{ x^2 (1-y)}
          - \frac{ 8 }{x (1+y)^3}
           +  \frac{ 12 }{x (1+y)^2}   \nn\\
& &
          - \frac{ 22}{ x (1+y)}
           +  \frac{ 10}{ x (1-y)}
           +  \frac{ 4}{ x}
          - \frac{ 8 x }{(1+y)^3}
           +  \frac{ 12 x}{ (1+y)^2} \nn\\
& &
          - \frac{ 22 x }{(1+y)}  
           +  \frac{ 10 x}{ (1-y)}
          +  4 x
           +  \frac{ 2 x^2}{ (1+y)}
          - \frac{ 2 x^2}{ (1-y)}
          - \frac{ 2}{ y}
          +  2 y \nn\\
& &
           +  \frac{ 16}{ (1+y)^3}
          - \frac{ 24 }{(1+y)^2}  
           +  \frac{ 52}{ (1+y)}
          - \frac{ 12 }{(1-y)}
          \Biggr) H(0,0;y) \nn\\
& &
       +   \Biggl(
          -  32
           +  \frac{ 4}{ x^2 (1+y)}
          - \frac{ 4}{ x^2 (1-y)}  
          - \frac{ 16}{ x (1+y)^3}
           +  \frac{ 24}{ x (1+y)^2} \nn\\
& &
          - \frac{ 44}{ x (1+y)}
           +  \frac{ 20}{ x (1-y)}
           +  \frac{ 8}{ x}
          - \frac{ 16 x}{ (1+y)^3}  
           +  \frac{ 24 x}{ (1+y)^2} \nn\\
& &
          - \frac{ 44 x }{(1+y)} 
           +  \frac{ 20 x}{ (1-y)}
          +  8 x
           +  \frac{ 4 x^2 }{(1+y)}
          - \frac{ 4 x^2}{ (1-y)} 
          - \frac{ 4}{ y}
          +  4 y \nn\\
& &
           +  \frac{ 32}{ (1+y)^3}
          - \frac{ 48}{ (1+y)^2}
           +  \frac{ 104}{ (1+y)}
          - \frac{ 24}{ (1-y)}
          \Biggr) H(0,1;y) \, , 
\label{1lB1}
\eea

\bea
B_2^{(1l,-1)}(-P^2,-Q^2) &=&  \Biggl(
          -  48
          - \frac{ 16}{ x^2 (1\!-\!y)^2}
          \! + \! \frac{ 16}{ x^2 (1\!-\!y)}
          - \frac{ 16}{ x}
          + 16 x
          \! + \! \frac{ 16 x^2}{ (1\!-\!y)^2} \nn\\
& &
          - \frac{ 16 x^2}{ (1\!-\!y)}
          - \frac{ 8}{ y (1+x)}
          - \frac{ 8}{ y (1-x)}
          \! + \! \frac{ 8}{ y}
          - \frac{ 8 y}{ (1+x)}\nn\\
& &
          - \frac{ 8 y}{ (1-x)} 
          + 8 y
          - \frac{ 64}{ (1+x) (1\!-\!y)^2}
          \! + \! \frac{ 64}{ (1+x) (1\!-\!y)}\nn\\
& &
          \! + \! \frac{ 80}{ (1+x)} 
          - \frac{ 64}{ (1-x) (1\!-\!y)^2}
          \! + \! \frac{ 64}{ (1-x) (1\!-\!y)}
          \! + \! \frac{ 16}{ (1-x)}\nn\\
& &
          \! + \! \frac{ 64}{ (1\!-\!y)^2} 
          - \frac{ 64}{ (1\!-\!y)}
          \Biggr) H(0;x) \ , 
\label{1lB2p}
\eea

\bea
B_2^{(1l,0)}(-P^2,-Q^2) &=&  \zeta(2) \Biggl( 
    \frac{16}{ x^2 (1+y)}
          - \frac{ 16}{ x^2 (1-y)}
          - \frac{ 64}{ x (1+y)^3}
           +  \frac{ 96}{ x (1+y)^2} \nn\\
& &
          - \frac{ 80}{ x (1+y)}
           +  \frac{ 16}{ x (1-y)}
           +  \frac{ 12}{ x}
          - \frac{ 64 x}{ (1+y)^3}
           +  \frac{ 96 x}{ (1+y)^2} \nn\\
& &
          - \frac{ 80 x}{ (1+y)}
           +  \frac{ 16 x}{ (1-y)}
          +  20 x
           +  \frac{ 16 x^2}{ (1+y)}
          - \frac{ 16 x^2}{ (1-y)}
          - \frac{ 2}{ y (1+x)} \nn\\
& &
          - \frac{ 2}{ y (1-x)}
          - \frac{ 6}{ y}
          - \frac{ 2 y }{(1+x)}
          - \frac{ 2 y }{(1-x)}
          + 10 y
           +  \frac{ 20 }{(1+x)} \nn\\
& &
           +  \frac{ 4 }{(1-x)}
           +  \frac{ 128 }{(1+y)^3}
          - \frac{ 192}{ (1+y)^2}
           +  \frac{ 256}{ (1+y)}
          - \frac{ 32 }{(1-y)}
          - 92 \Biggr) \nn\\
& &
       +   \Biggl(
          - 8
           +  \frac{ 2}{ x}
          - 2 x
          - \frac{ 4}{ y (1-x)}
           +  \frac{ 2}{ y}
          - \frac{ 4 y}{ (1-x)}
          +  2 y \nn\\
& &
           +  \frac{ 16}{ (1-x)}
          \Biggr) H(0;x)  
       +   \Biggl(
            12
          - \frac{ 16}{ x (1+y)^2}
           +  \frac{ 16}{ x (1+y)}
          - \frac{ 2}{ x} \nn\\
& &
          - \frac{ 16 x}{ (1+y)^2}
           +  \frac{ 16 x}{ (1+y)}
          -  2 x
          - \frac{ 2}{ y}  
          - 2 y
           +  \frac{ 32}{ (1+y)^2} \nn\\
& &
          - \frac{ 32}{ (1+y)}
          \Biggr) H(0;y)
       +   \Biggl(
           24
          - \frac{ 32}{ x (1+y)^2}
           +  \frac{ 32}{ x (1+y)}
          - \frac{ 4}{ x} \nn\\
& &
          - \frac{ 32 x }{(1+y)^2}
           +  \frac{ 32 x}{ (1+y)}
          - 4 x
          - \frac{ 4}{ y}  
          - 4 y
           +  \frac{ 64}{ (1+y)^2} \nn\\
& &
          - \frac{ 64}{ (1+y)}
          \Biggr) H(1;y) \!
       + \! \Biggl( \!
          - 24
          - \frac{ 8}{ x} 
          +  8 x
          - \frac{ 4}{ y (1+x)}
          - \frac{ 4}{ y (1-x)} \nn\\
& & 
          + \, \frac{ 4}{ y} \!
          - \! \frac{ 4 y}{ (1+x)} \!
          - \! \frac{ 4 y }{(1-x)} \!
          + \! 4 y \!
           + \! \frac{ 40}{ (1+x)} \!
           + \! \frac{ 8}{ (1-x)}
          \Biggr) H(-1,0;x) \nn\\
& &
       +   \Biggl(
            12
           +  \frac{ 8}{ x^2 (1-y)^2}
          - \frac{ 8}{ x^2 (1-y)}  
           +  \frac{ 4}{ x}
          -  4 x
          - \frac{ 8 x^2 }{(1-y)^2} \nn\\
& &
           +  \frac{ 8 x^2}{ (1-y)}
           +  \frac{ 2}{ y (1+x)}
           +  \frac{ 2}{ y (1-x)}
          - \frac{ 2}{ y}
           +  \frac{ 2 y}{ (1+x)}  
           +  \frac{ 2 y}{ (1-x)} \nn\\
& &
          -  2 y
           +  \frac{ 32}{ (1+x) (1-y)^2}
          - \frac{ 32}{ (1+x) (1-y)}
          - \frac{ 20}{ (1+x)}   \nn\\
& &
           +  \frac{ 32}{ (1-x) (1-y)^2}
          - \frac{ 32}{ (1-x) (1-y)}
          - \frac{ 4 }{(1-x)}
          - \frac{ 32}{ (1-y)^2}   \nn\\
& &
           +  \frac{ 32 }{(1-y)}
          \Biggr) H(0;x) H(0;y)
        +  \Biggl(
           24
           +  \frac{ 16}{ x^2 (1-y)^2}
          - \frac{ 16}{ x^2 (1-y)} \nn\\
& &
           +  \frac{ 8}{ x}
          -  8 x  
          - \frac{ 16 x^2}{ (1-y)^2}
           +  \frac{ 16 x^2}{ (1-y)}
           +  \frac{ 4}{ y (1+x)}
           +  \frac{ 4}{ y (1-x)} \nn\\
& &
          - \frac{ 4}{ y}
           +  \frac{ 4 y }{(1+x)}
           +  \frac{ 4 y}{ (1-x)}  
          - 4 y
           +  \frac{ 64}{ (1+x) (1-y)^2} \nn\\
& &
          - \frac{ 64}{ (1+x) (1-y)}
          - \frac{ 40}{ (1+x)}
           +  \frac{ 64 }{(1-x) (1-y)^2}   \nn\\
& &
          - \! \frac{ 64}{ (1\!-\!x) (1\!-\!y)} \!
          - \! \frac{ 8}{ (1\!-\!x)} \!
          - \! \frac{ 64}{ (1\!-\!y)^2} \!
           + \! \frac{ 64}{ (1\!-\!y)} \!
          \Biggr) H(0;x) H(1;y)    \nn\\
& &
       +   \Biggl(
           12
           +  \frac{ 4}{ x}
          -  4 x
           +  \frac{ 2}{ y (1+x)}  
           +  \frac{ 2}{ y (1-x)}
          - \frac{ 2}{ y}
           +  \frac{ 2 y}{ (1+x)} \nn\\
& &
           +  \frac{ 2 y}{ (1-x)}
          - 2 y
          - \frac{ 20}{ (1+x)}
          - \frac{ 4}{ (1-x)}
          \Biggr) H(0,0;x) \nn\\
& &
       +   \Biggl(
          - 20
           +  \frac{ 4 }{x^2 (1+y)}
          - \frac{ 4}{ x^2 (1-y)}
          - \frac{ 16}{ x (1+y)^3}
           +  \frac{ 24}{ x (1+y)^2} \nn\\
& &
          - \frac{ 20}{ x (1+y)}\!
           + \!\frac{ 4}{ x (1-y)}\!
           + \! \frac{ 4}{ x} \!
          - \frac{ 16 x }{(1+y)^3} \!
           +  \frac{ 24 x }{(1+y)^2} \!
          - \frac{ 20 x}{ (1+y)} \nn\\
& &
           +  \frac{ 4 x}{ (1-y)}
          + 4 x
           +  \frac{ 4 x^2}{ (1+y)}
          - \frac{ 4 x^2}{ (1-y)}
          - \frac{ 2}{ y}
          + 2 y
           +  \frac{ 32}{ (1+y)^3} \nn\\
& &
          - \frac{ 48}{ (1+y)^2}
           +  \frac{ 64}{ (1+y)}
          - \frac{ 8}{ (1-y)}
          \Biggr) H(0,0;y) \nn\\
& &
       +   \Biggl(
          -  40
           +  \frac{ 8}{ x^2 (1+y)}
          - \frac{ 8}{ x^2 (1-y)}
          - \frac{ 32}{ x (1+y)^3}
           +  \frac{ 48}{ x (1+y)^2} \nn\\
& &
          - \frac{ 40}{ x (1+y)}
           + \! \frac{ 8}{ x (1-y)} \!
           + \! \frac{ 8}{ x} \!
          - \frac{ 32 x}{ (1+y)^3}
           +  \frac{ 48 x}{ (1+y)^2}
          - \frac{ 40 x}{ (1+y)} \nn\\
& &
           +  \frac{ 8 x}{ (1-y)}
          + 8 x
           +  \frac{ 8 x^2}{ (1+y)}
          - \frac{ 8 x^2}{ (1-y)}
          - \frac{ 4}{ y}
          + 4 y
           +  \frac{ 64}{ (1+y)^3} \nn\\
& &
          - \frac{ 96}{ (1+y)^2}
           +  \frac{ 128}{ (1+y)}
          - \frac{ 16}{ (1-y)}
          \Biggr) H(0,1;y) \, ,
\label{1lB2} 
\eea

\bea
B_3^{(1l,-1)}(-P^2,-Q^2) &=&  \Biggl(
           \frac{ 8}{ x^2 (1\!-\!y)^2}
          - \frac{ 8}{ x^2 (1\!-\!y)}
          \! + \! \frac{ 32}{ x (1\!-\!y)^2}
          - \frac{ 32}{ x (1\!-\!y)} \nn\\
& &
          - \frac{ 32 x}{ (1\!-\!y)^2} 
          \! + \! \frac{ 32 x}{ (1\!-\!y)}
          - \frac{ 8 x^2}{ (1\!-\!y)^2}
          \! + \! \frac{ 8 x^2}{ (1\!-\!y)}\nn\\
& &
          - \frac{ 32}{ (1+x) (1\!-\!y)^2} 
          \! + \! \frac{ 32}{ (1+x) (1\!-\!y)}
          \! + \! \frac{ 96}{ (1-x) (1\!-\!y)^2} \nn\\
& &
          - \frac{ 96}{ (1-x) (1\!-\!y)} 
          - \frac{ 32}{ (1\!-\!y)^2}
          \! + \! \frac{ 32}{ (1\!-\!y)}
          \Biggr) H(0;x) \, ,
\label{1lB3p}
\eea

\bea
B_3^{(1l,0)}(-P^2,-Q^2) &=&  \zeta(2) \Biggl( 
          - \frac{ 8}{ x^2 (1+y)}
           +  \frac{ 8}{ x^2 (1-y)}
           +  \frac{ 32}{ x (1+y)^3}
          - \frac{ 48}{ x (1+y)^2} \nn\\
& &
          - \frac{ 8}{ x (1+y)}
           +  \frac{ 24}{ x (1-y)}
           +  \frac{ 32 x}{ (1+y)^3}
          - \frac{ 48 x }{(1+y)^2}
          - \frac{ 8 x}{ (1+y)} \nn\\
& &
           +  \frac{ 24 x}{ (1-y)}
          - \frac{ 8 x^2}{ (1+y)}
           +  \frac{ 8 x^2}{ (1-y)}
           +  \frac{ 2}{ y (1+x)}
          - \frac{ 2}{ y (1-x)} \nn\\
& &
           +  \frac{ 2 y}{ (1+x)}
          - \frac{ 2 y }{(1-x)}
           +  \frac{ 12}{ (1+x)}
          - \frac{ 4}{ (1-x)}
          - \frac{ 64 }{(1+y)^3} \nn\\
& &
           +  \frac{ 96 }{(1+y)^2}
          - \frac{ 48}{ (1+y)}
          - \frac{ 16}{ (1-y)}
          + 12 \Biggr) 
       +   \Biggl(
           8
          + \frac{ 4}{ x^2 (1-y)^2} \nn\\
& &
          - \frac{ 4}{ x^2 (1-y)}
           +  \frac{ 8}{ x (1-y)^2} 
          - \frac{ 8}{ x (1-y)}
           +  \frac{ 4}{ x}
          - \frac{ 8 x}{ (1-y)^2} \nn\\
& &
           +  \frac{ 8 x}{ (1-y)}
          - 4 x
          - \frac{ 4 x^2}{ (1-y)^2}
           +  \frac{ 4 x^2}{ (1-y)} 
           +  \frac{ 8 }{y (1+x)}
          - \frac{ 4 }{y} \nn\\
& &
           + \! \frac{ 8 y}{ (1+x)} \!
          - \! 4 y \!
          - \! \frac{ 16}{ (1+x)} \!
           + \! \frac{ 32}{ (1-x) (1-y)^2}
          - \frac{ 32}{ (1-x) (1-y)} \nn\\
& &
          - \! \frac{ 16}{ (1-y)^2}\!
           + \! \frac{ 16}{ (1-y)} \!
          \Biggr) H(0;x) \!
       + \! \Biggl(
          - \! 8 
           + \! \frac{ 8}{ x (1+y)^2} 
          - \frac{ 8}{ x (1+y)} \nn\\
& &
           +  \frac{ 8 x}{ (1+y)^2}
          - \frac{ 8 x}{ (1+y)}
          - \frac{ 16}{ (1+y)^2}
           +  \frac{ 16}{ (1+y)}
          \Biggr) H(0;y)  \nn\\
& &
       +   \Biggl(
          - 16
           +  \frac{ 16}{ x (1+y)^2} 
          - \frac{ 16}{ x (1+y)}
           +  \frac{ 16 x}{ (1+y)^2} 
          - \frac{ 16 x}{ (1+y)} \nn\\
& &
          - \frac{ 32}{ (1+y)^2}
           +  \frac{ 32}{ (1+y)}
          \Biggr) H(1;y) 
       +   \Biggl(
          -  8 
           +  \frac{ 4}{ y (1+x)} \nn\\
& &
          - \! \frac{ 4 }{y (1\!-\!x)}\!
           +  \! \frac{ 4 y}{ (1\!+\!x)}\!
          - \! \frac{ 4 y}{ (1\!-\!x)}\!
           + \! \frac{ 24}{ (1\!+\!x)}\!
          - \!\frac{ 8}{ (1\!-\!x)}\!
          \Biggr) H(-1,\!0;x) \nn\\
& &
       +   \Biggl(
          -  4
          - \frac{ 4}{ x^2 (1-y)^2}
           +  \frac{ 4}{ x^2 (1-y)}
          - \frac{ 16}{ x (1-y)^2} 
           +  \frac{ 16 }{x (1-y)} \nn\\
& &
           +  \frac{ 16 x}{ (1-y)^2}
          - \frac{ 16 x}{ (1-y)}
           +  \frac{ 4 x^2}{ (1-y)^2}
          - \frac{ 4 x^2}{ (1-y)}
           +  \frac{ 2}{ y (1+x)}  \nn\\
& &
          - \frac{ 2}{ y (1-x)}
           +  \frac{ 2 y}{ (1+x)}
          - \frac{ 2 y}{ (1-x)}
           +  \frac{ 16}{ (1+x) (1-y)^2} \nn\\
& &
          - \! \frac{ 16}{ (1\!+\!x) (1\!-\!y)}\! 
           + \! \frac{ 12}{ (1\!+\!x)} \!
          - \! \frac{ 48 }{(1\!-\!x) (1\!-\!y)^2} \!
           + \! \frac{ 48}{ (1\!-\!x) (1\!-\!y)} \nn\\
& &
          - \frac{ 4}{ (1-x)}
           +  \frac{ 16}{ (1-y)^2} 
          - \frac{ 16}{ (1-y)}
          \Biggr) H(0;x) H(0;y)
       +   \Biggl(
          - 8  \nn\\
& &
          - \! \frac{ 8}{ x^2 (1-y)^2} 
          + \! \frac{ 8}{ x^2 (1-y)} 
          - \! \frac{ 32}{ x (1-y)^2} 
           + \! \frac{ 32}{ x (1-y)}
           + \! \frac{ 32 x}{ (1-y)^2} \nn\\
& &
          - \frac{ 32 x}{ (1-y)}
           +  \frac{ 8 x^2}{ (1-y)^2}
          - \frac{ 8 x^2}{ (1-y)} 
           +  \frac{ 4}{ y (1+x)} 
          - \frac{ 4}{ y (1-x)} \nn\\
& &
           +  \frac{ 4 y}{ (1+x)}
          - \frac{ 4 y }{(1-x)}
           +  \frac{ 32}{ (1+x) (1-y)^2}
          - \frac{ 32}{ (1+x) (1-y)}  \nn\\
& &
           +  \frac{ 24}{ (1+x)}
          - \frac{ 96}{ (1-x) (1-y)^2}
           +  \frac{ 96}{ (1-x) (1-y)} 
          - \frac{ 8}{ (1-x)} \nn\\
& &
           +  \frac{ 32}{ (1-y)^2} 
          - \frac{ 32}{ (1-y)}
          \Biggr) H(0;x) H(1;y)
       +   \Biggl(
           4 
          - \frac{ 2}{ y (1+x)} \nn\\
& & 
           + \! \frac{ 2}{ y (1-x)} \!
          - \! \frac{ 2 y}{ (1+x)}\!
           + \! \frac{ 2 y}{ (1-x)} \!
          - \! \frac{ 12}{ (1+x)} \!
           + \! \frac{ 4}{ (1-x)}\!
          \Biggr) H(0,0;x)  \nn\\
& &
       +   \Biggl(
            4 
          - \frac{ 2}{ x^2 (1+y)}
           +  \frac{ 2}{ x^2 (1-y)}
           +  \frac{ 8}{ x (1+y)^3} 
          - \frac{ 12}{ x (1+y)^2} \nn\\
& &
          - \frac{ 2}{ x (1+y)} 
           +  \frac{ 6}{ x (1-y)}
           +  \frac{ 8 x}{ (1+y)^3}
          - \frac{ 12 x }{(1+y)^2}
          - \frac{ 2 x}{ (1+y)}  \nn\\
& &
           +  \frac{ 6 x}{ (1-y)}
          - \frac{ 2 x^2}{ (1+y)} 
           +  \frac{ 2 x^2}{ (1-y)}
          - \frac{ 16}{ (1+y)^3}
           +  \frac{ 24}{ (1+y)^2} \nn\\
& &
          - \frac{ 12}{ (1+y)} 
          - \frac{ 4}{ (1-y)}
          \Biggr) H(0,0;y) 
       +   \Biggl(
            8 
          - \frac{ 4}{ x^2 (1+y)} \nn\\
& &
           + \! \frac{ 4}{ x^2 (1-y)}\!
           + \! \frac{ 16}{ x (1+y)^3} \!
          - \frac{ 24 }{x (1+y)^2}\!
          - \frac{ 4 }{x (1+y)} \!
           + \! \frac{ 12 }{x (1-y)} \nn\\
& &
           +  \frac{ 16 x }{(1+y)^3}
          - \frac{ 24 x}{ (1+y)^2}
          - \frac{ 4 x }{(1+y)} 
           +  \frac{ 12 x }{(1-y)} 
          - \frac{ 4 x^2}{ (1+y)}  \nn\\
& &
           +  \frac{ 4 x^2}{ (1-y)}
          - \frac{ 32}{ (1+y)^3}
           +  \frac{ 48}{ (1+y)^2}
          - \frac{ 24}{ (1+y)}  \nn\\
& &
          - \frac{ 8}{ (1-y)}
          \Biggr) H(0,1;y) \ . 
\label{1lB3} 
\eea

\subsection{Auxiliary Functions for the Two-Loop Cross Section}

\bea
\Pi_0^{(2l)}(-P^2) &=& 
          - \frac{5}{24}
          + \frac{13 }{6 (1-x)^2}
          - \frac{13 }{6 (1-x)}
          + \Bigg(
                      1
                   -  \frac{4 }{(1-x)^4}
                   +  \frac{8 }{(1-x)^3}
                   \nn \\
                   & & 
                   -  \frac{4 }{(1-x)^2}
            \Bigg) \zeta(3) \!
          + \! \Bigg(
                      \frac{1}{4}\!
                   + \! \frac{3 }{(1-x)^3}
                   -  \frac{9 }{2 (1-x)^2}\!
                   + \! \frac{1 }{(1-x)}
            \Bigg) H(0;x)
            \nn \\
            & & 
          + \Bigg(
                   -  \frac{4}{3}
                   -  \frac{16 }{3 (1-x)^3}
                   +  \frac{8 }{(1-x)^2}
            \Bigg) H(-1,0;x)
          + \Bigg(
                      2
                   -  \frac{7 }{3 (1-x)^4}
                   \nn \\
                   & & 
                   +  \frac{26 }{3 (1-x)^3}
                   -  \frac{23 }{3 (1-x)^2}
                   -  \frac{2 }{3 (1-x)}
            \Bigg)
          + \Bigg(
                      \frac{2}{3}
                   +  \frac{8 }{3 (1-x)^3}
                   \nn \\
                   & & 
                   -  \frac{4 }{(1-x)^2}
            \Bigg) H(1,0;x)
          + \Bigg(
                      \frac{4}{3}
                   -  \frac{16 }{3 (1-x)^4}
                   +  \frac{32 }{3 (1-x)^3}
                   \nn \\
                   & & 
                   -  \frac{16 }{3 (1-x)^2}
            \Bigg) \Big[ H(0,-1,0;x) - H(-1,0,0;x) \Big]
          + \Bigg(
                      \frac{2}{3}
                   -  \frac{8 }{3 (1-x)^4}
                   \nn \\
                   & & 
                   +  \frac{16 }{3 (1-x)^3}
                   -  \frac{8 }{3 (1-x)^2}
            \Bigg) \Big[ H(1,0,0;x) - H(0,1,0;x) \Big] \ ,
\label{2lPi0} 
\eea

\bea
F_1^{(2l)}(-P^2) &=& 
            \frac{383}{108}
          + \frac{49 }{9 (1+x)^2}
          - \frac{49 }{9 (1+x)}
          + \Bigg(
                   -  \frac{1}{4}
                   +  \frac{98 }{3 (1+x)^4}
                   -  \frac{196 }{3 (1+x)^3}
                   \nn \\
                   & & 
                   +  \frac{229 }{6 (1+x)^2}
                   -  \frac{11 }{2 (1+x)}
            \Bigg) \zeta(2)
          + \Bigg(
                   -  \frac{265}{216}
                   +  \frac{89 }{9 (1+x)^3}
                   -  \frac{89 }{6 (1+x)^2}
                   \nn \\
                   & & 
                   +  \frac{563 }{108 (1+x)}
                   +  \frac{59 }{27 (1-x)}
            \Bigg) H(0;x) 
          + \Bigg(
                     \frac{19}{36}
                   +  \frac{62 }{9 (1+x)^4}
                   -  \frac{124 }{9 (1+x)^3}
                   \nn \\
                   & & 
                   +  \frac{163 }{18 (1+x)^2}
                   -  \frac{13 }{6 (1+x)}
            \Bigg) H(0,0;x)
          + \Bigg(
                   -  \frac{1 }{6}
                   -  \frac{6 }{(1+x)^5}
                   +  \frac{15 }{(1+x)^4}
                   \nn \\
                   & & 
                   -  \frac{11 }{(1+x)^3}
                   +  \frac{3 }{2 (1+x)^2}
                   +  \frac{5 }{12 (1+x)}
                   +  \frac{5 }{12 (1-x)}
            \Bigg) \Big[   \zeta(2) H(0;x) 
                         \nn \\
                         & & 
                         + H(0,0,0;x) \Big] \ , 
\label{2lF1} \\
F_2^{(2l)}(-P^2) &=&         
          - \frac{17 }{3 (1+x)^2}
          + \frac{17 }{3 (1+x)}
          + \Bigg(
                   -  \frac{34 }{(1+x)^4}
                   +  \frac{68 }{(1+x)^3}
                   -  \frac{33 }{(1+x)^2}
                   \nn \\
                   & & 
                   -  \frac{1 }{(1+x)}
            \Bigg) \zeta(2)
          + \Bigg(
                   -  \frac{31 }{3 (1+x)^3}
                   +  \frac{31 }{2 (1+x)^2}
                   -  \frac{59 }{18 (1+x)}
                   \nn \\
                   & & 
                   -  \frac{17 }{9 (1-x)}
            \Bigg) H(0;x)
          + \Bigg(
                   -  \frac{22 }{3 (1+x)^4}
                   +  \frac{44 }{3 (1+x)^3}
                   -  \frac{23 }{3 (1+x)^2}
                   \nn \\
                   & & 
                   +  \frac{1 }{3 (1+x)}
            \Bigg) H(0,0;x)
          + \Bigg(
                      \frac{6 }{(1+x)^5}
                   -  \frac{15 }{(1+x)^4}
                   +  \frac{21 }{2 (1+x)^3}
                   \nn \\
                   & & 
                   - \! \frac{3 }{4 (1+x)^2}\!
                   - \! \frac{3 }{8 (1+x)}\!
                   - \! \frac{3 }{8 (1-x)}\!
            \Bigg) \Big[   \zeta(2) H(0;x) \!
                         + \!H(0,0,0;x) \Big] , 
\label{2lF2}
\eea


\bea
B_1^{(2l,-1)}(-P^2,-Q^2) &=&-\Biggl[\Biggl(
          \frac{56}{3} 
          - \frac{16}{3 x^2 (1\!-\!y)^4}
          \! + \! \frac{32}{3 x^2 (1\!-\!y)^3}
          - \frac{28}{9 x^2 (1\!-\!y)^2} \nn \\
& &          
   - \frac{20}{9 x^2 (1\!-\!y)}
          \! + \! \frac{64}{3 x (1\!-\!y)^4}
          - \frac{128}{3 x (1\!-\!y)^3}
          \! + \! \frac{16}{9 x (1\!-\!y)^2}\nn \\
& &         
          + \frac{176}{9 x (1\!-\!y)}
          \! + \! \frac{40}{9 x}
          - \frac{64 x}{3(1\!-\!y)^4}
          \! + \! \frac{128 x}{3(1\!-\!y)^3}
          - \frac{16 x}{9(1\!-\!y)^2}\nn \\
& &         
          - \frac{176 x}{9(1\!-\!y)}
          - \frac{40 x}{9} 
          \! + \! \frac{16 x^2}{3 (1\!-\!y)^4}
          - \frac{32 x^2}{3 (1\!-\!y)^3}
          \! + \! \frac{28 x^2}{9 (1\!-\!y)^2}\nn \\
& &
          + \frac{20 x^2}{9(1\!-\!y)}
          \! + \! \frac{20}{9 y (1-x)}
          \! + \! \frac{20}{9 y (x+1)}
          - \frac{20}{9 y}
          \! + \! \frac{20 y}{9 (1-x)}\nn \\
& &        
          + \frac{20 y}{9 (x+1)}
          - \frac{20 y}{9}
          \! + \! \frac{64}{3 (1-x) (1\!-\!y)^4}
          - \frac{128}{3 (1-x) (1\!-\!y)^3}\nn \\
& &  
          + \frac{112}{9 (1-x) (1\!-\!y)^2}
          \! + \! \frac{80}{9 (1-x) (1\!-\!y)}
          - \frac{88}{9 (1-x)}\nn \\
& &         
          - \frac{64}{(1\!-\!y)^4 (x+1)}
          \! + \! \frac{64}{3 (1\!-\!y)^4}
          \! + \! \frac{128}{(1\!-\!y)^3 (x+1)} \nn \\
& &
          - \frac{128}{3(1\!-\!y)^3}
          \! + \! \frac{16}{3 (1\!-\!y)^2 (x+1)}
          - \frac{80}{9 (1\!-\!y)^2} \nn \\
& &
          - \frac{208}{3 (1\!-\!y) (x+1)}
          \! + \! \frac{272}{9(1\!-\!y)}
          - \frac{248}{9(x+1)} \Biggr)  H(0;x) 
          + \Biggl(
   - 8 \nn \\
& &
          - \frac{16}{3 x^2 (1\!-\!y)^5}
          \! + \! \frac{40}{3 x^2 (1\!-\!y)^4}
          -   \frac{8}{x^2 (1\!-\!y)^3}
          - \frac{4}{3 x^2 (1\!-\!y)^2} \nn \\
& & 
          + \frac{4}{3 x^2 (1\!-\!y)}
          \! + \! \frac{64}{3  x (1\!-\!y)^5}
          - \frac{160}{3 x (1\!-\!y)^4}
          \! + \! \frac{64}{3 x (1\!-\!y)^3}\nn \\
& & 
          + \frac{64}{3 x (1\!-\!y)^2}
          - \frac{16}{3 x (1\!-\!y)}
          - \frac{8}{3 x}
          - \frac{64}{3  (1\!-\!y)^5}
          \! + \! \frac{160}{3 (1\!-\!y)^4} \nn \\
& & 
          - \frac{64}{3 (1\!-\!y)^3}
          - \frac{64}{3  (1\!-\!y)^2}
          \! + \! \frac{16}{3 (1\!-\!y)}
          \! + \! \frac{8 x}{3}  
          \! + \! \frac{16 x^2}{3 (1\!-\!y)^5} \nn \\
& & 
          - \frac{40 x^2}{3  (1\!-\!y)^4}
          \! + \! \frac{8 x^2}{ (1\!-\!y)^3}
          \! + \! \frac{4 x^2}{3 (1\!-\!y)^2}
          - \frac{4 x^2}{3 (1\!-\!y)}
          \! + \! \frac{4}{3  y (1-x)} \nn \\
& & 
          + \frac{4}{3  y (x+1)}
          - \frac{4}{3  y}
          - \frac{4 y}{3 (1-x)}
          - \frac{4 y}{3 (x+1)}
          \! + \! \frac{4 y}{3 } \nn \\
& & 
          + \frac{64}{3  (1-x) (1\!-\!y)^5}
          - \frac{160}{3  (1-x) (1\!-\!y)^4}
          \! + \! \frac{32}{(1-x) (1\!-\!y)^3} \nn \\
& & 
          + \frac{16}{3  (1-x) (1\!-\!y)^2}
          - \frac{32}{3  (1-x) (1\!-\!y)}
          \! + \! \frac{8}{3  (1-x)} \nn \\
& & 
          - \frac{64}{(1\!-\!y)^5 (x+1)}
          \! + \! \frac{64}{3 (1\!-\!y)^5}
          \! + \! \frac{160}{ (1\!-\!y)^4 (x+1)}
          - \frac{160}{3  (1\!-\!y)^4} \nn \\
& & 
          - \frac{160}{3  (1\!-\!y)^3 (x+1)}
          \! + \! \frac{32}{3 (1\!-\!y)^3}
          - \frac{80}{(1\!-\!y)^2 (x+1)}
          + \frac{112}{3 (1\!-\!y)^2} \nn\\
& &
          \! + \! \frac{32}{3 (1\!-\!y) (x+1)}
          \! + \! \frac{40}{3 (x+1)}
          \Biggr) H(0;x) H(0;y) \Biggr]\, , 
\label{B12lM1}
\eea


\bea
B_1^{(2l,0)}(-P^2,-Q^2) &=&  -\Biggl\{
          - \frac{16}{9}
          - \frac{16}{45 x^2 (1-y)^4}
          + \frac{32}{45 x^2 (1-y)^3}
          + \frac{104}{135 x^2 (1-y)^2} \nn\\
& &
          - \! \frac{152}{135 x^2 (1-y)} \!
          + \! \frac{64}{45 x (1-y)^4} \!
          - \! \frac{128}{45 x (1-y)^3} \!
          - \! \frac{512}{135 x (1-y)^2} \nn\\
& &
          + \frac{704}{135 x (1-y)}
          + \frac{304}{135 x}        
          + \frac{64 x}{45  (1-y)^4}
          - \frac{128 x}{45  (1-y)^3} \nn\\
& &
          - \frac{512 x}{135  (1-y)^2}
           +  \frac{704 x}{135  (1-y)}
           +  \frac{304 x}{135} 
          - \frac{16 x^2}{45 (1-y)^4} \nn\\
& &
           +  \frac{32 x^2}{45   (1-y)^3}
           +  \frac{104 x^2}{135   (1-y)^2}
          - \frac{152 x^2}{135   (1-y)}
          - \frac{56}{9 y (x+1)^2}  \nn\\
& &
          +  \frac{56}{9 y (x+1)}
           +  \frac{152}{135 y}
          - \frac{56 y}{9 (x+1)^2}
           +  \frac{56 y}{9 (x+1)}
           +  \frac{152 y}{135} \nn\\
& &
          - \frac{32}{45 (1-y)^4} 
          + \frac{64}{45 (1-y)^3}
          - \frac{32}{3 (1-y)^2 (x+1)^2} \nn\\
& &
           +  \frac{32}{3 (1-y)^2 (x+1)}
           +  \frac{224}{27 (1-y)^2} 
          + \frac{32}{3 (1-y) (x+1)^2}  \nn\\
& &
          - \frac{32}{3 (1-y) (x+1)}
          - \frac{1216}{135 (1-y)}
           +  \frac{16}{3 (x+1)^2} 
          - \frac{16}{3 (x+1)} \nn\\
& &
+   \Biggl(
            \frac{64}{3}  
          - \frac{32}{3 x^2 (1-y)^4}  
           +  \frac{64}{3 x^2 (1-y)^3}  
          - \frac{56}{9 x^2 (1-y)^2} \nn\\
& &
          - \frac{40}{9 x^2 (1-y)} 
          + \frac{176}{3 x (1-y)^4}  
          - \frac{352}{3 x (1-y)^3}  
           +  \frac{140}{9 x (1-y)^2}   \nn\\
& &
           +  \frac{388}{9 x (1-y)}           
           +  \frac{16}{3 x (y+1)^2} 
           - \frac{16}{3 x (y+1)}  
           +  \frac{14}{9 x}  
           +  \frac{112 x}{3 (1-y)^4}   \nn\\
& &
           - \frac{224 x}{3   (1-y)^3}  
           +  \frac{124 x}{9   (1-y)^2}           
           +  \frac{212 x}{9   (1-y)}  
          + \frac{16 x}{3   (y+1)^2}  
          - \frac{16 x}{3   (y+1)}   \nn\\
& &
          - \frac{26 x}{9}    
          - \frac{16 x^2 }{3 (1-y)^4}  
          +  \frac{32 x^2 }{3  (1-y)^3}  
          - \frac{28 x^2}{9 (1-y)^2}  
          - \frac{20 x^2 }{9(1-y)}   \nn\\
& &
           +  \frac{10}{9 y (1-x)}  
          - \frac{112}{3 y (x+1)^4}  
           +  \frac{224}{3 y (x+1)^3}  
          - \frac{116}{3 y (x+1)^2}  \nn\\
& &
          + \frac{22}{9 y (x+1)}  
           +  \frac{2}{9 y}  
           +  \frac{10 y}{9 (1-x)}  
          - \frac{112 y}{3  (x+1)^4}  
           +  \frac{224 y}{3   (x+1)^3}   \nn\\
& &          
          - \frac{116 y}{3  (x+1)^2} 
          + \frac{22 y}{9   (x+1)}  
          +  \frac{2 y}{9 }  
          +  \frac{32}{3 (1-x) (1-y)^4}   \nn\\
& &
          - \frac{64}{3 (1-x) (1-y)^3}            
          + \frac{56}{9 (1-x) (1-y)^2}  
	  +  \frac{40}{9 (1-x) (1-y)}   \nn\\
& &
          - \! \frac{44}{9 (1-x)}  \!
          - \! \frac{192}{ (1-y)^4 (x+1)^2} \!            
          + \! \frac{160 }{(1-y)^4 (x+1)} \!
          - \! \frac{304}{3 (1-y)^4 }  \nn\\
& &
          +  \frac{384}{ (1-y)^3 (x+1)^2}  
          - \frac{320}{ (1-y)^3 (x+1)} 
          + \frac{608}{3 (1-y)^3}   \nn\\
& &
          - \frac{64}{ (1-y)^2 (x+1)^4}  
          +  \frac{128}{ (1-y)^2 (x+1)^3}  
          -  \frac{368}{3 (1-y)^2 (x+1)^2}   \nn\\
& &          
          + \frac{184}{3 (1-y)^2 (x+1)}  
          - \frac{244}{9 (1-y)^2}  
           +  \frac{64}{ (1-y) (x+1)^4}  \nn\\
& & 
          - \frac{128 }{(1-y) (x+1)^3} 
          - \frac{208}{3 (1-y) (x+1)^2}  
           +  \frac{296}{3 (1-y) (x+1)}   \nn\\
& &
          - \frac{668}{9 (1-y)}  
           +  \frac{32}{ (x+1)^4}
          - \frac{64}{ (x+1)^3} 
           +  \frac{344}{3 (x+1)^2}  
          - \frac{868}{9 (x+1)}   \nn\\
& &
          - \frac{32}{3 (y+1)^2}  
           +  \frac{32}{3 (y+1)} 
\Biggr) \zeta(2)
+ \Biggl(
          - \frac{124}{3} 
           +  \frac{40}{3  x^2 (1-y)^4} \nn\\
& &
          - \frac{80}{3  x^2 (1-y)^3}
          + \frac{278}{27  x^2 (1-y)^2}
          +  \frac{82}{27  x^2 (1-y)}            
          - \frac{176}{3  x (1-y)^4} \nn\\
& &
          +  \frac{352}{3  x (1-y)^-3}
          - \frac{452}{27  x (1-y)^2}
          - \frac{1132}{27  x (1-y)}
          - \frac{242}{27  x} \nn\\
& &       
          + \! \frac{176  x}{3 (1-y)^4} \!
          - \! \frac{352  x}{3  (1-y)^3} \!
          + \! \frac{452  x}{27 (1-y)^2} \!
          + \! \frac{1132  x}{27  (1-y)} \!
          + \! \frac{242  x}{27 } \nn\\
& &
          - \frac{40  x^2}{3 (1-y)^4}
          + \frac{80 x^2}{3  (1-y)^3}
          - \frac{278  x^2 }{27(1-y)^2}
          - \frac{82 x^2}{27  (1-y)} \nn\\
& &
          - \frac{70}{27  y (1-x)}
          - \frac{112}{9  y (x+1)^3} 
          + \frac{56}{3  y (x+1)^2} 
          - \frac{358}{27  y (x+1)} \nn\\
& &
           +  \frac{130}{27  y}
          - \frac{70  y }{27(1-x)}
          - \frac{112  y }{9 (x+1)^3} 
          + \frac{56  y }{3 (x+1)^2} 
          - \frac{358  y }{27 (x+1)} \nn\\
& &
           +  \frac{130  y }{27}
          - \frac{64}{ (1-x) (1-y)^4}
           +  \frac{128}{  (1-x) (1-y)^3} \nn\\
& &
          - \frac{1664}{27  (1-x) (1-y)^2}
          - \frac{64}{27  (1-x)(1-y)}
           +  \frac{464}{27  (1-x)} \nn\\
& &
           +  \frac{512}{3  (1-y)^4 (x+1)}
          - \frac{160}{3  (1-y)^4}
          - \frac{1024}{3  (1-y)^3 (x+1)} \nn\\
& &
           +  \frac{320}{3  (1-y)^3}
          - \frac{64}{3  (1-y)^2 (x+1)^3}
          + \frac{32}{  (1-y)^2 (x+1)^2} \nn\\
& &
           +  \frac{16}{3  (1-y)^2 (x+1)}
           +  \frac{616}{27  (1-y)^2}
           +  \frac{64}{3  (1-y)(x+1)^3} \nn\\
& &
          - \frac{32}{(1-y) (x+1)^2}
           +  \frac{496}{3  (1-y) (x+1)}
          - \frac{2056}{27  (1-y)} \nn\\
& &
           +  \frac{32}{3  (x+1)^3}
          - \frac{16}{  (x+1)^2}
           +  \frac{1912}{27  (x+1)}
\Biggr) H(0;x) +
\Biggl(      
            \frac{46}{45}   \nn\\
& &
          - \frac{16}{45   x^2 (1-y)^5}
           +  \frac{8}{9   x^2 (1-y)^4}
          - \frac{8}{9   x^2 (1-y)^2}
           +  \frac{2}{9   x^2 (1-y)} \nn\\
& &
           +  \frac{2}{15   x^2}
           +  \frac{4 y}{15   x} 
           +  \frac{64}{45   x (1-y)^5}
          - \frac{32}{9   x (1-y)^4}
          - \frac{32}{45   x (1-y)^3} \nn\\
& &
           +  \frac{208}{45   x (1-y)^2}
           +  \frac{16}{15   x (1-y)}
           +  \frac{8}{3   x (y+1)}
          - \frac{148}{45   x}
           +  \frac{4 x y}{15} \nn\\
& &
           +  \frac{64   x}{45 (1-y)^5}
          - \frac{32  x}{9  (1-y)^4}
          - \frac{32  x}{45  (1-y)^3}
           +  \frac{208   x}{45 (1-y)^2} \nn\\
& &
           +  \frac{16   x}{15 (1-y)}
           +  \frac{8  x}{3  (y+1)}
          - \frac{148  x}{45} 
          - \frac{16  x^2}{45  (1-y)^5}
           +  \frac{8  x^2 }{9 (1-y)^4} \nn\\
& &
          - \frac{8   x^2}{9 (1-y)^2}\!
          + \! \frac{2  x^2}{9  (1-y)}\!
          + \! \frac{2   x^2}{15}\!
          + \! \frac{2}{y}\!
          - \frac{46 y}{15}  \!
          + \! \frac{2  y^2}{15} 
          - \frac{32}{45   (1-y)^5} \nn\\
& &
           +  \frac{16}{9   (1-y)^4}
           +  \frac{128}{15   (1-y)^3}
           - \frac{656}{45   (1-y)^2}
           +  \frac{52}{5   (1-y)} \nn\\
& &
          - \frac{16}{3   (y+1)}
\Biggr) H(0;y) +   
\Biggl(   
          - \frac{28}{3}  
          - \frac{8}{3  x^2 (1-y)^5} 
          +  \frac{20}{3  x^2 (1-y)^4}  \nn\\
& &
          - \frac{4}{  x^2 (1-y)^3} 
          - \frac{2}{3  x^2 (1-y)^2} 
          +  \frac{2}{3  x^2 (1-y)} 
          +  \frac{32}{3  x (1-y)^5}  \nn\\
& &
          - \frac{80}{3  x (1-y)^4} \!
          + \! \frac{32}{3  x (1-y)^3} \!
          + \! \frac{32}{3  x (1-y)^2} \!
          - \frac{8}{3  x  (1-y)} 
          - \frac{8}{3  x} \nn\\
& &
          - \frac{32  x }{3 (1-y)^5} \! 
          + \! \frac{80  x }{3 (1-y)^4} \!
          - \! \frac{32  x }{3 (1-y)^3}\!
          - \! \frac{32  x }{3 (1-y)^2} \!
          + \! \frac{8  x }{3 (1-y)} \nn\\
& &
           +  \frac{8  x }{3} 
           +  \frac{8 x^2}{3 (1-y)^5} 
          - \frac{20 x^2}{3  (1-y)^4} 
           +  \frac{4  x^2}{(1-y)^3}  
           +  \frac{2 x^2}{3  (1-y)^2}  \nn\\
& &
          - \frac{2 x^2 }{3  (1-y)} 
          - \frac{2}{3  y (1-x)}
           +  \frac{2}{3  y (x+1)} 
          - \frac{2  y}{ (1-x)}
          - \frac{2 y}{3  (x+1) }  \nn\\
& &
           + \! \frac{4 y}{3}\!
           + \! \frac{32}{3  (1-x) (1-y)^5 }\!
          - \! \frac{80}{3  (1-x) (1-y)^4 }\!
          + \! \frac{16}{  (1-x) (1-y)^3}  \nn\\
& &
           +  \frac{8}{3  (1-x) (1-y)^2}  
          - \frac{16}{3  (1-x) (1-y)} 
           +  \frac{4}{3  (1-x)}  \nn\\
& &
          - \frac{32 }{ (1-y)^5 (x+1)} \! 
           + \! \frac{32}{3  (1-y)^5} \!
           + \! \frac{80}{  (1-y)^4 (x+1)}\!
          - \! \frac{80}{3  (1-y)^4}   \nn\\
& &
          - \! \frac{80}{3  (1-y)^3 (x+1)}\!
          + \! \frac{16}{3  (1-y)^3} \!
          - \! \frac{40 }{ (1-y)^2 (x+1)}  \!
          + \! \frac{56}{3  (1-y)^2}  \nn\\
& &
          + \frac{16}{3  (1-y) (x+1)}
           +  \frac{52}{3  (x+1)}
\Biggr) \zeta(2) H(0;x)
+ \Biggl(  
          - \frac{16}{3}  \nn\\
& &
          - \frac{32}{3   x^2 (1-y)^5} 
          +  \frac{80}{3   x^2 (1-y)^4} 
          - \frac{16}{   x^2 (1-y)^3} 
          - \frac{8}{3   x^2 (1-y)^2}  \nn\\
& &
           +  \frac{4}{   x^2 (1-y)} 
          - \frac{4}{3   x^2 (y+1)} 
          + \frac{128}{3   x (1-y)^5} 
          - \frac{320}{3   x (1-y)^4}  \nn\\
& &
           +  \frac{128}{3   x (1-y)^3} 
           +  \frac{128}{3   x (1-y)^2} 
          - \frac{52}{3   x (1-y)} 
          + \frac{16}{3   x (y+1)^3}  \nn\\
& &
          - \frac{8  }{ x (y+1)^2} 
           +  \frac{44}{3   x (y+1)}
          - \frac{8}{   x} 
          - \frac{128  x }{3 (1-y)^5 }
           +  \frac{320   x }{3(1-y)^4}  \nn\\
& &
          - \frac{128   x}{3 (1-y)^3} 
          - \frac{128  x}{3  (1-y)^2} 
           +  \frac{4   x }{(1-y)} 
           +  \frac{16  x }{3 (y+1)^3}  \nn\\
& &
          - \frac{8   x}{ (y+1)^2} 
           +  \frac{44  x}{3  (y+1)} 
          + \frac{8   x }{3}
           +  \frac{32  x^2}{3  (1-y)^5 }
          - \frac{80   x^2}{3 (1-y)^4}  \nn\\
& &
           +  \frac{16   x^2}{ (1-y)^3} 
           +  \frac{8  x^2 }{3 (1-y)^2} 
          - \frac{4  x^2}{3  (1-y)} 
          - \frac{4  x^2}{3  (y+1)} 
           +  \frac{8}{3   y (1-x)}  \nn\\
& &
           +  \frac{8}{3   y (x+1)} 
          - \frac{4}{3   y} 
          - \frac{8  y }{3 (1-x) }
          - \frac{8   y }{3(x+1)} 
          + \frac{4  y }{3}  \nn\\
& &
           +  \frac{128}{3   (1-x) (1-y)^5} 
          - \frac{320}{3   (1-x) (1-y)^4} 
           +  \frac{64}{ (1-x) (1-y)^3} \nn\\
& &
          + \frac{32}{3   (1-x) (1-y)^2} 
          - \frac{64}{3   (1-x) (1-y)} 
           +  \frac{16}{3   (1-x)}  \nn\\
& &
          - \frac{128}{ (1-y)^5 (x+1)} 
          + \frac{128}{3   (1-y)^5} 
           +  \frac{320}{(1-y)^4 (x+1)}  \nn\\
& &
          - \! \frac{320}{3(1\!-\!y)^4} \!
          - \! \frac{320}{3(1\!-\!y)^3 (x\!+\!1)} \!
          + \! \frac{64}{3(1\!-\!y)^3} \!
          - \! \frac{160}{(1\!-\!y)^2 (x\!+\!1)}  \nn\\
& &
           +  \frac{224}{3   (1-y)^2} 
           +  \frac{64}{3   (1-y) (x+1)} 
           +  \frac{8}{(1-y)} 
          + \frac{80}{3   (x+1)}  \nn\\
& &
          - \! \frac{32}{3(y+1)^3} \!
          + \! \frac{16}{(y+1)^2} \!
          - \! \frac{104}{3 (y+1)}
\Biggr) \zeta(2) H(0;y) \!
+ \! \Biggl(   
            \frac{32}{15} \!
          + \! \frac{2}{15x^2}  \nn\\
& &
          + \frac{4}{15  x y}
           +  \frac{4 y }{15  x}
          - \frac{16}{15  x}
           +  \frac{4 x }{15 y}
           +  \frac{4 x  y}{15} 
          - \frac{16  x}{15}
           +  \frac{2 x^2}{15 }
           +  \frac{2}{15  y^2}  \nn\\
& &
          - \frac{16}{15  y}
          - \frac{16 y}{15}  
          + \frac{2 y^2}{15} 
\Biggr)  H(1;y) 
+ \Biggl(
            \frac{112}{9}  
           +  \frac{56}{3   x^2 (1-y)^5} \nn\\
& &
          - \frac{140}{3   x^2 (1-y)^4}
           +  \frac{32}{   x^2 (1-y)^3}
          - \frac{4}{3   x^2 (1-y)^2}
          - \frac{8}{3   x^2 (1-y)} \nn\\
& &
          - \frac{80}{x (1-y)^5}
           +  \frac{200}{x (1-y)^4}
          - \frac{896}{9   x (1-y)^3}
          - \frac{152}{3   x (1-y)^2} \nn\\
& &
           +  \frac{68}{3   x (1-y)}
           +  \frac{34}{9   x}
           +  \frac{80   x}{ (1-y)^5}
          - \frac{200  x}{ (1-y)^4}
           +  \frac{896  x }{9 (1-y)^3}  \nn\\
& &
          + \frac{152   x}{3 (1-y)^2}
          - \frac{68  x }{3 (1-y)}
          - \frac{34 x}{9}  
          - \frac{56  x^2 }{3 (1-y)^5}
           +  \frac{140  x^2}{3  (1-y)^4} \nn\\
& &
          - \frac{32   x^2}{ (1-y)^3} 
          + \frac{4 x^2}{3   (1-y)^2}
           +  \frac{8  x^2 }{3 (1-y)}
          - \frac{4}{9   y (1-x)}
          - \frac{4}{y (x+1)} \nn\\
& &
           +  \frac{20}{9   y}
           +  \frac{4  y }{9 (1-x)} 
           +  \frac{4   y}{ (x+1)} 
          - \frac{20  y}{9} 
          - \frac{256}{3   (1-x) (1-y)^5} \nn\\
& &
           +  \frac{640}{3   (1-x) (1-y)^4}
          - \frac{1408}{9  (1-x)(1-y)^3}
          + \frac{64}{3   (1-x) (1-y)^2} \nn\\
& &
           + \! \frac{304}{9   (1-x) (1-y)} \!
          - \! \frac{40}{3   (1-x)} \!
           + \! \frac{704}{3   (1-y)^5 (x+1)}\!
          - \! \frac{224}{3   (1-y)^5} \nn\\
& &
          - \! \frac{1760}{3(1\!-\!y)^4 (x\!+\!1)} \!
           + \! \frac{560}{3(1\!-\!y)^4} \!
           + \! \frac{2368}{9(1\!-\!y)^3 (x\!+\!1)} \!
          - \! \frac{160}{3(1\!-\!y)^3} \nn\\
& &
           +  \frac{192}{(1-y)^2 (x+1)}
          - \frac{320}{3   (1-y)^2}
          - \frac{80}{ (1-y) (x+1)}
          + \frac{208}{9   (1-y)} \nn\\
& &
          - \frac{104}{9   (x+1)}
\Biggr) H(0;x) H(0;y) + \Biggl(        
          - \frac{40}{9} 
           +  \frac{16}{3  x (1-y)^4} \nn\\
& &
          - \frac{32}{3  x (1-y)^3}
           +  \frac{4}{  x (1-y)^2}
           +  \frac{4}{3  x (1-y)}
          - \frac{26}{9  x}
           +  \frac{80 x }{3 (1-y)^4} \nn\\
& &
          - \frac{160  x }{3(1-y)^3}
          + \frac{52 x }{9 (1-y)^2}
           +  \frac{188 x}{9  (1-y)}
           +  \frac{14 x}{9} 
          - \frac{16 x^2}{3  (1-y)^4} \nn\\
& &
           + \! \frac{32 x^2}{3  (1-y)^3}\!
          - \! \frac{28 x^2}{9  (1-y)^2}\!
          - \! \frac{20 x^2}{9  (1-y)}\!
          - \! \frac{10}{9  y (1-x)}\!
          - \! \frac{112}{9  y (x\!+\!1)^4} \nn\\
& &
           +  \frac{224}{9  y (x+1)^3} 
          -  \frac{116}{9  y (x+1)^2}
          -  \frac{2}{3  y (x+1)}
           +  \frac{10}{9  y}
          -  \frac{10  y}{9 (1-x)} \nn\\
& &
          - \frac{112  y }{9 (x+1)^4}
           +  \frac{224  y}{9 (x+1)^3}
          - \frac{116 y}{9  (x+1)^2}
          - \frac{2 y}{3  (x+1)}
           +  \frac{10  y}{9} \nn\\
& &
          - \frac{32}{3  (1-x) (1-y)^4}
           +  \frac{64}{3  (1-x) (1-y)^3}
          - \frac{56}{9  (1-x) (1-y)^2} \nn\\
& &
          - \frac{40}{9  (1-x) (1-y)}
           +  \frac{44}{9  (1-x)}
          - \frac{64}{(1-y)^4 (x+1)^2} \nn\\
& &
          + \frac{96}{(1-y)^4 (x+1)}
          - \frac{48}{(1-y)^4}
           +  \frac{128}{(1-y)^3 (x+1)^2} \nn\\
& &
          - \frac{192}{(1-y)^3 (x+1)}
          + \frac{96}{(1-y)^3}
          - \frac{64}{3  (1-y)^2 (x+1)^4} \nn\\
& &
           +  \frac{128}{3 (1-y)^2 (x+1)^3}
          - \frac{368}{9 (1-y)^2 (x+1)^2}
          + \frac{152}{9 (1-y)^2 (x+1)} \nn\\
& &
          - \frac{28}{9  (1-y)^2}
           +  \frac{64}{3  (1-y) (x+1)^4}
          - \frac{ 128}{3  (1-y) (x+1)^3} \nn\\
& &
          - \! \frac{208}{9  (1-y) (x+1)^2}\!
           + \! \frac{712}{9  (1-y) (x+1)}\!
          - \! \frac{404}{9  (1-y)}\!
           + \! \frac{32}{3  (x+1)^4} \nn\\
& &
          - \frac{64}{3  (x+1)^3}
           +  \frac{344}{9 (x+1)^2}
          - \frac{124}{9 (x+1)}
\Biggr) H(0,0;x) 
+ \Biggl(
          - \frac{74}{9}  \nn\\
& &
          - \frac{16}{9  x^2 (1-y)^4}
          + \frac{32}{9  x^2 (1-y)^3}
           +  \frac{7}{9  x^2 (1-y)^2}
          - \frac{23}{9  x^2 (1-y)} \nn\\
& &
           +  \frac{80}{9  x (1-y)^4}
          - \frac{160}{9  x (1-y)^3}
          - \frac{56}{9  x (1-y)^2}
           +  \frac{136}{9  x (1-y)} \nn\\
& &
           +  \frac{16}{3  x (y+1)^2}
          - \frac{16}{3  x (y+1)}
           +  \frac{32}{9  x}
           +  \frac{80  x }{9 (1-y)^4}
          - \frac{160 x }{9 (1-y)^3} \nn\\
& &
          - \frac{56  x }{9 (1-y)^2}
           +  \frac{136 x }{9 (1-y)}
           +  \frac{16  x }{3 (y+1)^2}
          - \frac{16  x}{3 (y+1)} \nn\\
& &
           +  \frac{32  x}{9}
          - \frac{16  x^2 }{9 (1-y)^4}
           +  \frac{32  x^2 }{9 (1-y)^3}
           +  \frac{7   x^2 }{9 (1-y)^2}
          - \frac{23 x^2 }{9 (1-y)} \nn\\
& &
           + \! \frac{25}{9  y}\!
           + \! \frac{25  y}{9}
          - \frac{32}{3  (1-y)^4}\!
          + \! \frac{64}{3  (1-y)^3}\!
           + \! \frac{14}{(1-y)^2}
          - \frac{74}{3  (1-y)} \nn\\
& &
          - \frac{32}{3  (y+1)^2}
           +  \frac{32}{3  (y+1)}
\Biggr) H(0,0;y) + 
\Biggl(
            \frac{56}{3}  
          - \frac{16}{3   x^2 (1-y)^4} \nn\\
& &
           +  \frac{32}{3   x^2 (1-y)^3}
          - \frac{28}{9   x^2 (1-y)^2}
          - \frac{20}{9   x^2 (1-y)}
           +  \frac{64}{3   x (1-y)^4} \nn\\
& &
          - \frac{128}{3   x (1-y)^3}
           +  \frac{16}{9   x (1-y)^2}
           +  \frac{176}{9   x (1-y)}
           +  \frac{40}{9   x}
          - \frac{64 x}{3  (1-y)^4} \nn\\
& &
           +  \frac{128 x}{3  (1-y)^3}
          - \frac{16 x}{9  (1-y)^2}
          - \frac{176 x}{9   (1-y)}
          - \frac{40 x}{9}  
           +  \frac{16  x^2}{3  (1-y)^4} \nn\\
& &
          - \! \frac{32   x^2 }{3(1-y)^3} \!
           + \! \frac{28   x^2}{9 (1-y)^2} \!
           + \! \frac{20   x^2}{9 (1-y)} \!
           + \! \frac{20}{9   y (1-x)} \!
           + \! \frac{20}{9   y (x+1)} \nn\\
& &
          - \frac{20}{9   y}
          +  \frac{20   y }{9 (1-x)}
           +  \frac{20   y }{9 (x+1)}
          - \frac{20   y }{9}
           +  \frac{64}{3 (1-x)(1-y)^4} \nn\\
& &
          - \frac{128}{3(1-x) (1-y)^3}
          + \frac{112}{9   (1-x) (1-y)^2}
           +  \frac{80}{9   (1-x) (1-y)} \nn\\
& &
          - \! \frac{88}{9   (1-x)} \!
          - \! \frac{64}{   (1-y)^4 (x+1)}\!
          + \! \frac{64}{3   (1-y)^4} \!
           + \! \frac{128 }{  (1-y)^3 (x+1)} \nn\\
& &
          - \! \frac{128}{3(1\!-\!y)^3} \!
          + \! \frac{16}{3(1\!-\!y)^2 (x\!+\!1)} \!
          - \! \frac{80}{9(1\!-\!y)^2} \!
          - \! \frac{208}{3(1\!-\!y) (x\!+\!1)} \nn\\
& &
           +  \frac{272}{9(1-y)}
          - \frac{248}{9(x+1)}
\Biggr)H(-1,0;x) + 
\Biggl(   
          - \frac{16}{3} 
          - \frac{4}{3  x}
           +  \frac{4  x}{3} \nn\\
& &
          - \! \frac{4}{3  y (1-x)}
           + \! \frac{2}{3  y}
          - \! \frac{4  y }{3 (1-x)}
           + \! \frac{2  y}{3}
           + \! \frac{32}{3  (x+1)}
\Biggr) H(0,0,0;x)    \nn\\
& & + \Biggl(
            \frac{20}{3} 
          - \frac{8}{3  x^2 (1-y)^5}
           +  \frac{20}{3  x^2 (1-y)^4}
          - \frac{4}{  x^2 (1-y)^3} \nn\\
& &           
          - \frac{2}{3  x^2 (1-y)^2}
           +  \frac{2}{  x^2 (1-y)}
          - \frac{4}{3  x^2 (y+1)}
           +  \frac{32}{3  x (1-y)^5} \nn\\
& &
          - \frac{80}{3  x (1-y)^4}
          + \frac{32}{3  x (1-y)^3}
           +  \frac{32}{3  x (1-y)^2}
          - \frac{28}{3  x (1-y)} \nn\\
& &
           +  \frac{16}{3  x (y+1)^3}
          - \frac{8}{  x (y+1)^2}
          + \frac{44}{3  x (y+1)}
          - \frac{4}{  x}
          - \frac{32  x }{3(1-y)^5} \nn\\
& &
           +  \frac{80  x }{3(1-y)^4}
          - \frac{32  x }{3 (1-y)^3}
          - \frac{32  x }{3(1-y)^2}
          - \frac{4  x }{(1-y)}
           +  \frac{16 x}{3  (y+1)^3} \nn\\
& &
          - \frac{8  x}{   (y+1)^2}
           +  \frac{44  x}{3 (y+1)}
          - \frac{4  x}{3}
           +  \frac{8 x^2}{3  (1-y)^5}
          - \frac{20 x^2}{3  (1-y)^4} \nn\\
& &
           +  \frac{4  x^2}{ (1-y)^3}
           +  \frac{2 x^2 }{3 (1-y)^2}
           +  \frac{2 x^2 }{3 (1-y)}
          - \frac{4  x^2 }{3(y+1)} \nn\\
& &
           + \! \frac{2}{3  y (1-x)} \!
          + \! \frac{2}{3  y (x+1)} \!
           + \! \frac{2}{3  y} \!
          - \frac{2  y}{3 (1-x)}
          - \frac{2  y}{3 (x+1)}
          - \frac{2  y}{3}  \nn\\
& &
           +  \frac{32}{3  (1-x) (1-y)^5}
          - \frac{80}{3  (1-x) (1-y)^4}
           +  \frac{16}{(1-x) (1-y)^3} \nn\\
& &
           +  \frac{8}{3  (1-x) (1-y)^2}
          - \frac{16}{3  (1-x) (1-y)}
          + \frac{4}{3  (1-x)} \nn\\
& &
          - \! \frac{32}{(1-y)^5 (x+1)}\!
           + \! \frac{32}{3  (1-y)^5}\!
           + \! \frac{80}{(1-y)^4 (x+1)}\!
          - \! \frac{80}{3  (1-y)^4} \nn\\
& &
          - \frac{80}{3  (1-y)^3 (x+1)}
           +  \frac{16}{3  (1-y)^3}
          - \frac{40}{(1-y)^2 (x+1)} \nn\\
& &
           +  \frac{56}{3  (1-y)^2}
          + \frac{16}{3  (1-y) (x+1)}
           +  \frac{8}{(1-y)}
           +  \frac{20}{3  (x+1)} \nn\\
& &
          - \frac{32}{3  (y+1)^3}
           +  \frac{16}{ (y+1)^2}
          - \frac{104}{3  (y+1)}
\Biggr) H(0,0,0;y) + 
\Biggl(       
            \frac{64}{3}   \nn\\
& &
           +  \frac{32}{3   x^2 (1-y)^5}
          - \frac{80}{3   x^2 (1-y)^4}
          + \frac{16 }{  x^2 (1-y)^3}
           +  \frac{8}{3   x^2 (1-y)^2} \nn\\
& &
          - \frac{8}{3   x^2 (1-y)}
          - \frac{128}{3   x (1-y)^5}
           +  \frac{320}{3   x (1-y)^4}
          - \frac{128}{3   x (1-y)^3} \nn\\
& &
          - \frac{128}{3   x (1-y)^2}
           +  \frac{32}{3   x (1-y)}
           +  \frac{20}{3   x}
           +  \frac{128   x }{3 (1-y)^5}
          - \frac{320  x }{3 (1-y)^4}  \nn\\
& &
          + \frac{128  x}{3  (1-y)^3}
           +  \frac{128  x}{3  (1-y)^2}
          - \frac{32   x }{3 (1-y)}
          - \frac{20  x}{3} 
          - \frac{32   x^2}{3 (1-y)^5} \nn\\
& &
           + \! \frac{80  x^2}{3  (1-y)^4} \!
          - \! \frac{16   x^2}{ (1-y)^3} \!
          - \! \frac{8  x^2 }{3 (1-y)^2} \!
           + \! \frac{8  x^2}{3  (1-y)} \!
          - \! \frac{4}{3   y (1-x)} \nn\\
& &
          - \frac{8}{3   y (x+1)}
           +  \frac{2}{y} 
          + \frac{4   y}{ (1-x)}
           +  \frac{8 y}{3  (x+1)}
          - \frac{10  y}{3}  \nn\\
& &
          - \frac{128}{3   (1-x)(1-y)^5}
           +  \frac{320}{3   (1-x) (1-y)^4}
          - \frac{64 }{  (1-x) (1-y)^3} \nn\\
& &
          - \frac{32}{3   (1-x) (1-y)^2}
           +  \frac{64}{3   (1-x)(1-y)}
          - \frac{16}{3   (1-x)} \nn\\
& &
          + \! \frac{128}{   (1-y)^5 (x+1)}
          - \frac{128}{3   (1-y)^5}
          - \frac{320 }{  (1-y)^4 (x+1)}\!
           + \! \frac{320}{3   (1-y)^4} \nn\\
& &
          + \! \frac{320}{3   (1-y)^3 (x\!+\!1)} \!
          - \! \frac{64}{3   (1-y)^3} \!
           + \! \frac{160}{   (1-y)^2 (x\!+\!1)} \!
          - \! \frac{224}{3   (1-y)^2} \nn\\
& &
          - \frac{64}{3   (1-y) (x+1)}
          - \frac{112}{3   (x+1)}
\Biggr) H(0,0;y) H(0;x) + 
\Biggl(
            4  \nn\\
& &
           +  \frac{8}{3  x^2 (1-y)^5}
          - \frac{20}{3  x^2 (1-y)^4}
           +  \frac{4}{  x^2 (1-y)^3}
           +  \frac{2}{3  x^2 (1-y)^2} \nn\\
& &
          - \frac{2}{3  x^2 (1-y)}
          - \frac{32}{3  x (1-y)^5}
          + \frac{80}{3  x (1-y)^4}
          - \frac{32}{3  x (1-y)^3} \nn\\
& &
          - \frac{32}{3  x (1-y)^2}
           +  \frac{8}{3  x (1-y)}
           +  \frac{4}{3  x}
           +  \frac{32  x}{3 (1-y)^5}
          - \frac{80  x}{3 (1-y)^4} \nn\\
& &
           +  \frac{32  x}{3 (1-y)^3}
           +  \frac{32  x}{3 (1-y)^2}
          - \frac{8  x }{3 (1-y)}
          - \frac{4  x}{3}
          - \frac{8 x^2}{3  (1-y)^5} \nn\\
& &
          + \frac{20  x^2}{3 (1-y)^4}
          - \frac{4  x^2}{ (1-y)^3}
          - \frac{2 x^2 }{3 (1-y)^2}
           +  \frac{2  x^2}{3 (1-y)} \nn\\
& &
          - \! \frac{2}{3  y (1-x)}\!
          - \! \frac{2}{3  y (x+1)}\!
          + \! \frac{2}{3  y}\!
           + \! \frac{2  y}{3 (1-x)}\!
           + \! \frac{2 y}{3  (x+1)}\!
          - \! \frac{2  y}{3} \nn\\
& &
          - \frac{32}{3  (1-x) (1-y)^5}
           +  \frac{80}{3  (1-x) (1-y)^4}
          - \frac{16  }{(1-x) (1-y)^3} \nn\\
& &
          - \frac{8}{3  (1-x) (1-y)^2}
           +  \frac{16}{3  (1-x) (1-y)}
          - \frac{4}{3  (1-x)} \nn\\
& &
          + \frac{32}{  (1-y)^5 (x+1)}
          - \frac{32}{3  (1-y)^5}
          - \frac{80}{  (1-y)^4 (x+1)} \nn\\
& &
           +  \frac{80}{3  (1-y)^4}
          + \frac{80}{3  (1-y)^3 (x+1)}
          - \frac{16}{3  (1-y)^3} \nn\\
& &
           +  \frac{40}{  (1-y)^2 (x+1)}
          - \frac{56}{3  (1-y)^2}
          - \frac{16}{3  (1-y) (x+1)} \nn\\
& &
          - \frac{20}{3  (x+1)}
\Biggr)      
\Biggl[             
\Bigl( G(-y,0,0;x)  - G(-1/y,0,0;x) \Bigr)  \nn\\
& & +    
\Bigl(G(-y,0;x) + G(-1/y,0,x)\Bigr) H(0;y) +
\Bigl(G(-1/y;x)  \nn\\
& & - G(-y;x)\Bigr)   
\Bigl( H(0,0;y) + 3 \zeta(2)
\Bigr) + 2 H(0;x) H(1,0;y)   \nn\\
& &- 2 H(-1,0;x) H(0;y)
              - 6 H(-1,0;y) H(0;x)
\Biggr] \Biggr\}\, . \label{2lB1} 
\eea


\bea
B_2^{(2l,-1)}(-P^2,-Q^2)\!\!\! &=&\!\!\! -\!\Biggl[  
\Biggl(
            \frac{56}{3} 
          - \frac{32}{3  x^2 (1\!-\!y)^4}
          \! + \! \frac{64}{3  x^2 (1\!-\!y)^3}
          - \frac{56}{9  x^2 (1\!-\!y)^2}
          - \frac{40}{9  x^2 (1\!-\!y)}\nn \\
& &                     
          - \frac{32}{3  x (1\!-\!y)^-2}
          \! + \! \frac{32}{3  x (1\!-\!y)}
          \! + \! \frac{40}{9  x}
          \! + \! \frac{32  x }{3(1\!-\!y)^2}
          - \frac{32  x }{3(1\!-\!y)}
          - \frac{40  x}{9}\nn \\
& &
          + \frac{32  x^2}{3 (1\!-\!y)^4}
          - \frac{64  x^2}{3 (1\!-\!y)^3}
          \! + \! \frac{56 x^2}{9  (1\!-\!y)^2}
          \! + \! \frac{40 x^2}{9  (1\!-\!y)}
          \! + \! \frac{20}{9  y (1\!-\!x)} \nn \\
& &
          + \frac{20}{9  y (x\!+\!1)}
          - \frac{20}{9  y}
          \! + \! \frac{20  y }{9(1\!-\!x)}
          \! + \! \frac{20  y }{9(x\!+\!1)}
          - \frac{20  y}{9}\nn \\
& &
          - \frac{128}{3  (1\!-\!x) (1\!-\!y)^4}
          \! + \! \frac{256}{3  (1\!-\!x) (1\!-\!y)^3}
          - \frac{224}{9  (1\!-\!x) (1\!-\!y)^2} \nn \\
& &
          - \frac{160}{9  (1\!-\!x) (1\!-\!y)}
          - \frac{88}{9  (1\!-\!x)}
          - \frac{128}{3  (1\!-\!y)^4 (x\!+\!1)}
          \! + \! \frac{128}{3  (1\!-\!y)^4}\nn \\
& &
          + \frac{256}{3  (1\!-\!y)^3 (x\!+\!1)}
          - \frac{256}{3  (1\!-\!y)^3}
          \! + \! \frac{160}{9  (1\!-\!y)^2 (x\!+\!1)}
          \! + \! \frac{32}{9  (1\!-\!y)^2}\nn \\
& &
          - \frac{544}{9  (1\!-\!y) (x\!+\!1)}
          \! + \! \frac{352}{9  (1\!-\!y)}
          - \frac{248}{9  (x\!+\!1)}
\Biggr) H(0;x) 
+ \Biggl(
          - 8 \nn \\
& &
          - \frac{32}{3   x^2 (1\!-\!y)^5}
          \! + \! \frac{80}{3   x^2 (1\!-\!y)^4}
          - \frac{16}{   x^2 (1\!-\!y)^3}
          - \frac{8}{3   x^2 (1\!-\!y)^2}\nn \\
& &          
          + \frac{8}{3   x^2 (1\!-\!y)}
          - \frac{32}{3   x (1\!-\!y)^3}
          \! + \! \frac{16}{   x (1\!-\!y)^2}
          - \frac{8}{3   x}
          \! + \! \frac{32  x }{3 (1\!-\!y)^3}\nn \\
& &    
          - \frac{16   x}{ (1\!-\!y)^2}
          \! + \! \frac{8  x}{3 }
          \! + \! \frac{32  x^2}{3  (1\!-\!y)^5}
          - \frac{80 x^2}{3   (1\!-\!y)^4}
          \! + \! \frac{16   x^2}{ (1\!-\!y)^3}
          \! + \! \frac{8  x^2}{3  (1\!-\!y)^2}\nn \\
& &  
          - \frac{8  x^2 }{3 (1\!-\!y)}
          \! + \! \frac{4}{3   y (1\!-\!x)}
          \! + \! \frac{4}{3   y (x\!+\!1)}
          - \frac{4}{3   y}
          - \frac{4   y }{3(1\!-\!x)}\nn \\
& &           
          - \frac{4   y}{3 (x\!+\!1)}
          \! + \! \frac{4  y}{3 }
          - \frac{128}{3   (1\!-\!x) (1\!-\!y)}
          \! + \! \frac{320}{3   (1\!-\!x) (1\!-\!y)^4}\nn \\
& &           
          - \frac{64}{   (1\!-\!x) (1\!-\!y)^3}
          - \frac{32}{3   (1\!-\!x) (1\!-\!y)^2}
          \! + \! \frac{16}{3   (1\!-\!x) (1\!-\!y)} \nn \\
& &           
          + \frac{8}{3   (1\!-\!x)}
          - \frac{128}{3   (1\!-\!y)^5 (x\!+\!1)}
          \! + \! \frac{128}{3   (1\!-\!y)^5}
          \! + \! \frac{320}{3   (1\!-\!y)^4 (x\!+\!1)}\nn \\
& &    
          - \frac{320}{3   (1\!-\!y)^4}
          - \frac{64}{3   (1\!-\!y)^3 (x\!+\!1)}
          \! + \! \frac{128}{3   (1\!-\!y)^3}
          - \frac{224}{3   (1\!-\!y)^2 (x\!+\!1)}\nn \\
& &
          +  \frac{128}{3(1-y)^2}
           +  \frac{16}{3(1-y) (x+1)} 
          -  \frac{16}{3(1-y)} \nn\\
& &
           +  \frac{40}{3(x+1)}
\Biggr) H(0;x) H(0;y) \Biggr],
\label{B22lM1}
\eea


\bea
B_2^{(2l,0)}(-P^2,-Q^2)  &=& -\Biggl\{
          - \frac{64}{9}
          - \frac{32}{45 x^2 (1-y)^4}
           +  \frac{64}{45 x^2 (1-y)^3}
           +  \frac{208}{135 x^2 (1-y)^2} \nn\\
& &
          - \frac{304}{135 x^2 (1-y)}
          - \frac{64}{15 x (1-y)^2}
           +  \frac{64}{15 x (1-y)}
           +  \frac{304}{135 x} \nn\\
& &
          - \frac{64 x }{15(1-y)^2}
           +  \frac{64 x }{15(1-y)}
           +  \frac{304 x}{135}
          - \frac{32 x^2}{45 (1-y)^4} \nn\\
& &
           +  \frac{64 x^2 }{45(1-y)^3}
           +  \frac{208 x^2 }{135(1-y)^2}
          - \frac{304 x^2 }{135(1-y)}
          - \frac{8}{3 y (x+1)^2} \nn\\
& &
           +  \frac{8}{3 y (x+1)}
           +  \frac{152}{135 y}
          - \frac{8 y }{3(x+1)^2}
           +  \frac{8 y }{3(x+1)}
           +  \frac{152 y}{135} \nn\\
& &
          - \frac{64}{45 (1-y)^4}
           +  \frac{128}{45 (1-y)^3}
          - \frac{64}{9 (1-y)^2 (x+1)^2} \nn\\
& &
           +  \frac{64}{9 (1-y)^2 (x+1)}
           +  \frac{128}{135 (1-y)^2}
           +  \frac{64}{9 (1-y) (x+1)^2} \nn\\
& &
          - \frac{64}{9 (1-y) (x+1)}
          - \frac{64}{27 (1-y)}
           +  \frac{16}{3 (x+1)^2}
          - \frac{16}{3 (x+1)} \nn\\
& &
+ \Biggl(
           \frac{4}{3}
          - \frac{64}{3 x^2 (1-y)^4}
           +  \frac{128}{3 x^2 (1-y)^3}
          - \frac{112}{9 x^2 (1-y)^2} \nn\\
& &
          - \frac{80}{9 x^2 (1-y)}
           +  \frac{32}{ x (1-y)^4}
          - \frac{64}{ x (1-y)^3}
           +  \frac{8}{3 x (1-y)^2} \nn\\
& &
           +  \frac{88}{3 x (1-y)}
           +  \frac{32}{3 x (y+1)^2}
          - \frac{32}{3 x (y+1)}
           +  \frac{14}{9 x}
           +  \frac{32 x}{ (1-y)^4} \nn\\
& &
          - \frac{64 x}{ (1-y)^3}
           +  \frac{40 x }{3(1-y)^2}
           +  \frac{56 x }{3(1-y)}
           +  \frac{32 x }{3(y+1)^2} \nn\\
& &
          - \frac{32 x }{3(y+1)}
          - \frac{26 x}{9}
          - \frac{32 x^2}{3 (1-y)^4}
           +  \frac{64 x^2}{3 (1-y)^3}
          - \frac{56 x^2 }{9(1-y)^2} \nn\\
& &
          - \frac{40x^2}{9  (1-y)}
           +  \frac{10}{9 y (1-x)}
          - \frac{16}{ y (x+1)^4}
           +  \frac{32}{ y (x+1)^3} \nn\\
& &
          - \frac{20}{ y (x+1)^2}
           +  \frac{46}{9 y (x+1)}
           +  \frac{2}{9 y}
           +  \frac{10 y }{9(1-x)}
          - \frac{16 y }{(x+1)^4} \nn\\
& &
           + \! \frac{32 y }{(x+1)^3}\! 
          - \! \frac{20 y }{(x+1)^2}\! 
           + \!  \frac{46 y }{9(x+1)}\! 
           + \!  \frac{2 y}{9}\! 
          - \! \frac{64}{3 (1-x) (1-y)^4} \nn\\
& &
           + \!  \frac{128}{3 (1-x) (1-y)^3}\! 
          - \! \frac{112}{9 (1-x) (1-y)^2}\! 
          - \! \frac{80}{9 (1-x) (1-y)} \nn\\
& &
          - \frac{44}{9 (1-x)}
          - \frac{128}{ (1-y)^4 (x+1)^2}
           +  \frac{320}{3 (1-y)^4 (x+1)} \nn\\
& &
          - \frac{224}{3 (1-y)^4}
           +  \frac{256}{ (1-y)^3 (x+1)^2}
          - \frac{640}{3 (1-y)^3 (x+1)} \nn\\
& &
           +  \frac{448}{3 (1-y)^3}
          - \frac{128}{3 (1-y)^2 (x+1)^4}
           +  \frac{256}{3 (1-y)^2 (x+1)^3} \nn\\
& &
          - \frac{160}{3 (1-y)^2 (x+1)^2}
           +  \frac{176}{9 (1-y)^2 (x+1)}
          - \frac{104}{9 (1-y)^2} \nn\\
& &
           +  \frac{128}{3 (1-y) (x+1)^4}
          - \frac{256}{3 (1-y) (x+1)^3}
          - \frac{224}{3 (1-y) (x+1)^2} \nn\\
& &
           +  \frac{784}{9 (1-y) (x+1)}
          - \frac{568}{9 (1-y)}
           +  \frac{32}{ (x+1)^4}
          - \frac{64}{ (x+1)^3} \nn\\
& &
           +  \frac{40}{ (x+1)^2}
          - \frac{196}{9 (x+1)}
          - \frac{64}{3 (y+1)^2}
           +  \frac{64}{3 (y+1)}
          \Biggr) \zeta(2)  \nn\\
& &
+ \Biggl(
          - \frac{308}{9}
           +  \frac{32}{ x^2 (1-y)^4}
          - \frac{64}{ x^2 (1-y)^3}
           +  \frac{640}{27 x^2 (1-y)^2} \nn\\
& &
           +  \frac{224}{27 x^2 (1-y)}
           +  \frac{32}{ x (1-y)^2}
          - \frac{32}{ x (1-y)}
          - \frac{272}{27 x}
          - \frac{32 x }{(1-y)^2} \nn\\
& &
           +  \frac{32 x }{(1-y)}
           +  \frac{272 x}{27}
          - \frac{32 x^2}{ (1-y)^4}
           +  \frac{64 x^2}{ (1-y)^3}
          - \frac{640 x^2 }{27(1-y)^2} \nn\\
& &
          - \frac{224 x^2 }{27(1-y)}
          - \frac{118}{27 y (1-x)}
          - \frac{16}{3 y (x+1)^3}
           +  \frac{8}{ y (x+1)^2} \nn\\
& &
          - \frac{214}{27 y (x+1)}
           +  \frac{130}{27 y}
          - \frac{118 y }{27(1-x)}
          - \frac{16 y }{3(x+1)^3}
           +  \frac{8 y }{(x+1)^2} \nn\\
& &
          - \frac{214 y }{27(x+1)}
           +  \frac{130 y}{27}
           +  \frac{128}{ (1-x) (1-y)^4}
          - \frac{256}{ (1-x) (1-y)^3} \nn\\
& &
           +  \frac{2656}{27 (1-x) (1-y)^2}
           +  \frac{800}{27 (1-x) (1-y)}
           +  \frac{356}{27 (1-x)} \nn\\
& &
           +  \frac{128}{ (1-y)^4 (x+1)}
          - \frac{128}{ (1-y)^4}
          - \frac{256}{ (1-y)^3 (x+1)} \nn\\
& &
           +  \frac{256}{ (1-y)^3}
          - \frac{128}{9 (1-y)^2 (x+1)^3}
           +  \frac{64}{3 (1-y)^2 (x+1)^2} \nn\\
& &
          - \frac{1184}{27 (1-y)^2 (x+1)}
          - \frac{832}{27 (1-y)^2}
           +  \frac{128}{9 (1-y) (x+1)^3} \nn\\
& &
          - \frac{64}{3 (1-y) (x+1)^2}
           +  \frac{4640}{27 (1-y)(x+1)}
          - \frac{2624}{27 (1-y)} \nn\\
& &
           +  \frac{32}{3 (x+1)^3}
          - \frac{16}{ (x+1)^2}
           +  \frac{1636}{27 (x+1)}
          \Biggr) H(0;x) 
+ \Biggl(  
            \frac{42}{5} \nn\\
& &
          - \frac{32}{45 x^2 (1-y)^5}
           +  \frac{16}{9 x^2 (1-y)^4}
          - \frac{16}{9 x^2 (1-y)^2}
           +  \frac{4}{9 x^2 (1-y)} \nn\\
& &
           +  \frac{4}{15 x^2}
           +  \frac{4 y}{15 x}
          - \frac{64}{15 x (1-y)^3}
           +  \frac{32}{5 x (1-y)^2}
          - \frac{112}{45 x (1-y)} \nn\\
& &
           +  \frac{16}{3 x (y+1)}
          - \frac{124}{45 x}
           +  \frac{4 x y}{15}
          - \frac{64 x }{15(1-y)^3}
           +  \frac{32 x }{5(1-y)^2} \nn\\
& &
          - \frac{112 x }{45(1-y)}
           +  \frac{16 x }{3(y+1)}
          - \frac{124 x}{45}
          - \frac{32 x^2 }{45(1-y)^5}
           +  \frac{16 x^2}{9 (1-y)^4} \nn\\
& &
          - \! \frac{16 x^2}{9 (1-y)^2}\! 
           + \!  \frac{4 x^2 }{9(1-y)}\! 
           + \!  \frac{4 x^2}{15}\! 
           + \!  \frac{2}{ y}\! 
          - \! \frac{46 y}{15}\! 
           + \!  \frac{2 y^2}{15}\! 
          - \! \frac{64}{45 (1-y)^5} \nn\\
& &
           +  \frac{32}{9 (1-y)^4}
           +  \frac{64}{45 (1-y)^3}
          - \frac{256}{45 (1-y)^2}
          - \frac{8}{5 (1-y)} \nn\\
& &
          - \frac{32}{3 (y+1)}
\Biggr)   H(0;y) +
\Biggl(    - 8
          - \frac{16}{3 x^2 (1-y)^5}
           +  \frac{40}{3 x^2 (1-y)^4} \nn\\
& &
          - \frac{8 }{x^2 (1-y)^3}
          - \frac{4}{3 x^2 (1-y)^2}
           +  \frac{4}{3 x^2 (1-y)}
          - \frac{16}{3 x (1-y)^3} \nn\\
& &
           +  \frac{8}{ x (1-y)^2}
          - \frac{8}{3 x}
           +  \frac{16 x }{3(1-y)^3}
          - \frac{8 x}{ (1-y)^2}
           +  \frac{8 x}{3}
           +  \frac{16 x^2 }{3(1-y)^5} \nn\\
& &
          - \frac{40 x^2 }{3(1-y)^4}
           +  \frac{8 x^2}{ (1-y)^3}
           +  \frac{4 x^2 }{3(1-y)^2}
          - \frac{4 x^2}{3 (1-y)}
          - \frac{4 y }{3(1-x)} \nn\\
& &
          - \frac{4 y }{3(x+1)}
           +  \frac{4 y}{3}
          - \frac{64}{3 (1-x) (1-y)^5}
           +  \frac{160}{3 (1-x) (1-y)^4} \nn\\
& &
          - \frac{32 }{(1-x) (1-y)^3}
          - \frac{16}{3 (1-x) (1-y)^2}
           +  \frac{8}{3 (1-x) (1-y)} \nn\\
& &
           + \! \frac{8}{3 (1-x)}\! 
          -\!  \frac{64}{3 (1-y)^5 (x+1)}\! 
           + \!  \frac{64}{3 (1-y)^5}\! 
           + \!  \frac{160}{3 (1-y)^4 (x+1)} \nn\\
& &
          - \! \frac{160}{3 (1-y)^4}\! 
          - \! \frac{32}{3 (1-y)^3 (x\! +\! 1)}\! 
           + \!  \frac{64}{3 (1-y)^3}\! 
          - \! \frac{112}{3 (1-y)^2 (x\! +\! 1)} \nn\\
& &
           + \! \frac{64}{3 (1\!-\!y)^2}\!
           + \! \frac{8}{3 (1\!-\!y) (x\!+\!1)}\!
          - \!\frac{8}{3 (1\!-\!y)}\!
           + \! \frac{40}{3 (x\!+\!1)}\!
\Biggr) \zeta(2) H(0;x)  \nn\\
& &
+ \Biggl(
          - \!\frac{8}{3}\!
          - \!\frac{64}{3 x^2 (1\!-\!y)^5}\!
           + \! \frac{160}{3 x^2 (1\!-\!y)^4}\!
          - \!\frac{32}{ x^2 (1\!-\!y)^3}\!
          - \!\frac{16}{3 x^2 (1\!-\!y)^2} \nn\\
& &
           +  \frac{8}{ x^2 (1-y)}
          - \frac{8}{3 x^2 (y+1)}
          - \frac{64}{3 x (1-y)^3}
           +  \frac{32}{ x (1-y)^2} \nn\\
& &
          - \frac{8}{3 x (1-y)}
           +  \frac{32}{3 x (y+1)^3}
          - \frac{16}{ x (y+1)^2}
           +  \frac{40}{3 x (y+1)} \nn\\
& &
          - \frac{8}{ x}
           +  \frac{64 x }{3(1-y)^3}
          - \frac{32 x}{ (1-y)^2}
          - \frac{8 x }{3(1-y)}
           +  \frac{32 x }{3(y+1)^3} \nn\\
& &
          - \frac{16 x }{(y+1)^2}
           +  \frac{40 x }{3(y+1)}
           +  \frac{8 x}{3}
           +  \frac{64 x^2}{3 (1-y)^5}
          - \frac{160 x^2 }{3(1-y)^4} \nn\\
& &
           +  \frac{32 x^2 }{(1-y)^3}
           +  \frac{16 x^2}{3 (1-y)^2}
          - \frac{8 x^2 }{3(1-y)}
          - \frac{8 x^2}{3 (y+1)}
           +  \frac{8}{3 y (1-x)} \nn\\
& &
           + \! \frac{8}{3 y (x\!+\!1)}\!
          - \!\frac{4}{3 y}\!
          - \!\frac{8 y }{3(1\!-\!x)}\!
          - \!\frac{8 y }{3(x\!+\!1)}\!
           +\!  \frac{4 y}{3}\!
          - \!\frac{256}{3 (1\!-\!x) (1\!-\!y)^5} \nn\\
& &
           +  \frac{640}{3 (1-x) (1-y)^4}
          - \frac{128}{ (1-x) (1-y)^3}
          - \frac{64}{3 (1-x) (1-y)^2} \nn\\
& &
           +  \frac{32}{3 (1-x) (1-y)}
           +  \frac{16}{3 (1-x)}
          - \frac{256}{3 (1-y)^5 (x+1)} \nn\\
& &
           + \! \frac{256}{3 (1-y)^5}\!
           + \! \frac{640}{3 (1-y)^4 (x\!+\!1)}\!
          - \!\frac{640}{3 (1-y)^4}\!
          - \!\frac{128}{3 (1-y)^3 (x\!+\!1)} \nn\\
& &
           + \! \frac{256}{3 (1-y)^3}\!
          - \!\frac{448}{3 (1-y)^2 (x\!+\!1)}\!
           + \! \frac{256}{3 (1-y)^2}\!
           + \! \frac{32}{3 (1-y) (x\!+\!1)} \nn\\
& &
          - \frac{16}{3 (1-y)}
           +  \frac{80}{3 (x+1)}
          - \frac{64}{3 (y+1)^3}
           +  \frac{32}{ (y+1)^2} \nn\\
& &
          - \frac{128}{3 (y+1)}
\Biggr) \zeta(2) H(0;y)
 + \Biggl( \frac{12}{5}
           +  \frac{4}{15 x^2}
           +  \frac{4}{15 x y}
           +  \frac{4 y}{15 x} \nn\\
& &
          - \frac{8}{15 x}
           +  \frac{4 x }{15 y}
           +  \frac{4 x y}{15}
          - \frac{8 x}{15}
           +  \frac{4 x^2}{15}
           +  \frac{2}{15 y^2}
          - \frac{16}{15 y}
          - \frac{16 y}{15} \nn\\
& &
           +  \frac{2 y^2}{15}
\Biggr) H(1;y) 
+ \Biggl(      
  \frac{44}{3}
           +  \frac{128}{3 x^2 (1-y)^5}
          - \frac{320}{3 x^2 (1-y)^4}\nn\\
& &
           +  \frac{72}{ x^2 (1-y)^3} 
          - \frac{4}{3 x^2 (1-y)^2}
          - \frac{20}{3 x^2 (1-y)}
           +  \frac{352}{9 x (1-y)^3} \nn\\
& &
          - \frac{176}{3 x (1-y)^2}
           +  \frac{32}{3 x (1-y)}
           +  \frac{40}{9 x}
          - \frac{352 x }{9(1-y)^3}
           +  \frac{176 x}{3 (1-y)^2} \nn\\
& &
          - \frac{32 x }{3(1-y)}
          - \frac{40 x}{9}
          - \frac{128 x^2}{3 (1-y)^5}
           +  \frac{320 x^2 }{3(1-y)^4}
          - \frac{72 x^2}{ (1-y)^3} \nn\\
& &
           +  \frac{4 x^2 }{3(1-y)^2}
           +  \frac{20x^2 }{3 (1-y)}
          - \frac{20}{9 y (1-x)}
          - \frac{20}{9 y (x+1)} \nn\\
& &
           +  \frac{20}{9 y}
           +  \frac{20 y }{9(1-x)}
           +  \frac{20 y }{9(x+1)}
          - \frac{20 y}{9}
           +  \frac{512}{3 (1-x) (1-y)^5} \nn\\
& &
          - \frac{1280}{3 (1-x) (1-y)^4}
           +  \frac{288}{ (1-x) (1-y)^3}
          - \frac{16}{3 (1-x) (1-y)^2} \nn\\
& &
          - \! \frac{136}{9 (1-x) (1-y)}\! 
          - \! \frac{52}{9 (1-x)}\! 
           + \!  \frac{512}{3 (1-y)^5 (x\! +\! 1)}\! 
          - \! \frac{512}{3 (1-y)^5} \nn\\
& &
          - \frac{1280}{3 (1-y)^4 (x+1)}
           +  \frac{1280}{3 (1-y)^4}
           +  \frac{1184}{9 (1-y)^3 (x+1)} \nn\\
& &
          - \!\frac{1888}{9 (1-y)^3}\!
           + \! \frac{688}{3 (1-y)^2 (x\!+\!1)}\!
          - \!\frac{112 }{(1-y)^2}\!
          - \!\frac{520}{9 (1-y)(x\!+\!1)} \nn\\
& &
           + \! \frac{328}{9 (1-y)}\!
          - \!\frac{212}{9 (x+1)}
\Biggr) H(0;y) H(0;x) 
+ \Biggl(
          - \frac{28}{3}\!
           + \! \frac{32}{3 x (1-y)^4} \nn\\
& &
          - \frac{64}{3 x (1-y)^3}
           +  \frac{8 }{x (1-y)^2}
           +  \frac{8}{3 x (1-y)}
          - \frac{26}{9 x}
           +  \frac{32 x }{3(1-y)^4} \nn\\
& &
          - \frac{64 x }{3(1-y)^3}
          - \frac{8 x }{3(1-y)^2}
           +  \frac{40 x }{3(1-y)}
           +  \frac{14 x}{9}
          - \frac{32x^2 }{3 (1-y)^4} \nn\\
& &
           +  \frac{64x^2 }{3 (1-y)^3}
          - \frac{56 x^2 }{9(1-y)^2}
          - \frac{40 x^2 }{9(1-y)}
          - \frac{10}{9 y (1-x)} \nn\\
& &
          - \frac{16}{3 y (x+1)^4}
           +  \frac{32}{3 y (x+1)^3}
          - \frac{20}{3 y (x+1)^2}
           +  \frac{2}{9 y (x+1)} \nn\\
& &
           +  \frac{10}{9 y}
          - \frac{10 y }{9(1-x)}
          - \frac{16 y }{3 (x+1)^4}
           +  \frac{32 y }{3(x+1)^3}
          - \frac{20 y }{3(x+1)^2} \nn\\
& &
           +  \frac{2 y }{9 (x+1)}
           +  \frac{10 y}{9}
           +  \frac{64}{3 (1-x) (1-y)^4}
          - \frac{128}{3 (1-x) (1-y)^3} \nn\\
& &
           +  \frac{112}{9 (1-x) (1-y)^2}
           +  \frac{80}{9 (1-x) (1-y)}
           +  \frac{44}{9 (1-x)} \nn\\
& &
          - \frac{128}{3 (1-y)^4 (x+1)^2}
           +  \frac{64}{ (1-y)^4 (x+1)}
          - \frac{160}{3 (1-y)^4} \nn\\
& &
           +  \frac{256}{3 (1-y)^3 (x+1)^2}
          - \frac{128}{ (1-y)^3 (x+1)}
           +  \frac{320}{3 (1-y)^3} \nn\\
& &
          - \! \frac{128}{9 (1-y)^2 (x\!+\!1)^4}\!
           + \! \frac{256}{9 (1-y)^2 (x\!+\!1)^3}\!
          - \!\frac{160}{9 (1-y)^2 (x\!+\!1)^2} \nn\\
& &
          - \frac{16}{3 (1-y)^2 (x+1)}
          - \frac{56}{9 (1-y)^2}
           +  \frac{128}{9 (1-y) (x+1)^4} \nn\\
& &
          - \frac{256}{9 (1-y) (x+1)^3}
          - \frac{224}{9 (1-y) (x+1)^2}
           +  \frac{208}{3 (1-y) (x+1)} \nn\\
& &
          - \frac{424}{9 (1-y)}
           +  \frac{32}{3 (x+1)^4}
          - \frac{64}{3 (x+1)^3}
           +  \frac{40}{3 (x+1)^2} \nn\\
& &
           + \! \frac{100}{9 (x\!+\!1)}
\Biggr) H(0,0;x) \!
+ \!\Biggl(
          - \frac{142}{9}\!
          - \!\frac{32}{9 x^2 (1-y)^4}\!
           + \! \frac{64}{9 x^2 (1-y)^3} \nn\\
& &
           +  \frac{14}{9 x^2(1-y)^2}
          - \frac{46}{9 x^2 (1-y)}
           +  \frac{32}{9 x (1-y)^4}
          - \frac{64}{9 x (1-y)^3} \nn\\
& &
          - \frac{40}{9 x (1-y)^2}
           +  \frac{8 }{x (1-y)}
           +  \frac{32}{3 x (y+1)^2}
          - \frac{32}{3 x (y+1)} \nn\\
& &
           +  \frac{32}{9 x}
           +  \frac{32 x }{9(1-y)^4}
          - \frac{64 x }{9(1-y)^3}
          - \frac{40 x }{9(1-y)^2}
           +  \frac{8 x }{(1-y)} \nn\\
& &
           +  \frac{32 x }{3(y+1)^2}
          - \frac{32 x }{3(y+1)}
           +  \frac{32 x}{9}
          - \frac{32 x^2}{9 (1-y)^4}
           +  \frac{64 x^2 }{9(1-y)^3} \nn\\
& &
           + \! \frac{14 x^2}{9 (1-y)^2}\!
          - \!\frac{46 x^2 }{9(1-y)}\!
           + \! \frac{25}{9 y}\!
           + \! \frac{25 y}{9}\!
          - \!\frac{64}{9 (1-y)^4}\!
           + \! \frac{128}{9 (1-y)^3} \nn\\
& &
           + \! \frac{20}{3 (1-y)^2}\!
          - \!\frac{124}{9 (1-y)}\!
          - \!\frac{64}{3 (y+1)^2}\!
           + \! \frac{64}{3 (y+1)}
\Biggr) H(0,0;y)  \nn\\
& &
+ \Biggl(
            \frac{56}{3}
          - \frac{32}{3 x^2 (1-y)^4}
           +  \frac{64}{3 x^2 (1-y)^3}
          - \frac{56}{9 x^2 (1-y)^2} \nn\\
& &
          - \frac{40}{9 x^2 (1-y)}
          - \frac{32}{3 x (1-y)^2}
           +  \frac{32}{3 x (1-y)}
           +  \frac{40}{9 x}
           +  \frac{32 x }{3(1-y)^2} \nn\\
& &
          - \frac{32 x }{3(1-y)}
          - \frac{40 x}{9}
           +  \frac{32 x^2}{3 (1-y)^4}
          - \frac{64 x^2 }{3(1-y)^3}
           +  \frac{56 x^2}{9 (1-y)^2} \nn\\
& &
           +  \frac{40 x^2}{9 (1-y)}
           +  \frac{20}{9 y (1-x)}
           +  \frac{20}{9 y (x+1)}
          - \frac{20}{9 y}
           +  \frac{20 y }{9(1-x)} \nn\\
& &
           +  \frac{20 y }{9(x+1)}
          - \frac{20 y}{9}
          - \frac{128}{3 (1-x) (1-y)^4}
           +  \frac{256}{3 (1-x) (1-y)^3} \nn\\
& &
          - \frac{224}{9 (1-x) (1-y)^2}
          - \frac{160}{9 (1-x) (1-y)}
          - \frac{88}{9 (1-x)} \nn\\
& &
          - \frac{128}{3 (1-y)^4 (x+1)}
           +  \frac{128}{3 (1-y)^4}
           +  \frac{256}{3 (1-y)^3 (x+1)} \nn\\
& &
          - \frac{256}{3 (1-y)^3}
           +  \frac{160}{9 (1-y)^2 (x+1)}
           +  \frac{32}{9 (1-y)^2} \nn\\
& &
          - \frac{544}{9 (1-y) (x+1)}
           +  \frac{352}{9 (1-y)}
          - \frac{248}{9 (x+1)}
\Biggr)H(-1,0;x)  \nn\\
& &
+ \Biggl(   - 4 \!
          - \!\frac{4}{3 x}\!
           + \! \frac{4 x}{3}\!
          - \!\frac{2}{3 y (1-x)}\!
          - \!\frac{2}{3 y (x+1)}\!
           + \! \frac{2}{3 y}\!
          - \!\frac{2 y }{3(1-x)} \nn\\
& &
          - \frac{2 y }{3(x+1)}
          +\frac{ 2 y}{3}
           +  \frac{4}{3 (1-x)}
           +  \frac{20}{3 (x+1)}
\Biggr) H(0,0,0;x)  \nn\\
& &
+ \Biggl(
            \frac{28}{3}
          - \frac{16}{3 x^2 (1-y)^5}
           +  \frac{40}{3 x^2 (1-y)^4}
          - \frac{8}{ x^2 (1-y)^3} \nn\\
& &
          - \frac{4}{3 x^2 (1-y)^2}
           +  \frac{4 }{x^2 (1-y)}
          - \frac{8}{3 x^2 (y+1)}
          - \frac{16}{3 x (1-y)^3} \nn\\
& &
           +  \frac{8}{ x (1-y)^2}
          - \frac{8}{3 x (1-y)}
           +  \frac{32}{3 x (y+1)^3}
          - \frac{16 x (y+1)^2} \nn\\
& &
           +  \frac{40}{3 x (y+1)}
          - \frac{4}{ x}
           +  \frac{16 x }{3(1-y)^3}
          - \frac{8 x }{(1-y)^2}
          - \frac{8 x }{3(1-y)} \nn\\
& &
           +  \frac{32 x }{3(y+1)^3}
          - \frac{16 x }{(y+1)^2}
           +  \frac{40 x }{3(y+1)}
          - \frac{4 x}{3}
           +  \frac{16 x^2}{3 (1-y)^5} \nn\\
& &
          - \frac{40 x^2 }{3(1-y)^4}
           +  \frac{8 x^2}{ (1-y)^3}
           +  \frac{4 x^2 }{3(1-y)^2}
           +  \frac{4 x^2 }{3(1-y)} \nn\\
& &
          - \frac{8 x^2}{3 (y+1)}
           +  \frac{2}{3 y (1-x)}
           +  \frac{2}{3 y (x+1)}
           +  \frac{2}{3 y}
          - \frac{2 y }{3(1-x)} \nn\\
& &
          - \frac{2 y }{3(x+1)}
          - \frac{2 y}{3}
          - \frac{64}{3 (1-x) (1-y)^5}
           +  \frac{160}{3 (1-x) (1-y)^4} \nn\\
& &
          - \frac{32 }{(1-x) (1-y)^3}
          - \frac{16}{3 (1-x) (1-y)^2}
           +  \frac{8}{3 (1-x) (1-y)} \nn\\
& &
           + \! \frac{4}{3 (1\!-\!x)}\!
          - \!\frac{64}{3 (1\!-\!y)^5 (x\!+\!1)}\!
           + \! \frac{64}{3 (1\!-\!y)^5}\!
           + \! \frac{160}{3 (1\!-\!y)^4 (x\!+\!1)} \nn\\
& &
          - \!\frac{160}{3 (1\!-\!y)^4}\!
          - \!\frac{32}{3 (1\!-\!y)^3 (x\!+\!1)}\!
           + \! \frac{64}{3 (1\!-\!y)^3}\!
          - \!\frac{112}{3 (1\!-\!y)^2 (x\!+\!1)} \nn\\
& &
           +  \frac{64}{3 (1-y)^2}
           +  \frac{8}{3 (1-y) (x+1)}
           +  \frac{8}{3 (1-y)}
           +  \frac{20}{3 (x+1)}\nn\\
& &
          - \frac{64}{3 (y+1)^3} 
           +  \frac{32 }{(y+1)^2}
          - \frac{128}{3 (y+1)}
\Biggr) H(0,0,0;y)  \nn\\
& &
+ \Biggl(
             20
           +  \frac{64}{3 x^2 (1-y)^5}
          - \frac{160}{3 x^2 (1-y)^4}
           +  \frac{32}{ x^2 (1-y)^3} \nn\\
& &
           +  \frac{16}{3 x^2 (1-y)^2}
          - \frac{16}{3 x^2 (1-y)}
           +  \frac{64}{3 x (1-y)^3}
          - \frac{32 }{x (1-y)^2}
           +  \frac{20}{3 x} \nn\\
& &
          - \frac{64 x }{3(1-y)^3}
           +  \frac{32 x}{ (1-y)^2}
          - \frac{20 x}{3}
          - \frac{64 x^2 }{3(1-y)^5}
           +  \frac{160 x^2}{3 (1-y)^4} \nn\\
& &
          - \frac{32 x^2}{ (1-y)^3}
          - \frac{16 x^2 }{3 (1-y)^2}
           +  \frac{16 x^2 }{3(1-y)}
          - \frac{2}{ y (1-x)}
          - \frac{2}{ y (x+1)} \nn\\
& &
           +  \frac{2}{ y}
           +  \frac{10 y }{3(1-x)}
           +  \frac{10 y }{3(x+1)}
          - \frac{10 y}{3}
           +  \frac{256}{3 (1-x) (1-y)^5} \nn\\
& &
          - \frac{640}{3 (1-x) (1-y)^4}
           +  \frac{128}{(1-x) (1-y)^3}
           +  \frac{64}{3 (1-x) (1-y)^2} \nn\\
& &
          - \! \frac{32}{3 (1-x) (1-y)}\!
          - \!\frac{20}{3 (1-x)}\!
           + \! \frac{256}{3 (1-y)^5 (x\!+\!1)} \!
          - \!\frac{256}{3 (1-y)^5}\nn\\
& &
          - \frac{640}{3 (1-y)^4 (x+1)}
           +  \frac{640}{3 (1-y)^4} 
           +  \frac{128}{3 (1-y)^3 (x+1)} \nn\\
& &
          - \frac{256}{3 (1-y)^3}
           +  \frac{448}{3 (1-y)^2 (x+1)} 
          - \frac{256}{3 (1-y)^2} \nn\\
& &
          - \frac{32}{3 (1-y) (x+1)}
           +  \frac{32}{3 (1-y)}
          - \frac{100}{3 (x+1)}
\Biggr) H(0,0;y) H(0;x)  \nn\\
& &
+ \Biggl(
           4\!
           + \! \frac{16}{3 x^2 (1-y)^5}\!
          - \!\frac{40}{3 x^2 (1-y)^4}\!
           + \! \frac{8}{ x^2 (1-y)^3}\!
           + \! \frac{4}{3 x^2 (1-y)^2} \nn\\
& &
          - \frac{4}{3 x^2 (1-y)}
           +  \frac{16}{3 x (1-y)^3}
          - \frac{8}{ x (1-y)^2}
           +  \frac{4}{3 x}
          - \frac{16 x }{3(1-y)^3} \nn\\
& &
           +  \frac{8 x}{ (1-y)^2}
          - \frac{4 x}{3}
          - \frac{16 x^2 }{3(1-y)^5}
           +  \frac{40 x^2 }{3(1-y)^4}
          - \frac{8 x^2 }{(1-y)^3} \nn\\
& &
          - \frac{4 x^2 }{3(1-y)^2}
           +  \frac{4 x^2 }{3(1-y)}
          - \frac{2}{3 y (1-x)}
          - \frac{2}{3 y (x+1)}
           +  \frac{2}{3 y} \nn\\
& &
           +  \frac{2 y }{3(1-x)}
           +  \frac{2 y }{3(x+1)}
          - \frac{2 y}{3}
           +  \frac{64}{3 (1-x) (1-y)^5} \nn\\
& &
          - \frac{160}{3 (1-x) (1-y)^4}
           +  \frac{32}{ (1-x) (1-y)^3}
           +  \frac{16}{3 (1-x) (1-y)^2} \nn\\
& &
          - \! \frac{8}{3 (1-x) (1-y)}\!
          - \!\frac{4}{3 (1-x)}\!
           + \! \frac{64}{3 (1-y)^5 (x\!+\!1)}\!
          - \!\frac{64}{3 (1-y)^5} \nn\\
& &
          - \frac{160}{3 (1-y)^4 (x+1)}
           +  \frac{160}{3 (1-y)^4}
           +  \frac{32}{3 (1-y)^3 (x+1)} \nn\\
& &
          - \!\frac{64}{3 (1-y)^3}\!
           + \! \frac{112}{3 (1-y)^2 (x\!+\!1)}\!
          - \!\frac{64}{3 (1-y)^2}\!
          - \!\frac{8}{3 (1-y) (x\!+\!1)} \nn\\
& &
          +\frac{ 8}{3 (1-y)}
          - \frac{20}{3 (x+1)}
\Biggr)      
\Biggl[             
\Bigl( G(-y,0,0;x) 
              - G(-1/y,0,0;x) \Bigr)  \nn\\
& & + \Bigl(G(-y,0;x) + G(-1/y,0,x)\Bigr) H(0;y)
              + \Bigl(G(-1/y;x)   \nn\\
& &- G(-y;x)\Bigr)  
\Bigl( H(0,0;y) + 3 \zeta(2)
\Bigr) 
        + 2 H(0;x) H(1,0;y) \nn\\
& &
        - 2 H(-1,0;x) H(0;y)        
	- 6 H(-1,0;y) H(0;x)
\Biggr] \Biggr\}\, .  
\label{2lB2} 
\eea
%

\bea
B_3^{(2l,-1)}(-P^2,-Q^2)\!\! &=&\!\! -\Biggl[ \Biggl(
            \frac{16}{3  x^2 (1-y)^4}
          - \frac{32}{3  x^2 (1-y)^3}
           +  \frac{28}{9  x^2 (1-y)^2}
           +  \frac{20}{9  x^2 (1-y)} \nn\\
& &
           +  \frac{64}{3  x (1-y)^4}
          - \frac{128}{3  x (1-y)^3}
           +  \frac{112}{9  x (1-y)^2}
           +  \frac{80}{9  x (1-y)} \nn\\
& &
          - \frac{64 x }{3 (1-y)^4}
           +  \frac{128  x}{3 (1-y)^3}
          - \frac{112  x }{9(1-y)^2}
          - \frac{80 x }{9 (1-y)} \nn\\
& &
          - \frac{16  x^2}{3 (1-y)^4}
           +  \frac{32 x^2}{3  (1-y)^3}
          - \frac{28  x^2}{9 (1-y)^2}
          - \frac{20  x^2}{9 (1-y)} \nn\\
& &
           +  \frac{64}{  (1-x) (1-y)^4}
          - \frac{128}{  (1-x) (1-y)^3}
           +  \frac{112}{3  (1-x) (1-y)^2} \nn\\
& &
           +  \frac{80}{3  (1-x) (1-y)}
          - \frac{64}{3  (1-y)^4 (x+1)}
          - \frac{64}{3  (1-y)^4} \nn\\
& &
           +  \frac{128}{3  (1-y)^3 (x+1)}
           +  \frac{128}{3  (1-y)^3}
          - \frac{112}{9  (1-y)^2 (x+1)} \nn\\
& &
          - \frac{112}{9  (1-y)^2}
          - \frac{80}{9  (1-y) (x+1)}
          - \frac{80}{9  (1-y)}
\Biggr) H(0;x)  \nn\\
& & + \Biggl(
            \frac{16}{3   x^2 (1-y)^5} \!
          - \!\frac{40}{3   x^2 (1-y)^4}\!
           + \! \frac{8}{   x^2 (1-y)^3}\!
           + \! \frac{4}{3   x^2 (1-y)^2} \nn\\
& &
          - \frac{4}{3   x^2 (1-y)}
           +  \frac{64}{3   x (1-y)^5}
          - \frac{160}{3   x (1-y)^4}
           +  \frac{32}{   x (1-y)^3} \nn\\
& &
           +  \frac{16}{3   x (1-y)^2}
          - \frac{16}{3   x (1-y)}
          - \frac{64  x}{3  (1-y)^5}
           +  \frac{160  x }{3 (1-y)^4} \nn\\
& &
          - \! \frac{32   x }{(1-y)^3}\!
          - \!\frac{16  x }{3 (1-y)^2}\!
           +  \!\frac{16   x}{3 (1-y)}\!
          -\! \frac{16   x^2 }{3(1-y)^5}\!
           + \! \frac{40  x^2}{3  (1-y)^4} \nn\\
& &
          - \frac{8   x^2 }{(1-y)^3}
          - \frac{4  x^2 }{3 (1-y)^2}
           +  \frac{4  x^2}{3  (1-y)}
           +  \frac{64  }{ (1-x) (1-y)^5} \nn\\
& &
          - \frac{160}{   (1-x) (1-y)^4}
           +  \frac{96 }{  (1-x) (1-y)^3}
           +  \frac{16 }{  (1-x) (1-y)^2} \nn\\
& &
          - \frac{16 }{  (1-x) (1-y)}
          - \frac{64}{3   (1-y)^5 (x+1)}
          - \frac{64}{3   (1-y)^5} \nn\\
& &
           +  \frac{160}{3   (1-y)^4 (x+1)}
           +  \frac{160}{3   (1-y)^4}
          - \frac{32 }{  (1-y)^3 (x+1)} \nn\\
& &
          - \frac{32 }{  (1-y)^3}
          - \frac{16}{3   (1-y)^2 (x+1)}
          - \frac{16}{3   (1-y)^2} \nn\\
& &
           +  \frac{16}{3   (1-y) (x+1)}
           +  \frac{16}{3   (1-y)}
         \Biggr)  H(0;x) H(0;y) \Biggr]
\, ,
\label{B32lM1}
\eea


\bea
B_3^{(2l,0)}(-P^2,-Q^2) &=& - \Biggl\{ \frac{16}{3}
           +  \frac{16}{45 x^2 (1-y)^4}
          - \frac{32}{45 x^2 (1-y)^3}
          - \frac{104}{135 x^2 (1-y)^2} \nn\\
& &
           +  \frac{152}{135 x^2 (1-y)}
           +  \frac{64}{45 x (1-y)^4}
          - \frac{128}{45 x(1-y)^3} \nn\\
& &
           +  \frac{64}{135 x (1-y)^2}
           +  \frac{128}{135 x (1-y)}
           +  \frac{64 x }{45(1-y)^4}
          - \frac{128 x }{45(1-y)^3} \nn\\
& &
           +  \frac{64 x }{135(1-y)^2}
           +  \frac{128 x }{135(1-y)}
           +  \frac{16 x^2}{45 (1-y)^4}
          - \frac{32 x^2}{45 (1-y)^3} \nn\\
& &
          - \frac{104 x^2}{135 (1-y)^2}
           +  \frac{152 x^2}{135 (1-y)}
          - \frac{32}{9 y (x+1)^2}
           +  \frac{32}{9 y (x+1)} \nn\\
& &
          - \frac{32 y }{9(x+1)^2}
           +  \frac{32 y }{9(x+1)}
           +  \frac{32}{45 (1-y)^4}
          - \frac{64}{45 (1-y)^3} \nn\\
& &
          - \frac{32}{9 (1-y)^2 (x+1)^2}
           +  \frac{32}{9 (1-y)^2 (x+1)}
           +  \frac{992}{135 (1-y)^2} \nn\\
& &
           +  \frac{32}{9 (1-y) (x+1)^2}
          - \frac{32}{9 (1-y) (x+1)}
          - \frac{896}{135 (1-y)} \nn\\
& &
+ \Biggl(    20
           +  \frac{32}{3 x^2 (1-y)^4}
          - \frac{64}{3 x^2 (1-y)^3}
           +  \frac{56}{9 x^2 (1-y)^2} \nn\\
& &
           +  \frac{40}{9 x^2 (1-y)}
           +  \frac{80}{3 x (1-y)^4}
          - \frac{160}{3 x (1-y)^3}
           +  \frac{116}{9 x (1-y)^2} \nn\\
& &
           +  \frac{124}{9 x (1-y)}
          - \frac{16}{3 x (y+1)^2}
           +  \frac{16}{3 x (y+1)}
           +  \frac{16 x }{3(1-y)^4} \nn\\
& &
          - \frac{32 x }{3(1-y)^3}
           +  \frac{4 x }{9(1-y)^2}
           +  \frac{44 x }{9(1-y)}
          - \frac{16 x }{3(y+1)^2} \nn\\
& &
           +  \frac{16 x }{3(y+1)}
           +  \frac{16 x^2}{3 (1-y)^4}
          - \frac{32 x^2}{3 (1-y)^3}
           +  \frac{28x^2}{9 (1-y)^2} \nn\\
& &
           +  \frac{20x^2}{9 (1-y)}
          - \frac{64}{3 y (x+1)^4}
           +  \frac{128}{3 y (x+1)^3}
          - \frac{56}{3 y (x+1)^2} \nn\\
& &
          - \frac{8}{3 y (x+1)}
          - \frac{64 y }{3(x+1)^4}
           +  \frac{128 y }{3(x+1)^3}
          - \frac{56 y }{3(x+1)^2} \nn\\
& &
          - \frac{8 y }{3(x+1)}
           +  \frac{32 }{(1-x) (1-y)^4}
          - \frac{64 }{(1-x) (1-y)^3} \nn\\
& &
           +  \frac{56}{3 (1-x) (1-y)^2}
           +  \frac{40}{3 (1-x) (1-y)}
          - \frac{64}{ (1-y)^4 (x+1)^2} \nn\\
& &
           +  \frac{160}{3 (1-y)^4 (x+1)}
          - \frac{80}{3 (1-y)^4}
           +  \frac{128}{ (1-y)^3 (x+1)^2} \nn\\
& &
          - \frac{320}{3 (1-y)^3 (x+1)}
           +  \frac{160}{3 (1-y)^3}
          - \frac{64}{3 (1-y)^2 (x+1)^4} \nn\\
& &
           +  \frac{128}{3 (1-y)^2 (x+1)^3}
          - \frac{208}{3 (1-y)^2 (x+1)^2}
           +  \frac{376}{9 (1-y)^2 (x+1)} \nn\\
& &
          - \frac{140}{9 (1-y)^2}
           +  \frac{64}{3 (1-y) (x+1)^4}
          - \frac{128}{3 (1-y) (x+1)^3} \nn\\
& &
           + \! \frac{16}{3 (1-y) (x\!+\!1)^2}\!
           + \! \frac{104}{9 (1-y) (x+1)}\!
          - \!\frac{100}{9 (1-y)}\!
           + \! \frac{224}{3 (x\!+\!1)^2} \nn\\
& &
          - \!\frac{224}{3 (x+1)}\!
           + \! \frac{32}{3 (y+1)^2}\!
          - \!\frac{32}{3 (y\!+\!1)}
\Biggr)   \zeta(2) \!
+\! \Biggl(
          - \frac{88}{9}\!
          - \!\frac{40}{3 x^2 (1-y)^4} \nn\\
& &
           +  \frac{80}{3 x^2 (1-y)^3}
          - \frac{278}{27 x^2 (1-y)^2}
          - \frac{82}{27 x^2 (1-y)}
          - \frac{176}{3 x (1-y)^4} \nn\\
& &
           +  \frac{352}{3 x (1-y)^3}
          - \frac{1172}{27 x (1-y)^2}
          - \frac{412}{27 x (1-y)}
          - \frac{10}{9 x}
           +  \frac{176 x }{3(1-y)^4} \nn\\
& &
          - \frac{352 x }{3(1-y)^3}
           +  \frac{1172 x }{27(1-y)^2}
           +  \frac{412 x }{27(1-y)}
           +  \frac{10 x}{9}
           +  \frac{40 x^2 }{3(1-y)^4} \nn\\
& &
          - \frac{80 x^2 }{3(1-y)^3}
           +  \frac{278 x^2 }{27(1-y)^2}
           +  \frac{82 x^2}{27 (1-y)}
           +  \frac{2}{3 y (1-x)} \nn\\
& &
          - \frac{64}{9 y (x+1)^3}
           +  \frac{32}{3 y (x+1)^2}
          - \frac{58}{9 y (x+1)}
           +  \frac{10}{9 y}
           +  \frac{2 y }{3(1-x)} \nn\\
& &
          - \! \frac{64 y }{9(x+1)^3}\!
           + \! \frac{32 y }{3(x+1)^2}\!
          - \!\frac{58 y }{9(x\!+\!1)}\!
           + \! \frac{10 y}{9}\!
          - \!\frac{512}{3 (1-x) (1-y)^4} \nn\\
& &
           +  \frac{1024}{3 (1-x) (1-y)^3}
          - \frac{1184}{9 (1-x) (1-y)^2}
          - \frac{352}{9 (1-x) (1-y)} \nn\\
& &
           +  \frac{20}{9 (1-x)}
           +  \frac{64 }{(1-y)^4 (x+1)}
           +  \frac{160}{3 (1-y)^4} \nn\\
& &
          - \frac{128}{ (1-y)^3 (x+1)}
          - \frac{320}{3 (1-y)^3}
          - \frac{64}{9 (1-y)^2 (x+1)^3} \nn\\
& &
           +  \frac{32}{3 (1-y)^2 (x+1)^2}
           +  \frac{1520}{27 (1-y)^2 (x+1)}
           +  \frac{968}{27 (1-y)^2} \nn\\
& &
           +  \frac{64}{9 (1-y) (x+1)^3}
          - \frac{32}{3 (1-y) (x+1)^2}
           +  \frac{208}{27 (1-y) (x+1)} \nn\\
& &
          +\frac{ 472}{27 (1-y)}
          +\frac{ 52}{3 (x+1)}
\Biggr) H(0;x) 
+ \Biggl(
           - \frac{332}{45}
           +  \frac{16}{45 x^2 (1-y)^5} \nn\\
& &
          - \frac{8}{9 x^2 (1-y)^4}
           +  \frac{8}{9 x^2 (1-y)^2}
          - \frac{2}{9 x^2 (1-y)}
          - \frac{2}{15 x^2} \nn\\
& &
           +  \frac{64}{45 x (1-y)^5}
          - \frac{32}{9 x (1-y)^4}
           +  \frac{32}{9 x (1-y)^3}
          - \frac{16}{9 x (1-y)^2} \nn\\
& &
           +  \frac{32}{9 x (1-y)}
          - \frac{8}{3 x (y+1)}
          - \frac{8}{15 x}
           +  \frac{64 x }{45(1-y)^5}
          - \frac{32 x }{9(1-y)^4} \nn\\
& &
           +  \frac{32 x }{9(1-y)^3}
          - \frac{16 x }{9(1-y)^2}
           +  \frac{32 x }{9(1-y)}
          - \frac{8 x }{3(y+1)}
          - \frac{8 x}{15} \nn\\
& &
           +  \frac{16 x^2}{45 (1-y)^5}
          - \frac{8 x^2}{9 (1-y)^4}
           +  \frac{8 x^2}{9 (1-y)^2}
          - \frac{2 x^2}{9 (1-y)}
          - \frac{2 x^2}{15} \nn\\
& &
           + \! \frac{32}{45 (1-y)^5}\!
          - \frac{16}{9 (1-y)^4}\!
           + \! \frac{64}{9 (1-y)^3}\!
          - \frac{80}{9 (1-y)^2}\!
           + \! \frac{12 }{(1-y)} \nn\\
& &
           +  \frac{16}{3 (y+1)}
\Biggr) H(0;y) 
+ \Biggl(
          - \frac{4}{3}
          +  \frac{8}{3 x^2 (1-y)^5}
          - \frac{20}{3 x^2 (1-y)^4} \nn\\
& &
           +  \frac{4}{ x^2 (1-y)^3}
           +  \frac{2}{3 x^2 (1-y)^2}
          - \frac{2}{3 x^2 (1-y)}
           +  \frac{32}{3 x (1-y)^5} \nn\\
& &
          - \frac{80}{3 x (1-y)^4}
           +  \frac{16}{ x (1-y)^3}
           +  \frac{8}{3 x (1-y)^2}
          - \frac{8}{3 x (1-y)} \nn\\
& &
          - \frac{32 x}{3 (1-y)^5}
           +  \frac{80 x }{3(1-y)^4}
          - \frac{16 x}{ (1-y)^3}
          - \frac{8 x }{3(1-y)^2}
           +  \frac{8 x }{3(1-y)} \nn\\
& &
          - \frac{8 x^2}{3 (1-y)^5}
           +  \frac{20 x^2}{3 (1-y)^4}
          - \frac{4 x^2}{ (1-y)^3}
          - \frac{2 x^2}{3 (1-y)^2}
           +  \frac{2 x^2 }{3(1-y)} \nn\\
& &
          - \frac{2}{3 y (1-x)}
           +  \frac{2}{3 y (x+1)}
          - \frac{2 y }{3(1-x)}
           +  \frac{2 y }{3(x+1)} \nn\\
& &
           +  \frac{32}{ (1-x) (1-y)^5}
          - \frac{80}{ (1-x) (1-y)^4}
           +  \frac{48}{ (1-x) (1-y)^3} \nn\\
& &
           + \! \frac{8}{ (1\!-\!x) (1\!-\!y)^2}\!
          - \!\frac{8}{ (1\!-\!x) (1\!-\!y)}\!
          - \!\frac{4}{3 (1\!-\!x)}\!
          - \!\frac{32}{3 (1\!-\!y)^5 (x\!+\!1)} \nn\\
& &
          - \!\frac{32}{3 (1-y)^5}\!
           + \! \frac{80}{3 (1-y)^4 (x\!+\!1)}\!
           + \! \frac{80}{3 (1-y)^4}\!
          - \!\frac{16}{ (1-y)^3 (x\!+\!1)} \nn\\
& &
          - \!\frac{16}{ (1-y)^3}\!
          - \!\frac{8}{3 (1-y)^2 (x+1)}\!
          - \!\frac{8}{3 (1-y)^2}\!
           + \! \frac{8}{3 (1-y) (x+1)} \nn\\
& &
           +  \frac{8}{3 (1-y)}
           +  \frac{4}{ (x+1)}
\Biggr) \zeta(2)  H(0;x)
+ \Biggl(
          - \frac{8}{3}
           +  \frac{32}{3 x^2 (1-y)^5} \nn\\
& &
          - \frac{80}{3 x^2 (1-y)^4}
           +  \frac{16}{ x^2 (1-y)^3}
           +  \frac{8}{3 x^2 (1-y)^2}
          - \frac{4}{ x^2 (1-y)} \nn\\
& &
           +  \frac{4}{3 x^2 (y+1)}
           +  \frac{128}{3 x (1-y)^5}
          - \frac{320}{3 x (1-y)^4}
           +  \frac{64}{ x (1-y)^3} \nn\\
& &
           +  \frac{32}{3 x (1-y)^2}
          - \frac{44}{3 x (1-y)}
          - \frac{16}{3 x (y+1)^3}
           +  \frac{8}{ x (y+1)^2} \nn\\
& &
           +  \frac{4}{3 x (y+1)}
          - \frac{128 x}{3 (1-y)^5}
           +  \frac{320 x}{3 (1-y)^4}
          - \frac{64 x }{(1-y)^3} \nn\\
& &
          - \frac{32 x }{3(1-y)^2}
           +  \frac{20 x }{3(1-y)}
          - \frac{16 x }{3 (y+1)^3}
           +  \frac{8 x }{(y+1)^2} \nn\\
& &
           +  \frac{4 x }{3(y+1)}
          - \frac{32 x^2}{3 (1-y)^5}
           +  \frac{80 x^2}{3 (1-y)^4}
          - \frac{16 x^2 }{(1-y)^3} \nn\\
& &
          - \frac{8 x^2 }{3(1-y)^2}
           +  \frac{4 x^2}{3 (1-y)}
           +  \frac{4 x^2}{3 (y+1)}
           +  \frac{128}{ (1-x) (1-y)^5} \nn\\
& &
          - \frac{320}{ (1-x) (1-y)^4}
           +  \frac{192}{ (1-x) (1-y)^3}
           +  \frac{32}{ (1-x) (1-y)^2} \nn\\
& &
          - \frac{32}{ (1-x) (1-y)}
          - \frac{128}{3 (1-y)^5 (x+1)}
          - \frac{128}{3 (1-y)^5} \nn\\
& &
           +  \frac{320}{3 (1-y)^4 (x+1)}
           +  \frac{320}{3 (1-y)^4}
          - \frac{64}{ (1-y)^3 (x+1)} \nn\\
& &
          - \!\frac{64}{ (1-y)^3}\!
          - \!\frac{32}{3 (1-y)^2 (x+1)}\!
          - \!\frac{32}{3 (1-y)^2}\!
           + \! \frac{32}{3 (1-y) (x+1)} \nn\\
& &
           +  \frac{40}{3 (1-y)}
           +  \frac{32}{3 (y+1)^3}
          - \frac{16}{ (y+1)^2}
           +  \frac{8}{ (y+1)}
\Biggr) \zeta(2) H(0;y)  \nn\\
& &
- \Biggl(
              \frac{4}{15}
           +  \frac{2}{15 x^2}
           +  \frac{8}{15 x}
           +  \frac{8 x}{15}
           +  \frac{2 x^2}{15}
\Biggr)  H(1;y)
+ \Biggl(
          -  \frac{20}{9}\nn\\
& &
          - \frac{56}{3 x^2 (1-y)^5} 
           +  \frac{140}{3 x^2 (1-y)^4}
          - \frac{32}{ x^2 (1-y)^3}
           +  \frac{4}{3 x^2 (1-y)^2} \nn\\
& &
           +  \frac{8}{3 x^2 (1-y)}
          - \frac{80}{ x (1-y)^5}
           +  \frac{200}{ x (1-y)^4}
          - \frac{400}{3 x (1-y)^3} \nn\\
& &
           +  \frac{12}{ x (1-y)}
           +  \frac{2}{3 x}
           +  \frac{80 x }{(1-y)^5}
          - \frac{200 x}{(1-y)^4}
           +  \frac{400 x}{3 (1-y)^3} \nn\\
& &
          - \frac{12 x}{(1-y)}
          - \frac{2 x}{3}
           +  \frac{56 x^2 }{3(1-y)^5}
          - \frac{140 x^2}{3 (1-y)^4}
           +  \frac{32 x^2}{ (1-y)^3} \nn\\
& &
          - \frac{4 x^2}{3 (1-y)^2}
          - \frac{8 x^2}{3 (1-y)}
           +  \frac{10}{9 y (1-x)}
          - \frac{22}{9 y (x+1)}
           +  \frac{2}{3 y} \nn\\
& &
          - \frac{10 y }{9(1-x)}
           +  \frac{22 y }{9(x+1)}
          - \frac{2 y}{3}
          - \frac{704}{3 (1-x) (1-y)^5} \nn\\
& &
           +  \frac{1760}{3 (1-x) (1-y)^4}
          - \frac{3568}{9 (1-x) (1-y)^3}
           +  \frac{8 }{(1-x) (1-y)^2} \nn\\
& &
           +  \frac{416}{9 (1-x) (1-y)}
          - \frac{44}{9 (1-x)}
           +  \frac{256}{3 (1-y)^5 (x+1)} \nn\\
& &
           + \! \frac{224}{3 (1-y)^5}\!
          - \!\frac{640}{3 (1-y)^4 (x\!+\!1)}\!
          - \!\frac{560}{3 (1-y)^4}\!
           + \! \frac{1424}{9 (1-y)^3 (x\!+\!1)} \nn\\
& &
           +  \frac{1072}{9 (1-y)^3}
          - \frac{24}{ (1-y)^2 (x+1)}
           +  \frac{8}{ (1-y)^2}
          - \frac{224}{9 (1-y) (x+1)} \nn\\
& &
          - \frac{32}{3 (1-y)} \!
           + \! \frac{28}{3 (x+1)}
\Biggr)H(0;x) H(0;y) \!
+ \! \Biggl(
            \frac{44}{9}
          - \frac{16}{3 x (1-y)^4} \nn\\
& &
           + \! \frac{32}{3 x (1-y)^3}\!
          - \!\frac{4}{ x (1-y)^2}\!
          - \!\frac{4}{3 x (1-y)}\!
           + \! \frac{16 x}{ (1-y)^4}\!
          -\! \frac{32 x}{ (1-y)^3} \nn\\
& &
           + \! \frac{76 x }{9(1-y)^2}\!
           + \! \frac{68 x }{9(1-y)}\!
           + \! \frac{16 x^2}{3 (1-y)^4}\!
          - \!\frac{32 x^2 }{3(1-y)^3}\!
           + \! \frac{28 x^2 }{9(1-y)^2} \nn\\
& &
           +  \frac{20 x^2 }{9(1-y)}
          - \frac{64}{9 y (x+1)^4}
           +  \frac{128}{9 y (x+1)^3}
          - \frac{56}{9 y (x+1)^2} \nn\\
& &
          - \! \frac{8}{9 y (x+1)}\!
          - \!\frac{64 y}{9 (x+1)^4}\!
           +\!  \frac{128 y }{9(x+1)^3}\!
          - \!\frac{56 y}{9 (x+1)^2} \!
          - \!\frac{8 y }{9(x\!+\!1)}\nn\\
& &
          - \frac{32}{ (1-x) (1-y)^4}
           +  \frac{64}{ (1-x) (1-y)^3}
          - \frac{56}{3 (1-x) (1-y)^2}  \nn\\
& &
          - \frac{40}{3 (1-x) (1-y)}
          - \frac{64}{3 (1-y)^4 (x+1)}
           +  \frac{32}{ (1-y)^4 (x+1)} \nn\\
& &
           +  \frac{16}{3 (1-y)^4}
           +  \frac{128}{3 (1-y)^3 (x+1)^2}
          - \frac{64}{ (1-y)^3 (x+1)} \nn\\
& &
          - \frac{32}{3 (1-y)^3}
          - \frac{64}{9 (1-y)^2 (x+1)^4}
           +  \frac{128}{9 (1-y)^2 (x+1)^3} \nn\\
& &
          - \frac{208}{9 (1-y)^2 (x+1)^2}
           +  \frac{200}{9 (1-y)^2 (x+1)}
           +  \frac{28}{9 (1-y)^2} \nn\\
& &
           +  \frac{64}{9 (1-y) (x+1)^4}
          - \frac{128}{9 (1-y) (x+1)^3}
           +  \frac{16}{9 (1-y) (x+1)^2} \nn\\
& &
           + \! \frac{88}{9 (1-y) (x\!+\!1)}\!
           + \! \frac{20}{9 (1-y)}\!
           + \! \frac{ 224}{9 (x\!+\!1)^2}\!
          - \!\frac{224}{9 (x\!+\!1)}\!
\Biggr) H(0,\!0;x)  \nn\\
& &
+ \Biggl(
            \frac{68}{9}
           +  \frac{16}{9 x^2 (1-y)^4}
          - \frac{32}{9 x^2 (1-y)^3}
          - \frac{7}{9 x^2 (1-y)^2} \nn\\
& &
           +  \frac{23}{9 x^2 (1-y)}
           +  \frac{16}{3 x (1-y)^4}
          - \frac{32}{3 x (1-y)^3}
          - \frac{16}{9 x (1-y)^2} \nn\\
& &
           + \! \frac{64}{9 x (1-y)}\!
          - \!\frac{16}{3 x (y+1)^2}\!
           + \! \frac{16}{3 x (y+1)}\!
           + \! \frac{16 x}{3 (1-y)^4}\!
          - \!\frac{32 x}{3 (1-y)^3} \nn\\
& &
          - \!\frac{16 x}{9 (1-y)^2}\!
           + \! \frac{64 x}{9 (1-y)}\!
          - \!\frac{16 x }{3(y+1)^2}\!
           + \! \frac{16 x}{3 (y+1)}\!
           + \! \frac{16 x^2}{9 (1-y)^4} \nn\\
& &
          - \!\frac{32 x^2}{9 (1-y)^3}\!
          - \!\frac{7 x^2}{9 (1-y)^2}\!
           + \! \frac{23 x^2}{9 (1-y)}\!
          - \!\frac{32}{9 (1-y)^4}\!
           + \! \frac{64}{9 (1-y)^3} \nn\\
& &
           +  \frac{22}{3 (1-y)^2}
          - \frac{98}{9 (1-y)}
           +  \frac{32}{3 (y+1)^2}
          - \frac{32}{3 (y+1)}
\Biggr) H(0,0;y) \nn\\
& &
+ \Biggl(     \frac{16}{3 x^2 (1-y)^4}
          - \frac{32}{3 x^2 (1-y)^3}
           +  \frac{28}{9 x^2 (1-y)^2}
           +  \frac{20}{9 x^2 (1-y)} \nn\\
& &
           +  \frac{64}{3 x (1-y)^4}
          - \frac{128}{3 x (1-y)^3}
           +  \frac{112}{9 x (1-y)^2}
           +  \frac{80}{9 x (1-y)} \nn\\
& &
          - \frac{64 x }{3(1-y)^4}
           +  \frac{128 x }{3(1-y)^3}
          - \frac{112 x }{9(1-y)^2}
          - \frac{80 x }{9(1-y)} \nn\\
& &
          - \frac{16 x^2}{3 (1-y)^4}
           +  \frac{32 x^2}{3 (1-y)^3}
          - \frac{28 x^2}{9 (1-y)^2}
          - \frac{20 x^2}{9 (1-y)} \nn\\
& &
           +  \frac{64 }{(1-x) (1-y)^4}
          - \frac{128}{ (1-x) (1-y)^3}
           +  \frac{112}{3 (1-x) (1-y)^2} \nn\\
& &
           +  \frac{80}{3 (1-x) (1-y)}
          - \frac{64}{3 (1-y)^4 (x+1)}
          - \frac{64}{3 (1-y)^4} \nn\\
& &
           +  \frac{128}{3 (1-y)^3 (x+1)}
           +  \frac{128}{3 (1-y)^3}
          - \frac{112}{9 (1-y)^2 (x+1)} \nn\\
& &
          - \frac{112}{9 (1-y)^2}
          - \frac{80}{9 (1-y) (x+1)}
          - \frac{80}{9 (1-y)}
\Biggr) H(-1,0;x)  \nn\\
& &
- \Biggl(
            \frac{4}{3}
           +  \frac{2}{3 y (1-x)}
          - \frac{2}{3 y (x+1)}
           +  \frac{2 y }{3(1-x)}
          - \frac{2 y }{3(x+1)} \nn\\
& &
           +  \frac{4}{3 (1-x)}
          - \frac{4}{ (x+1)}
\Biggr) H(0,0,0;x) 
+ \Biggl(
          - \frac{8}{3}
           +  \frac{8}{3 x^2 (1-y)^5} \nn\\
& &
          - \frac{20}{3 x^2 (1-y)^4}
           +  \frac{4}{ x^2 (1-y)^3}
           +  \frac{2}{3 x^2 (1-y)^2}
          - \frac{2}{ x^2 (1-y)} \nn\\
& &
           +  \frac{4}{3 x^2 (y+1)}
           +  \frac{32}{3 x (1-y)^5}
          - \frac{80}{3 x (1-y)^4}
           +  \frac{16}{ x (1-y)^3} \nn\\
& &
           +  \frac{8}{3 x (1-y)^2}
          - \frac{20}{3 x (1-y)}
          - \frac{16}{3 x (y+1)^3}
           +  \frac{8}{ x (y+1)^2} \nn\\
& &
           +  \frac{4}{3 x (y+1)}
          - \frac{32 x }{3(1-y)^5}
           +  \frac{80 x }{3(1-y)^4}
          - \frac{16 x}{ (1-y)^3} \nn\\
& &
          - \frac{8 x }{3(1-y)^2}
          - \frac{4 x }{3(1-y)}
          - \frac{16 x }{3(y+1)^3}
           +  \frac{8 x}{ (y+1)^2}
           +  \frac{4x }{3 (y+1)} \nn\\
& &
          - \frac{8 x^2}{3 (1-y)^5}
           +  \frac{20 x^2}{3 (1-y)^4}
          - \frac{4 x^2 }{(1-y)^3}
          - \frac{2 x^2}{3 (1-y)^2}
          - \frac{2 x^2 }{3(1-y)} \nn\\
& &
           +  \frac{4 x^2}{3 (y+1)}
           +  \frac{32}{ (1-x) (1-y)^5}
          - \frac{80}{ (1-x) (1-y)^4} \nn\\
& &
           +  \frac{48}{ (1-x) (1-y)^3}
           +  \frac{8}{ (1-x) (1-y)^2}
          - \frac{8}{ (1-x) (1-y)} \nn\\
& &
          - \frac{32}{3 (1-y)^5 (x+1)}
          - \frac{32}{3 (1-y)^5}
           +  \frac{80}{3 (1-y)^4 (x+1)} \nn\\
& &
           + \! \frac{80}{3 (1-y)^4}\!
          - \!\frac{16}{ (1-y)^3 (x+1)}\!
          - \!\frac{16}{ (1-y)^3}\!
          - \!\frac{8}{3 (1-y)^2 (x+1)} \nn\\
& &
          - \frac{8}{3 (1-y)^2}
           +  \frac{8}{3 (1-y) (x+1)}
           +  \frac{16}{3 (1-y)}
           +  \frac{32}{3 (y+1)^3} \nn\\
& &
          - \frac{16}{ (y+1)^2}
           +  \frac{8}{ (y+1)}
\Biggr) H(0,0,0;y) 
+ \Biggl(     
            \frac{4}{3}
          - \frac{32}{3 x^2 (1-y)^5} \nn\\
& &
           +  \frac{80}{3 x^2 (1-y)^4}
          - \frac{16}{ x^2 (1-y)^3}
          - \frac{8}{3 x^2 (1-y)^2}
           +  \frac{8}{3 x^2 (1-y)} \nn\\
& &
          - \frac{128}{3 x (1-y)^5}
           +  \frac{320}{3 x (1-y)^4}
          - \frac{64}{ x (1-y)^3}
          - \frac{32}{3 x (1-y)^2} \nn\\
& &
           + \! \frac{32}{3 x (1-y)}\!
           + \! \frac{128 x}{3 (1-y)^5}\!
          - \!\frac{320 x }{3(1-y)^4}\!
           + \! \frac{64 x}{ (1-y)^3}\!
           + \! \frac{32 x }{3(1-y)^2} \nn\\
& &
          - \frac{32 x }{3(1-y)}
           +  \frac{32 x^2}{3 (1-y)^5}
          - \frac{80 x^2 }{3(1-y)^4}
           +  \frac{16 x^2 }{(1-y)^3}
           +  \frac{8 x^2}{3 (1-y)^2} \nn\\
& &
          - \frac{8 x^2 }{3(1-y)}
           +  \frac{2}{3 y (1-x)}
          - \frac{2}{3 y (x+1)}
           +  \frac{2 y }{3(1-x)}
          - \frac{2 y }{3(x+1)} \nn\\
& &
          - \frac{128}{ (1-x) (1-y)^5}
           +  \frac{320}{ (1-x) (1-y)^4}
          - \frac{192}{ (1-x) (1-y)^3} \nn\\
& &
          - \frac{32}{ (1-x) (1-y)^2}
           +  \frac{32}{ (1-x) (1-y)}
           +  \frac{4}{3 (1-x)} \nn\\
& &
           +  \frac{128}{3 (1-y)^5 (x+1)}
           +  \frac{128}{3 (1-y)^5}
          - \frac{320}{3 (1-y)^4 (x+1)} \nn\\
& &
          - \! \frac{320}{3 (1-y)^4}\!
           + \! \frac{64}{ (1-y)^3 (x\!+\!1)}\!
           + \! \frac{64}{ (1-y)^3}\!
           + \! \frac{32}{3 (1-y)^2 (x\!+\!1)} \nn\\
& &
           +  \frac{32}{3 (1-y)^2}
          - \frac{32}{3 (1-y) (x+1)}
          - \frac{32}{3 (1-y)}
          - \frac{4 }{(x+1)}
\Biggr) \times \nn\\
& & \times H(0,0;y) H(0;x) 
+ \Biggl(
          - \frac{8}{3 x^2 (1-y)^5}
           +  \frac{20}{3 x^2 (1-y)^4} \nn\\
& &
          - \frac{4}{ x^2 (1-y)^3}
          - \frac{2}{3 x^2 (1-y)^2}
           +  \frac{2}{3 x^2 (1-y)}
          - \frac{32}{3 x (1-y)^5} \nn\\
& &
           +  \frac{80}{3 x (1-y)^4}
          - \frac{16}{ x (1-y)^3}
          - \frac{8}{3 x (1-y)^2}
           +  \frac{8}{3 x (1-y)} \nn\\
& &
           +  \frac{32 x }{3(1-y)^5}
          - \frac{80 x }{3(1-y)^4}
           +  \frac{16 x }{(1-y)^3}
           +  \frac{8 x }{3(1-y)^2} \nn\\
& &
          - \frac{8 x }{3(1-y)}
           +  \frac{8 x^2}{3 (1-y)^5}
          - \frac{20 x^2 }{3(1-y)^4}
           +  \frac{4 x^2}{ (1-y)^3}
           +  \frac{2 x^2}{3 (1-y)^2} \nn\\
& &
          - \frac{2 x^2}{3 (1-y)}
          - \frac{32}{ (1-x) (1-y)^5}
           +  \frac{80}{ (1-x) (1-y)^4} \nn\\
& &
          - \frac{48}{ (1-x) (1-y)^3}
          - \frac{8}{ (1-x) (1-y)^2}
           +  \frac{8}{ (1-x) (1-y)} \nn\\
& &
           +  \frac{32}{3 (1-y)^5 (x+1)}
           +  \frac{32}{3 (1-y)^5}
          - \frac{80}{3 (1-y)^4 (x+1)} \nn\\
& &
          - \! \frac{80}{3 (1-y)^4}\!
           + \! \frac{16}{ (1-y)^3 (x\!+\!1)}\!
           + \! \frac{16}{ (1-y)^3}\!
           + \! \frac{8}{3 (1-y)^2 (x\!+\!1)} \nn\\
& &
           +  \frac{8}{3 (1-y)^2}
          - \frac{8}{3 (1-y) (x+1)}
          - \frac{8}{3 (1-y)}
\Biggr)    
\Biggl[             
\Bigl( G(-y,0,0;x)   \nn\\
& & - G(-1/y,0,0;x) \Bigr) 
+ \Bigl(G(-y,0;x) + G(-1/y,0,x)\Bigr) H(0;y)   \nn\\
& &  
+ \Bigl(G(-1/y;x)  - G(-y;x)\Bigr)  
\Bigl( H(0,0;y) + 3 \zeta(2)
\Bigr) \nn\\
& &
    + 2 H(0;x) H(1,0;y)    
    - 2 H(-1,0;x) H(0;y)      \nn\\
& &   
    - 6 H(-1,0;y) H(0;x)
\Biggr] \Biggr\}\, . 
\label{2lB3} 
\eea

\section{Asymptotic Expansions \label{EXPANSIONS}}

In this Appendix, we provide the asymptotic expansions of the auxiliary
functions introduced in the paper. For the functions related to the
self-energy and vertex correction, that depend on a single kinematic 
invariant, we give the expansion in the case in which the kinematic 
invariant is significantly larger than the electron mass $m$. For the 
functions $B_i^{(1l)}$ and $B_i^{(2l)}$ ($i=1,2,3$), related to the box graphs, 
that depend on two kinematic invariants, we discuss the expansion in two 
different cases:
\begin{itemize}
\item[i)] both arguments of the functions $B_i^{(1l)}$ and $B_i^{(2l)}$ are  
larger than $m^2$; no specific hierarchy between the two arguments is assumed;
\item[ii)] both arguments are much larger than the electron mass, but
the first (second) argument is also much larger than the second (first).
\end{itemize}

Employing these results, in combination with the expression for the cross
section in terms of the auxiliary functions, it is possible to obtain the 
Bhabha scattering cross section in the kinematic regimes in which
$s \sim t \sim  u\gg m^2$, $s \sim t \gg u \gg m^2$, and $s \sim u \gg t \gg m^2$.

The expansions of the auxiliary functions are truncated in such a way that
the full cross section can be recovered up to terms of order $m^2/R^2$ 
(where $R^2 = -s,-t,-u$) with $R^2 \gg m^2$. In particular, for the functions
$\Pi_0^{(1l)}$, $\Pi_0^{(2l)}$, $F_1^{(1l)}$, $F_1^{(2l)}$, $F_2^{(1l)}$ and
$F_2^{(2l)}$ one needs only the term of order $(m^2/R^2)^0$. On the contrary, 
for the functions $B_i^{(1l)}$ and $B_i^{(2l)}$ ($i=1,2,3$)
one needs only the terms proportional to $(m^2/R^2)^{-1}$, since these functions
appear in the cross section always multiplied by $m^2/s$ or $m^2/t$.

In the following equations we use the definitions
\be
L_{P^2} = \ln{\left( \frac{m^2}{P^2} \right)} \, , \quad 
L_{Q^2} = \ln{\left( \frac{m^2}{Q^2} \right)} \, ,
\ee

In the expansions we will keep $P^2$ and $Q^2$ both in the Euclidean
region; the continuation to timelike values is trivial; if, for instance,
$P^2 \to -(s+i\epsilon)$, then
\be 
L_{P^2} \to \ln\left(\frac{m^2}{s}\right) + i \pi \, .
\ee

\boldmath
\subsection{One- and Two-Loop Auxiliary Functions in the Small Mass Limit
$P^2, Q^2 \gg m^2$}
\unboldmath

The first region of interest is the one in which 
$P^2, Q^2 \gg m^2$, ($x, y \ll 1$), relevant for
the large angle Bhabha scattering.
We find:
\bea
\Pi_0^{(1l,0)}(-P^2) &=& -\Biggl( \frac{5}{9} + \frac{1}{3} L_{P^2} \Biggr) \, , 
\label{EXP1} \\
\Pi_0^{(1l,1)}(-P^2) &=& -\Biggl(\frac{14}{27}  - \frac{1}{6} \zeta(2) 
+ \frac{5}{18} L_{P^2} + \frac{1}{12} L^2_{P^2} \Biggr)\, ,  
\label{EXP2} 
\eea
\bea
F_1^{(1l,-1)}(-P^2) &=& 1+ L_{P^2} \, ,  
\label{EXP3} \\
F_1^{(1l,0)}(-P^2) &=& -1 + \frac{1}{2} \zeta(2) - \frac{3}{4} L_{P^2}  -
\frac{1}{4} L^2_{P^2} \, ,  
\label{EXP4} \\
F_2^{(1l,0)}(-P^2) &=& 0 \, , 
\label{EXP5} 
\eea
\bea
B_1^{(1l,-1)}(-P^2,-Q^2) &=& - 16 \, \frac{P^2}{m^2} L_{P^2}
                             - 8 \, \frac{Q^2}{m^2} L_{P^2}
                            - 8 \, \frac{P^4}{m^2 Q^2} L_{P^2} 
\, ,  
\label{EXP6} \\
B_1^{(1l,0)}(-P^2,-Q^2) &=&
        - \frac{P^2}{m^2}  \Bigl[
          20 \zeta(2)
          + 2 L_{P^2}
          - 2 L^2_{P^2}
          - 4 L_{P^2} L_{Q^2}
          + 2 L_{Q^2}
          + 2 L^2_{Q^2}
          \Bigr]  \nn \\
& &	  
	  -  \frac{Q^2}{m^2} \Bigl[
          10 \zeta(2)
          + 2 L_{P^2}
          - L^2_{P^2}
          - 2 L_{P^2} L_{Q^2}
          + 2 L_{Q^2}
          + L^2_{Q^2}
          \Bigr] \nn \\
& &	  
       - \frac{P^4}{m^2 Q^2}  \Bigl[
            16 \zeta(2)
          + 4 L_{P^2}
          - 4 L_{P^2} L_{Q^2}
          + 2 L^2_{Q^2}
          \Bigr] \, , 
\label{EXP7} 
\eea
\bea
B_2^{(1l,-1)}(-P^2,-Q^2) &=& - 16 \, \frac{P^2}{m^2} L_{P^2}
                             - 8 \, \frac{Q^2}{m^2} L_{P^2}
                             - 16 \, \frac{P^4}{m^2 Q^2} L_{P^2} 
\, ,  
\label{EXP8} \\
B_2^{(1l,0)}(-P^2,-Q^2) &=&
        -\frac{P^2}{m^2} \Biggl[
          20 \zeta(2)
          - 2 L_{P^2}
          - 2 L^2_{P^2}
          - 4 L_{P^2} L_{Q^2}
          + 2 L_{Q^2}
          + 2 L^2_{Q^2}
          \Biggr]\nn \\
& &	  
	- \frac{Q^2}{m^2}  \Biggl[
          10 \zeta(2)
          + 2 L_{P^2}
          - L^2_{P^2}
          - 2 L_{P^2} L_{Q^2}
          + 2 L_{Q^2}
          + L^2_{Q^2}
          \Biggr]\nn \\
& &	  
	- \frac{P^4}{m^2 Q^2}  \Biggl[
           32 \zeta(2)
          - 8 L_{P^2} L_{Q^2}
          + 4 L^2_{Q^2}
          \Biggr] \, ,  
\label{EXP9} 
\eea
\bea
B_3^{(1l,-1)}(-P^2,-Q^2) &=& 8 \frac{P^4}{m^2 Q^2} L_{P^2} 
\, ,  
\label{EXP10} \\
B_3^{(1l,0)}(-P^2,-Q^2) &=&
           4 \frac{P^2}{m^2} L_{P^2} 
	  + 4 \frac{Q^2}{m^2} L_{P^2} 
          + \frac{P^4}{m^2 Q^2} \biggl[
           16 \zeta(2)
          + 4 L_{P^2}
          - 4 L_{P^2} L_{Q^2} \nn \\
& &	  
          + 2 L^2_{Q^2} 
          \Biggr] \, , 
\label{EXP11} 
\eea
and for the two-loop functions:
\be
\Pi_0^{(2l,0)}(-P^2) = -\frac{5}{24} + \zeta(3) - \frac{1}{4} L_{P^2} \, ,  
\label{EXP12} 
\ee

\bea
\hspace*{-5mm} F_1^{(2l,0)}(-P^2) &=& \frac{383}{108}
          - \frac{1}{4} \zeta(2)
          + \frac{265}{216} L_{P^2}
          + \frac{1}{6} \zeta(2) L_{P^2}
          + \frac{19}{72} L^2_{P^2} 
          + \frac{1}{36} L^3_{P^2} \, ,  
\label{EXP13} \\
\hspace*{-5mm} F_2^{(2l,0)}(-P^2) &=& 0 \, ,  
\label{EXP14} 
\eea
\bea
B_1^{(2l,-1)}(-P^2,-Q^2) \!\!&=& \!\!-\Biggl\{
        \frac{P^2}{m^2}  \Biggl[
           \frac{40}{9} L_{P^2}
          + \frac{8}{3} L_{P^2} L_{Q^2}
          \Biggr]
       + \frac{Q^2}{m^2}  \Biggl[
          \frac{20}{9} L_{P^2}
          + \frac{4}{3} L_{P^2} L_{Q^2}
          \Biggr] \nn \\
& &
       + \frac{P^4}{m^2 Q^2}  \Biggl[
           \frac{20}{9} L_{P^2}
          + \frac{4}{3} L_{P^2} L_{Q^2}
          \Biggr] \Biggr\}\, ,  
\label{EXP15} \\
%
%
B_1^{(2l,0)}(-P^2,-Q^2) \!\!&=& \!\!-\Biggl\{
\frac{P^2}{m^2}   \Biggl[
           \frac{68}{27} \! 
           + \!\frac{14}{9} \zeta(2) \!
           + \!4 \zeta(2) \, \ln \left( 1 + \frac{Q^2}{P^2}\right) \!
           - \! \frac{242}{27} L_{P^2} \!
           - \! \frac{13}{9} L^2_{P^2}  \nn \\
& & \hspace*{8mm}
          +  \!\frac{2}{3} L^2_{P^2} \ln\left(1  +  \frac{Q^2}{P^2} \right) \!
          - \! \frac{2}{9} L^2_{P^2} \!
          - \!\frac{ 4}{3} L_{P^2} \, L_{Q^2} \ln\left(1  +  \frac{Q^2}{P^2} \right) \nn \\
& & \hspace*{8mm}
 	  - \! \frac{34}{9} L_{P^2} L_{Q^2} \!
           - \! 2 L_{P^2} L^2_{Q^2} \!
           + \!  \frac{4}{3} L_{P^2} \, {\rm Li}_2\left( - \frac{Q^2}{P^2}\right) \!
            + \!  \frac{20}{9} L_{Q^2} \nn \\
& & \hspace*{8mm}
            +  8 \zeta(2) L_{Q^2}
            +  \frac{16}{9} L^2_{Q^2}
            +  \frac{2}{3} L^2_{Q^2} \, \ln\left(1 + \frac{Q^2}{P^2}\right)
           + \frac{2}{3} L^3_{Q^2} \nn \\
& & \hspace*{8mm}
            - \frac{4}{3} L_{Q^2} \, {\rm Li}_2\left( - \frac{Q^2}{P^2}\right)
           + \frac{4}{3} {\rm Li}_2\left( - \frac{Q^2}{P^2}\right)
           \Biggr] \nn \\
& &	  
+ \frac{Q^2}{m^2} \Biggl[ 
            \frac{34}{27} \!
          + \! \frac{22}{9} \zeta(2) \!
          + \! 2 \zeta(2) \ln\left(1 + \frac{Q^2}{P^2}\right) \!
          - \! \frac{130}{27} L_{P^2} \!
          - \! \frac{5}{9} L^2_{P^2} \nn \\
& & \hspace*{8mm}
          + \!\frac{1}{3} L^2_{P^2} \, \ln\left(1 + \frac{Q^2}{P^2}\right)\!
          - \!\frac{1}{9} L^3_{P^2}\!
          - \!\frac{2}{3} L_{P^2} L_{Q^2} \ln\left(1 + \frac{Q^2}{P^2}\right) \nn \\
& & \hspace*{8mm}
          - \frac{20}{9} L_{P^2} L_{Q^2} \!
          - \! L_{P^2} L^2_{Q^2} \!
          + \! \frac{2}{3} L_{P^2} \, {\rm Li}_2\left( - \frac{Q^2}{P^2}\right) \!
          + \! 4 \zeta(2) L_{Q^2} \nn \\
& & \hspace*{8mm}
          + 2 L_{Q^2}
          + \frac{25}{18} L^2_{Q^2}
          + \frac{1}{3} L^2_{Q^2} \, \ln\left(1 + \frac{Q^2}{P^2}\right)
          + \frac{1}{3} L^3_{Q^2}  \nn \\
& & \hspace*{8mm}
          - \frac{2}{3} L_{Q^2} \, {\rm Li}_2\left( - \frac{Q^2}{P^2}\right)
          + \frac{2}{3} \, {\rm Li}_3\left( - \frac{Q^2}{P^2}\right)
           \Biggr] \nn \\
& &	  
 + \frac{P^4}{m^2 Q^2}  \Biggl[
            \frac{34}{27}
          + \frac{40}{9} \zeta(2)
          + 2 \zeta(2) \ln\left(1 + \frac{Q^2}{P^2}\right)
          - \frac{82}{27} L_{P^2} \nn \\
& & \hspace*{14mm}
          + \frac{2}{3} \zeta(2) L_{P^2}
          + \frac{1}{3} L^2_{P^2}\ln\left(1 + \frac{Q^2}{P^2}\right)
          - \frac{8}{3} L_{P^2} L_{Q^2} \nn \\
& & \hspace*{14mm}
          - \frac{2}{3} L_{P^2} L_{Q^2} \, \ln\left(1 + \frac{Q^2}{P^2}\right)
          + \frac{2}{3} L_{P^2} \, {\rm Li}_2\left( - \frac{Q^2}{P^2}\right) \nn \\
& & \hspace*{14mm}
          - \frac{4}{3} L_{P^2} L^2_{Q^2}
          + \frac{2}{9} L_{Q^2}
          + \frac{16}{3} \zeta(2) L_{Q^2}
          + \frac{23}{18} L^2_{Q^2}  \nn \\
& & \hspace*{14mm}
          + \frac{1}{3} L^2_{Q^2} \ln\left(1 + \frac{Q^2}{P^2}\right)
          + \frac{5}{9} L^3_{Q^2}
          - \frac{2}{3} L_{Q^2} {\rm Li}_2\left( - \frac{Q^2}{P^2}\right) \nn \\
& & \hspace*{14mm}
          + \frac{2}{3} {\rm Li}_3\left( - \frac{Q^2}{P^2}\right)
          \Biggr] \Biggr\}\, , 
\label{EXP16} 
\eea
\bea
B_2^{(2l,-1)}(-P^2,-Q^2)\!\!\! &=& \!\!\!-\Biggl\{
       \frac{P^2}{m^2}   \Bigl[
          \frac{40}{9} L_{P^2}
          + \frac{8}{3} L_{P^2} L_{Q^2}
          \Bigr]
       + \frac{Q^2}{m^2}  \Bigl[ 
           \frac{20}{9} L_{P^2}
          + \frac{4}{3} L_{P^2} L_{Q^2}
          \Bigl]\nn \\
& &	  
	  + \frac{P^4}{m^2 Q^2}   \Bigl[
           \frac{40}{9} L_{P^2}
          + \frac{8}{3} L_{P^2} L_{Q^2}
          \Bigl]\Biggr\} \, ,  
\label{EXP17} \\
B_2^{(2l,0)}(-P^2,-Q^2)\!\!\!  &=& \!\!\!-\Biggl\{\frac{P^2}{m^2}   \Biggl[
           \frac{68}{27}
          + \frac{14}{9} \zeta(2)
          + 4 \zeta(2)  \ln\left(1 + \frac{Q^2}{P^2}\right)
          - \frac{272}{27} L_{P^2}
          - \frac{13}{9} L^2_{P^2} \nn \\
& & \hspace*{8mm}
          + \frac{2}{3} L^2_{P^2} \ln\left(1 + \frac{Q^2}{P^2}\right)
          - \frac{2}{9} L^3_{P^2}
          - \frac{4}{3} L_{P^2} L_{Q^2} \ln\left(1 + \frac{Q^2}{P^2}\right) \nn \\
& & \hspace*{8mm}
          - \frac{40}{9} L_{P^2} L_{Q^2}
          - 2 L_{P^2} L^2_{Q^2}
          + \frac{4}{3} L_{P^2} {\rm Li}_2\left( - \frac{Q^2}{P^2}\right)
          + \frac{20}{9} L_{Q^2} \nn \\
& & \hspace*{8mm}
          + 8 \zeta(2) L_{Q^2}
          + \frac{16}{9} L^2_{Q^2} 
          + \frac{2}{3} L^2_{Q^2}  \ln\left(1 + \frac{Q^2}{P^2}\right)
          + \frac{2}{3} L^2_{Q^2} \nn \\
& & \hspace*{8mm}
          - \frac{4}{3} L_{Q^2} {\rm Li}_2\left( - \frac{Q^2}{P^2}\right)
          + \frac{4}{3} {\rm Li}_3\left( - \frac{Q^2}{P^2}\right)
          \Biggr]\nn \\
& &	 
+ \frac{Q^2}{m^2}   \Biggl[   
           \frac{34}{27}\!
          + \!\frac{22}{9} \zeta(2) \! 
          + \!2 \zeta(2) \, \ln\left(1 + \frac{Q^2}{P^2}\right)
          - \frac{130}{27} L_{P^2}
          - \frac{5}{9} L^2_{P^2} \nn \\
& & \hspace*{8mm}
          +  \! \frac{1}{3} L^2_{P^2} \, \ln\left(1 + \frac{Q^2}{P^2}\right)
          - \frac{1}{9} L^3_{P^2} 
          - \frac{2}{3} L_{P^2} L_{Q^2} \ln\left(1 + \frac{Q^2}{P^2}\right) \nn \\
& & \hspace*{8mm}
          - \frac{20}{9} L_{P^2} L_{Q^2}
          - L_{P^2} L^2_{Q^2} 
          + \frac{2}{3} L_{P^2} {\rm Li}_2\left( - \frac{Q^2}{P^2}\right)
          + 2 L_{Q^2} \nn \\
& & \hspace*{8mm}
          + 4 \zeta(2) L_{Q^2}
          + \frac{25}{18} L^2_{Q^2} 
          + \frac{1}{3} L^2_{Q^2}  \ln\left(1 + \frac{Q^2}{P^2}\right)
          + \frac{1}{3} L^3_{Q^2}  \nn \\
& & \hspace*{8mm}
          - \frac{2}{3} L_{Q^2} {\rm Li}_2\left( - \frac{Q^2}{P^2}\right)
          + \frac{2}{3} {\rm Li}_3\left( - \frac{Q^2}{P^2}\right)
          \Biggr]\nn \\
& &
+ \frac{P^4}{m^2 Q^2}   \Biggl[
           \frac{68}{27} \!
          + \! \frac{80}{9} \zeta(2) \!
          + \! 4 \zeta(2) \ln\left(1 + \frac{Q^2}{P^2}\right) \!
          + \frac{4}{3} \zeta(2) L_{P^2}  \nn \\
& & \hspace*{14mm}
          - \frac{224}{27} L_{P^2} \!
          + \! \frac{2}{3} L^2_{P^2} \, \ln\left(1 + \frac{Q^2}{P^2}\right)
          - \frac{20}{3} L_{P^2} L_{Q^2} \nn \\
& & \hspace*{14mm}
          - \frac{4}{3} L_{P^2} L_{Q^2} \, \ln\left(1 + \frac{Q^2}{P^2}\right)
          - \frac{8}{3} L_{P^2} L^2_{Q^2} \!
          + \! \frac{4}{9} L_{Q^2} \nn \\
& & \hspace*{14mm}
          + \frac{4}{3} L_{P^2} {\rm Li}_2\left( - \frac{Q^2}{P^2}\right)
          + \frac{32}{3} \zeta(2) L_{Q^2}
          + \frac{23}{9} L^2_{Q^2}  \nn \\
& & \hspace*{14mm}
          + \frac{2}{3} L^2_{Q^2}  \ln\left(1 + \frac{Q^2}{P^2}\right)
          + \frac{10}{9} L^3_{Q^2} 
          - \frac{4}{3} L_{Q^2} {\rm Li}_2\left( - \frac{Q^2}{P^2}\right) \nn \\
& & \hspace*{14mm}
          + \frac{4}{3} {\rm Li}_3\left( - \frac{Q^2}{P^2}\right)
         \Biggr] \Biggr\} \, , 
\label{EXP18} 
\eea
\bea
B_3^{(2l,-1)}(-P^2,-Q^2) &=& \frac{P^4}{m^2 Q^2}   \Biggl[ 
                              \frac{20}{3} L_{P^2} 
			    + \frac{4}{3}L_{P^2}L_{Q^2}
\Biggr] \, ,  
\label{EXP19} \\
B_3^{(2l,0)}(-P^2,-Q^2) &=& \frac{P^2}{m^2} \Biggl[ 
          \frac{10}{9} L_{P^2}
          + \frac{2}{3} L_{P^2} L_{Q^2}
          \Biggr] 
	+ \frac{Q^2}{m^2}  \Biggl[ 
          \frac{10}{9} L_{P^2}
          + \frac{2}{3} L_{P^2} L_{Q^2}
          \Biggr]\nn \\
& &  
          - \frac{P^4}{m^2 Q^2}  \Biggl[ 
          - \frac{34}{27}
          - \frac{40}{9} \zeta(2)
          - 2 \zeta(2) \ln\left(1 + \frac{Q^2}{P^2}\right)
          + \frac{82}{27} L_{P^2} \nn \\
& & \hspace*{14mm}
          - \frac{2}{3} \zeta(2) L_{P^2}
          - \frac{1}{3} L^2_{P^2} \ln\left(1 + \frac{Q^2}{P^2}\right)
          + \frac{8}{3} L_{P^2} L_{Q^2}  \nn \\
& & \hspace*{14mm}
          + \frac{2}{3} L_{P^2} L_{Q^2} \ln\left(1 + \frac{Q^2}{P^2}\right)
          + \frac{4}{3} L_{P^2} L^2_{Q^2}
          - \frac{2}{9} L_{Q^2} \nn \\
& & \hspace*{14mm}
          - \frac{2}{3} L_{P^2} {\rm Li}_2\left( - \frac{Q^2}{P^2}\right)
          - \frac{16}{3} \zeta(2) L_{Q^2}
          - \frac{23}{18} L^2_{Q^2}  \nn \\
& & \hspace*{14mm}
          - \frac{1}{3} L^2_{Q^2}\ln\left(1 + \frac{Q^2}{P^2}\right)
          - \frac{5}{9} L^3_{Q^2} 
          + \frac{2}{3} L_{Q^2} {\rm Li}_2\left( - \frac{Q^2}{P^2}\right) \nn \\
& & \hspace*{14mm}
          - \frac{2}{3} {\rm Li}_3\left( - \frac{Q^2}{P^2}\right)
          \Biggr] \, . 
\label{EXP20} 
\eea

\boldmath
\subsection{One- and Two-Loop Auxiliary Functions in the Limit
$P^2 \gg Q^2 \gg m^2$ and $Q^2 \gg P^2 \gg m^2$}
\unboldmath

In this limit, the Eqs.~(\ref{EXP1}--\ref{EXP15},\ref{EXP17},\ref{EXP19})
remain valid, while Eqs.~(\ref{EXP16},\ref{EXP18},\ref{EXP20}) must be replaced,
in the case in which $P^2 \gg Q^2 \gg m^2$, by:
\bea
B_1^{(2l,0)}(-P^2,-Q^2) &=&  -\Biggl\{
          \frac{P^4}{m^2 Q^2}   \Biggl[
            \frac{34}{27}
          + \frac{40}{9} \zeta(2)
          - \frac{82}{27} L_{P^2}
          + \frac{2}{3} \zeta(2) L_{P^2}
          - \frac{8}{3} L_{P^2} L_{Q^2} \nn\\
& & \hspace*{14mm}
          - \frac{4}{3} L_{P^2} L^2_{Q^2} \! 
          + \! \frac{2}{9} L_{Q^2} \! 
          + \! \frac{16}{3} \zeta(2) L_{Q^2} \! 
          + \! \frac{23}{18} L^2_{Q^2} \! 
          + \! \frac{5}{9} L^3_{Q^2}
          \Biggr] \nn\\
& & 
       + \frac{P^2}{m^2}   \Biggl[
            \frac{50}{27} \! 
          +  \! \frac{32}{9} \zeta(2) \! 
          -  \! \frac{260}{27} L_{P^2} \! 
          -  \! \frac{10}{9} L^2_{P^2} \! 
          -  \! \frac{2}{9} L^3_{P^2} \! 
          -  \! \frac{40}{9} L_{P^2} L_{Q^2} \nn\\
& & \hspace*{8mm}
          - 2 L_{P^2} L^2_{Q^2}
          + \frac{26}{9} L_{Q^2}
          + 8 \zeta(2) L_{Q^2}
          + \frac{19}{9} L^2_{Q^2}
          + \frac{2}{3} L^3_{Q^2}
          \Biggr] \nn\\
& & 
       - \frac{Q^4}{m^2 P^2}   \Biggl[
            \frac{85}{162}
          - \frac{2}{3} \zeta(2)
          + \frac{11}{27} L_{P^2}
          - \frac{1}{9} L^2_{P^2}
          + \frac{2}{9} L_{P^2} L_{Q^2} \nn\\
& & \hspace*{14mm}
          - \frac{11}{27} L_{Q^2}
          - \frac{1}{9} L^2_{Q^2}
          \Biggr] \nn\\
& & 
       + \frac{Q^6}{m^2 P^4}   \Biggl[
            \frac{115}{2592}
          - \frac{1}{6} \zeta(2)
          + \frac{13}{216} L_{P^2}
          - \frac{1}{36} L^2_{P^2}
          + \frac{1}{18} L_{P^2} L_{Q^2} \nn\\
& & \hspace*{14mm}
          - \frac{13}{216} L_{Q^2}
          - \frac{1}{36} L^2_{Q^2}
          \Biggr] \nn\\
& & 
       + \frac{Q^2}{m^2}   \Biggl[
            \frac{1}{108}
          + \frac{49}{9} \zeta(2)
          - \frac{323}{54} L_{P^2}
          - \frac{1}{18} L^2_{P^2}
          - \frac{1}{9} L^3_{P^2} \nn\\
& & \hspace*{8mm}
          - \frac{29}{9} L_{P^2} L_{Q^2}
          - L_{P^2} L^2_{Q^2}
          + \frac{19}{6} L_{Q^2}
          + 4 \zeta(2) L_{Q^2}
          + \frac{17}{9} L^2_{Q^2} \nn\\
& & \hspace*{8mm}
          + \frac{1}{3} L^3_{Q^2}
          \Biggr]\Biggr\}
\, ,
\eea

\bea
B_2^{(2l,0)}(-P^2,-Q^2) &=&  -\Biggl\{
         \frac{P^4}{m^2 Q^2}   \Biggl[
            \frac{68}{27}
          + \frac{80}{9} \zeta(2)
          - \frac{224}{27} L_{P^2}
          + \frac{4}{3} \zeta(2) L_{P^2}
          - \frac{20}{3} L_{P^2} L_{Q^2} \nn\\
& & \hspace*{14mm}
          - \frac{8}{3} L_{P^2} L^2_{Q^2}\! 
          + \! \frac{4}{9} L_{Q^2}\! 
          + \! \frac{32}{3} \zeta(2) L_{Q^2}\! 
          + \! \frac{23}{9} L^2_{Q^2}\! 
          +\!  \frac{10}{9} L^3_{Q^2}
          \Biggr] \nn\\
& & 
       + \frac{P^2}{m^2}   \Biggl[
            \frac{32}{27}\!
          + \!\frac{50}{9} \zeta(2)\!
          - \!\frac{308}{27} L_{P^2}
          - \frac{7}{9} L^2_{P^2}
          - \frac{2}{9} L^3_{P^2}
          - \frac{52}{9} L_{P^2} L_{Q^2} \nn\\
& & \hspace*{8mm}
          - 2 L_{P^2} L^2_{Q^2}
          + \frac{32}{9} L_{Q^2}
          + 8 \zeta(2) L_{Q^2}
          + \frac{22}{9} L^2_{Q^2}
          + \frac{2}{3} L^3_{Q^2}
          \Biggr] \nn\\
& & 
       - \frac{Q^4}{m^2 P^2}   \Biggl[
            \frac{89}{162}
          - \frac{4}{3} \zeta(2)
          + \frac{13}{27} L_{P^2}
          - \frac{2}{9} L^2_{P^2}
          + \frac{4}{9} L_{P^2} L_{Q^2} \nn\\
& & \hspace*{14mm}
          - \frac{13}{27} L_{Q^2}
          - \frac{2}{9} L^2_{Q^2}
          \Biggr] \nn\\
& & 
       + \frac{Q^6}{m^2 P^4}   \Biggl[
            \frac{71}{1296}
          - \frac{2}{3} \zeta(2)
          + \frac{11}{108} L_{P^2}
          - \frac{1}{9} L^2_{P^2}
          + \frac{2}{9} L_{P^2} L_{Q^2} \nn\\
& & \hspace*{14mm}
          - \frac{11}{108} L_{Q^2}
          - \frac{1}{9} L^2_{Q^2}
          \Biggr] \nn\\
& & 
       + \frac{Q^2}{m^2}   \Biggl[
            \frac{5}{54}\!
          + \!\frac{40}{9} \zeta(2)\!
          - \!\frac{157}{27} L_{P^2}\!
          - \!\frac{2}{9} L^2_{P^2}\!
          - \!\frac{1}{9} L^3_{P^2}\!
          - \!\frac{26}{9} L_{P^2} L_{Q^2} \nn\\
& & \hspace*{8mm}
          - L_{P^2} L^2_{Q^2}\!
          + \!3 L_{Q^2}\!
          +\! 4 \zeta(2) L_{Q^2}\!
          + \!\frac{31}{18} L^2_{Q^2}\!
          + \!\frac{1}{3} L^3_{Q^2}
          \Biggr]\Biggr\}
\, ,
\eea

\bea
B_3^{(2l,0)}(-P^2,-Q^2) &=& -\Biggl\{
       - \frac{P^4}{m^2 Q^2}   \Biggl[
            \frac{34}{27}
          + \frac{40}{9} \zeta(2)
          - \frac{82}{27} L_{P^2}
          + \frac{2}{3} \zeta(2) L_{P^2}
          - \frac{8}{3} L_{P^2} L_{Q^2} \nn\\
& & \hspace*{14mm}
          - \frac{4}{3} L_{P^2} L^2_{Q^2}\!
          +\! \frac{2}{9} L_{Q^2}\!
          + \!\frac{16}{3} \zeta(2) L_{Q^2}\!
          + \!\frac{23}{18} L^2_{Q^2}\!
          + \!\frac{5}{9} L^3_{Q^2}
          \Biggr] \nn\\
& & 
       + \frac{P^2}{m^2}   \Biggl[
            \frac{2}{3}
          - 2 \zeta(2)
          - \frac{4}{9} L_{P^2}
          - \frac{1}{3} L^2_{P^2}
          - \frac{2}{3} L_{Q^2}
          - \frac{1}{3} L^2_{Q^2}
          \Biggr] \nn\\
& & 
       + \frac{Q^4}{m^2 P^2}   \Biggl[
            \frac{2}{81}
          - \frac{2}{3} \zeta(2)
          + \frac{2}{27} L_{P^2}
          - \frac{1}{9} L^2_{P^2}
          + \frac{2}{9} L_{P^2} L_{Q^2} \nn\\
& & \hspace*{14mm}
          - \frac{2}{27} L_{Q^2}
          - \frac{1}{9} L^2_{Q^2}
          \Biggr] \nn\\
& & 
       - \frac{Q^6}{m^2 P^4}   \Biggl[
            \frac{1}{96}
          - \frac{1}{2} \zeta(2)
          + \frac{1}{24} L_{P^2}
          - \frac{1}{12} L^2_{P^2}
          + \frac{1}{6} L_{P^2} L_{Q^2} \nn\\
& & \hspace*{14mm}
          - \frac{1}{24} L_{Q^2}
          - \frac{1}{12} L^2_{Q^2}
          \Biggr] \nn\\
& & 
       - \frac{Q^2}{m^2}   \Biggl[
            \frac{1}{12}
          - \zeta(2)
          + \frac{23}{18} L_{P^2}
          - \frac{1}{6} L^2_{P^2}
          + L_{P^2} L_{Q^2}
          - \frac{1}{6} L_{Q^2} \nn\\
& & \hspace*{8mm}
          - \frac{1}{6} L^2_{Q^2}
          \Biggr]\Biggr\}
\, .
\eea

In the case in which $Q^2 \gg P^2 \gg m^2$, Eqs.~(\ref{EXP16},\ref{EXP18},\ref{EXP20}) 
must be replaced by:
\bea
B_1^{(2l,0)}(-P^2,-Q^2) &=& -\Biggl\{
         \frac{P^2}{m^2}   \Biggl[
            \frac{50}{27} \!
          + \!\frac{32}{9} \zeta(2)\!
          - \frac{224}{27} L_{P^2}\!
          + \!\frac{4}{3} \zeta(2) L_{P^2}\!
          - \frac{10}{9} L^2_{P^2}\!
          - \frac{4}{9} L^3_{P^2} \nn\\
& & \hspace*{8mm}
          + \frac{2}{3} L^2_{P^2}L_{Q^2}
          - \frac{40}{9} L_{P^2} L_{Q^2}
          - \frac{8}{3} L_{P^2} L^2_{Q^2}
          + \frac{14}{9} L_{Q^2} \nn\\
& & \hspace*{8mm}
          + \frac{20}{3} \zeta(2) L_{Q^2}
          + \frac{19}{9} L^2_{Q^2}
          + \frac{8}{9} L^3_{Q^2}
          \Biggr] \nn\\
& & 
       + \frac{Q^2}{m^2}   \Biggl[
            \frac{34}{27} \!
          + \! \frac{22}{9} \zeta(2)
          - \frac{130}{27} L_{P^2}
          + \frac{2}{3} \zeta(2) L_{P^2}
          - \frac{5}{9} L^2_{P^2}
          - \frac{2}{9} L^3_{P^2} \nn\\
& & \hspace*{8mm}
          + \frac{1}{3} L^2_{P^2} L_{Q^2}
          - \frac{20}{9} L_{P^2} L_{Q^2}
          - \frac{4}{3} L_{P^2} L^2_{Q^2}
          + 2 L_{Q^2} \nn\\
& & \hspace*{8mm}
          + \frac{10}{3} \zeta(2) L_{Q^2}
          + \frac{25}{18} L^2_{Q^2}
          + \frac{4}{9} L^3_{Q^2}
          \Biggr] \nn\\
& & 
       - \frac{P^6}{m^2 Q^2}   \Biggl[
            \frac{85}{162}
          - \frac{2}{3} \zeta(2)
          - \frac{11}{27} L_{P^2}
          - \frac{1}{9} L^2_{P^2}
          + \frac{2}{9} L_{P^2} L_{Q^2} \nn\\
& & \hspace*{14mm}
          + \frac{11}{27} L_{Q^2}
          - \frac{1}{9} L^2_{Q^2}
          \Biggr] \nn\\
& & 
       + \frac{P^4}{m^2 Q^2}   \Biggl[
            \frac{1}{108}
          + \frac{67}{9} \zeta(2)
          - \frac{101}{54} L_{P^2}
          + \frac{4}{3} \zeta(2) L_{P^2}
          + \frac{1}{2} L^2_{P^2} \nn\\
& & \hspace*{14mm}
          - \frac{1}{9} L^3_{P^2}
          + \frac{1}{3} L^2_{P^2} L_{Q^2}
          - \frac{11}{3} L_{P^2} L_{Q^2}
          - \frac{5}{3} L_{P^2} L^2_{Q^2} \nn\\
& & \hspace*{14mm}
          - \frac{17}{18} L_{Q^2}
          + \frac{14}{3} \zeta(2) L_{Q^2}
          + \frac{16}{9} L^2_{Q^2}
          + \frac{2}{3} L^3_{Q^2}
          \Biggr]\Biggr\}
\, , 
\eea

\bea
B_2^{(2l,0)}(-P^2,-Q^2) &=& -\Biggl\{
         \frac{P^2}{m^2}   \Biggl[
            \frac{50}{27} \!
          + \! \frac{32}{9} \zeta(2)\!
          - \!\frac{254}{27} L_{P^2}\!
          + \!\frac{4}{3} \zeta(2) L_{P^2}\!
          - \!\frac{10}{9} L^2_{P^2}\!
          - \!\frac{4}{9} L^3_{P^2} \nn\\
& & \hspace*{8mm}
          + \frac{2}{3} L^2_{P^2} L_{Q^2}
          - \frac{46}{9} L_{P^2} L_{Q^2}
          - \frac{8}{3} L_{P^2} L^2_{Q^2}
          + \frac{14}{9} L_{Q^2} \nn\\
& & \hspace*{8mm}
          + \frac{20}{3} \zeta(2) L_{Q^2}
          + \frac{19}{9} L^2_{Q^2}
          + \frac{8}{9} L^3_{Q^2}
          \Biggr] \nn\\
& & 
       + \frac{Q^2}{m^2}   \Biggl[
            \frac{34}{27}
          + \frac{22}{9} \zeta(2)
          - \frac{130}{27} L_{P^2}
          + \frac{2}{3} \zeta(2) L_{P^2}
          - \frac{5}{9} L^2_{P^2} \nn\\
& & \hspace*{8mm}
          - \frac{2}{9} L^3_{P^2}
          + \frac{1}{3} L^2_{P^2} L_{Q^2}
          - \frac{20}{9} L_{P^2} L_{Q^2}
          - \frac{4}{3} L_{P^2} L^2_{Q^2}
          + 2 L_{Q^2} \nn\\
& & \hspace*{8mm}
          + \frac{10}{3} \zeta(2) L_{Q^2}
          + \frac{25}{18} L^2_{Q^2}
          + \frac{4}{9} L^3_{Q^2}
          \Biggr] \nn\\
& & 
       - \frac{P^6}{m^2 Q^2}   \Biggl[
            \frac{193}{162}
          - \frac{8}{3} \zeta(2)
          - \frac{29}{27} L_{P^2}
          - \frac{4}{9} L^2_{P^2}
          + \frac{8}{9} L_{P^2} L_{Q^2} \nn\\
& & \hspace*{14mm}
          + \frac{29}{27} L_{Q^2}
          - \frac{4}{9} L^2_{Q^2}
          \Biggr] \nn\\
& & 
       + \frac{P^4}{m^2 Q^2}   \Biggl[
            \frac{137}{108}
          + \frac{107}{9} \zeta(2)
          - \frac{385}{54} L_{P^2}
          + \frac{8}{3} \zeta(2) L_{P^2}
          + \frac{1}{2} L^2_{P^2} \nn\\
& & \hspace*{14mm}
          - \frac{2}{9} L^3_{P^2}
          + \frac{2}{3} L^2_{P^2} L_{Q^2}
          - \frac{23}{3} L_{P^2} L_{Q^2}
          - \frac{10}{3} L_{P^2} L^2_{Q^2} \nn\\
& & \hspace*{14mm}
          - \frac{13}{18} L_{Q^2}
          + \frac{28}{3} \zeta(2) L_{Q^2}
          + \frac{55}{18} L^2_{Q^2}
          + \frac{4}{3} L^3_{Q^2}
          \Biggr]\Biggr\}
\, , 
\eea

\bea
B_3^{(2l,0)}(-P^2,-Q^2) &=& -\Biggl\{
       - \frac{P^2}{m^2}   \Biggl[
            \frac{10}{9} L_{P^2}
          + \frac{2}{3} L_{P^2} L_{Q^2}
          \Biggr]
       - \frac{Q^2}{m^2}   \Biggl[
            \frac{10}{9} L_{P^2}
          + \frac{2}{3} L_{P^2} L_{Q^2}
          \Biggr] \nn\\
& & 
       + \frac{P^6}{m^2 Q^2}   \Biggl[
            \frac{2}{3}
          - 2 \zeta(2)
          - \frac{2}{3} L_{P^2}
          - \frac{1}{3} L^2_{P^2}
          + \frac{2}{3} L_{P^2} L_{Q^2}
          + \frac{2}{3} L_{Q^2} \nn\\
& & \hspace*{14mm}
          - \frac{1}{3} L^2_{Q^2}
          \Biggr] \nn\\
& & 
       - \frac{P^4}{m^2 Q^2}   \Biggl[
            \frac{34}{27}
          + \frac{40}{9} \zeta(2)
          - \frac{82}{27} L_{P^2}
          + \frac{4}{3} \zeta(2) L_{P^2}
          - \frac{1}{9} L^3_{P^2} \nn\\
& & \hspace*{14mm}
          + \frac{1}{3} L^2_{P^2} L_{Q^2}
          - \frac{8}{3} L_{P^2} L_{Q^2}
          - \frac{5}{3} L_{P^2} L^2_{Q^2}
          + \frac{2}{9} L_{Q^2} \nn\\
& & \hspace*{14mm}
          + \frac{14}{3} \zeta(2) L_{Q^2}
          + \frac{23}{18} L^2_{Q^2}
          + \frac{2}{3} L^3_{Q^2}
          \Biggr]\Biggr\}
\, , 
\eea



\begin{thebibliography}{99}
\def    \np     #1#2#3{{\it Nucl. Phys.} {\bf #1} (19#2) #3}
\def    \nptwoth     #1#2#3{{\it Nucl. Phys.} {\bf #1} (20#2) #3}
\def    \prep   #1#2#3{{\it Phys. Rep.} {\bf #1}  (19#2) #3}
\def    \pl     #1#2#3{{\it Phys. Lett.} {\bf #1} (19#2) #3}
\def    \pltwoth     #1#2#3{{\it Phys. Lett.} {\bf #1} (20#2) #3}
\def    \plold  #1#2#3{{\it Phys. Lett.} {\bf #1B} (19#2) #3}
\def    \prl    #1#2#3{{\it Phys. Rev. Lett.} {\bf #1}  (19#2) #3} 
\def    \pr     #1#2#3{{\it Phys. Rev.} {\bf #1}  (19#2) #3}
\def    \prd    #1#2#3{{\it Phys. Rev.} {\bf D#1}  (19#2) #3} 
\def    \prdtwoth    #1#2#3{{\it Phys. Rev.} {\bf D#1}  (20#2) #3} 
\def    \zeit   #1#2#3{{\it Z. Phys.} {\bf C#1}  (19#2) #3}
\def    \cmp    #1#2#3{{\it Comm. Math. Phys.} {\bf #1}  (19#2) #3}
\def    \ibid   #1#2#3{{\it ibid.} {\bf #1} (19#2) #3}
\def    \nc     #1#2#3{{\it Nuovo Cim.} {\bf #1} (19#2) #3}
\def    \acta   #1#2#3{{\it Acta Phys. Polon.} {\bf #1} (19#2) #3}
\def    \tmp    #1#2#3{{\it Theor. Math. Phys.} {\bf #1} (19#2) #3}
\def    \comp    #1#2#3{{\it Comput. Phys. Commun.} {\bf #1} (20#2) #3}
\def    \hepph  #1 {{\tt hep-ph/#1}}
\def    \hepth  #1 {{\tt hep-th/#1}}
\def    \hepex  #1 {{\tt hep-ex/#1}}
\parskip 0pt
\itemsep=0pt


\bibitem{mont}
  G.~Montagna, O.~Nicrosini and F.~Piccinini, 
  {\it Riv. Nuovo Cim.} {\bf 21N9} (1998) 1, (\hepph{9802302}).

\bibitem{Bhabha1loop}
  M. Consoli, \np{B160}{79}{208};\\
  M.~Bohm, A.~Denner and W.~Hollik, \np{B304}{88}{687};\\ 
  M.~Greco, {\it Riv.\ Nuovo Cim.\ } {\bf 11N5} (1988) 1.


\bibitem{TUTTI}
G.~Faldt and P.~Osland, \np{B413}{94}{64};
(\hepph{9304301}). \\
A.~B.~Arbuzov, E.~A.~Kuraev and B.~G.~Shaikhatdenov,
{\it Mod. Phys. Lett.} A {\bf 13} (1998) 2305
(\hepph{9806215}). \\
A.~B.~Arbuzov, E.~A.~Kuraev, N.~P.~Merenkov and L.~Trentadue,
\np{B474}{96}{271}. \\
G.~Faldt and P.~Osland, \np{B413}{94}{16},
Erratum-ibid. {\bf B419} (1994) 404;
(\hepph{9304212}). \\
A.~B.~Arbuzov, E.~A.~Kuraev, N.~P.~Merenkov and L.~Trentadue,
{\it Phys. Atom. Nucl.}  {\bf 60} (1997) 591
({\it Yad. Fiz.}  {\bf 60N4} (1997) 673). 


\bibitem{Bhabha2loop}
  Z. Bern, L. Dixon and A. Ghinculov, \prdtwoth{63}{01}{053007}, 
  (\hepph{0010075}).

\bibitem{Bas}
  E.~W.~Glover, J.~B.~Tausk and J.~J.~van der Bij, \pltwoth{B516}{01}{33}, 
  (\hepph{0106052}).

%
\bibitem{us}
   R.~Bonciani, A.~Ferroglia, P.~Mastrolia, E.~Remiddi and J.~J.~van der Bij,
   \nptwoth{B681}{04}{261}, (\hepph{0310333}).
\bibitem{laporta}
  S.~Laporta and E.~Remiddi, \pl{B379}{96}{283} ({\tt hep-ph/9602417}).\\
  S.~Laporta, {\it Int.\ J.\ Mod.\ Phys.}  {\bf A 15} (2000) 5087,
(\hepph{0102033}).


\bibitem{Tkac}
  F.V. Tkachov, \pl{B100}{81}{65}.\\
  K.G. Chetyrkin and F.V. Tkachov, \np{B192}{81}{159}.

\bibitem{Rem3}
  T. Gehrmann and E. Remiddi, \nptwoth{B580}{00}{485}, 
  (\hepph{9912329}). 

\bibitem{RoPieRem1}
  R. Bonciani, P. Mastrolia and E. Remiddi, \nptwoth{B661}{03}{289}
  (\hepph{0301170}).


\bibitem{DimReg}
  G. 't Hooft and M. Veltman, \np{B44}{72}{189}.\\
  C. G. Bollini and J. J. Giambiagi, \plold{40}{72}{566}; 
  \nc{12B}{72}{20}. \\
  J. Ashmore, {\it Lett. Nuovo Cimento} {\bf 4} (1972) 289.\\
  G. M. Cicuta and E. Montaldi, {\it Lett. Nuovo Cimento} {\bf 4} 
  (1972) 289. \\
  R. Gastmans and R. Meuldermans, \np{B63}{73}{277}.



\bibitem{Kot}
  A. V. Kotikov, \pl{B254}{91}{158}.

\bibitem{Rem1} 
  E. Remiddi, \nc{110A}{97}{1435}, (\hepth{9711188}).

\bibitem{Rem2}
  M. Caffo, H. Czy\.{z}, S. Laporta and E. Remiddi, 
  \acta{B29}{98}{2627}, (\hepph{9807119}).\\
  M. Caffo, H. Czy\.{z}, S. Laporta and E. Remiddi, 
  \nc{A111}{98}{365}, (\hepph{9805118}).



\bibitem{Polylog}
  E. Remiddi and J. A. M. Vermaseren, {\it Int. J. Mod. Phys.} 
  {\bf A15} (2000) 725, (\hepph{9905237}). 

\bibitem{Polylog3}
  T. Gehrmann and E. Remiddi, \comp{141}{01}{296}, 
  (\hepph{0107173}).

\bibitem{Polylog2}
  T. Gehrmann and E. Remiddi, \nptwoth{B601}{01}{248}, 
  (\hepph{0008287}), appendix A.

\bibitem{Polylog4}
  T.~Gehrmann and E.~Remiddi, \nptwoth{B640}{02}{379}, 
  (\hepph{0207020}).

\bibitem{Polylog5}
  T.~Gehrmann and E.~Remiddi,\comp{144}{02}{200},
  (\hepph{0111255}).


\bibitem{paolo}
  G.~K\"all\'en and A. Sabry, {\it Dan. Mat. Fys. Medd.} {\bf 29} 17 (1955). \\
  A.~Djouadi and P.~Gambino, \pr{D49}{94}{3499};
  Erratum-ibid. {\bf D53} (1996) 4111, (\hepph{9309298}).

\bibitem{BMR}
  R. Barbieri, J. A. Mignaco and E. Remiddi,  \nc{11A}{72}{824}; 
  \nc{11A}{72}{865}.\\
  P. Mastrolia and E. Remiddi, \nptwoth{B664}{03}{341}, (\hepph{0302162}). 

\bibitem{RoPieRem2}
  R.~Bonciani, P.~Mastrolia and E.~Remiddi, \nptwoth{B676}{04}{399},
  (\hepph{0307295}).



\bibitem{VN}
  P.~Van Nieuwenhuizen, \np{B28}{71}{429}. 

\bibitem{BOX1l}
I.~B.~Khriplovich, {\it Yad. Fiz.}  {\bf 17} (1973) 576. \\
F.~A.~Berends, R.~Kleiss and S.~Jadach, \np{B202}{82}{63}. 

\bibitem{Scalar}
G.~'t Hooft and M.~J.~G.~Veltman, \np{B153}{79}{365}. 



\bibitem{FORM}
  J.A.M. Vermaseren, {\it Symbolic Manipulation with}
  {\tt FORM}, Version 2, CAN, Amsterdam, 1991; \\
  New features of {\tt FORM}, ({\tt math-ph/0010025}).


\bibitem{Mario}
  M. Argeri, PhD Thesis, University of Bologna 2002.


\bibitem{CP}

  C.~M.~Carloni Calame, C.~Lunardini, G.~Montagna, O.~Nicrosini and 
  F.~Piccinini,
  \nptwoth{B584}{00}{459} (\hepph{0003268}). \\
  C.~M.~Carloni Calame,
  \pltwoth{B520}{01}{16} (\hepph{0103117}). \\
  C.~M.~Carloni Calame, G.~Montagna, O.~Nicrosini and F.~Piccinini,
  {\it Nucl. Phys. Proc. Suppl.}  {\bf 131} (2004) 48
  (\hepph{0312014}).\\
  C.~M.~Carloni Calame, talk at the "International Workshop e+e- collisions
  from $\Phi$ to $\Psi$", BINP, Novosibirsk. To appear in the proceedings
  of the conference.\\
  G.~Balossini, C.~M.~Carloni~Calame, G.~Montagna, O.~Nicrosini and 
  F.~Piccinini, in preparation.
                                                                                                                                               



\end{thebibliography}
\end{document}